\documentclass[aip,jcp,amsmath,amssymb,preprint,floatfix,notitlepage]{revtex4-1}

\usepackage{graphicx}
\usepackage{dcolumn}
\usepackage{bm}
\usepackage{placeins} 
\usepackage{tabularx}
\usepackage[version=3]{mhchem} 
\usepackage{array}
\usepackage{natmove}
\usepackage{mciteplus}
\usepackage[caption=false]{subfig}
\usepackage{graphicx}
\usepackage{xr}
\usepackage{xcolor}
\usepackage[normalem]{ ulem }
\usepackage{soul}

\makeatletter
\providecommand*{\input@path}{}
\g@addto@macro\input@path{{./Figures/}}
\makeatother

\graphicspath{{.}{Figures/}} 

\begin{document}
\title{Atomistic simulations of molten carbonates: thermodynamic and transport properties of the \ce{Li2CO3}--\ce{Na2CO3}--\ce{K2CO3} system}

\author{Elsa Desmaele}\email{elsa.desmaele@gmail.com}
\affiliation{Sorbonne Universit\'e, CNRS, Laboratoire de Physique Th\'eorique de la Mati\`{e}re Condens\'ee, LPTMC, F75005, Paris, France}
\author{Nicolas Sator}
\affiliation{Sorbonne Universit\'e, CNRS, Laboratoire de Physique Th\'eorique de la Mati\`{e}re Condens\'ee, LPTMC, F75005, Paris, France}
\author{Rodolphe Vuilleumier}
\affiliation{PASTEUR, D\'epartement de chimie, \'Ecole normale sup\'erieure, PSL University, Sorbonne Universit\'e, CNRS, 75005 Paris, France}
\author{Bertrand Guillot}\email{guillot@lptmc.jussieu.fr}
\affiliation{Sorbonne Universit\'e, CNRS, Laboratoire de Physique Th\'eorique de la Mati\`{e}re Condens\'ee, LPTMC, F75005, Paris, France}

\date{\today}

\begin{abstract}
Although molten carbonates only represent, at most, a very minor phase in the Earth's mantle, they are thought to be implied in anomalous high-conductivity zones in its upper part (70--350~km). Besides the high electrical conductivity of these molten salts is also exploitable in fuel cells. Here we report quantitative calculations of their properties, over a large range of thermodynamic conditions and chemical compositions, that are a requisite to develop technological devices and to provide a better understanding of a number of geochemical processes.\\
To model molten carbonates by atomistic simulations, we have developed an optimized classical force field based on experimental data of the literature and on the liquid structure issued from ab initio molecular dynamics simulations performed by ourselves. In implementing this force field into a molecular dynamics simulation code, we have evaluated the thermodynamics (equation of state and surface tension), the microscopic liquid structure and the transport properties (diffusion coefficients, electrical conductivity and viscosity) of molten alkali carbonates (\ce{Li2CO3}, \ce{Na2CO3}, \ce{K2CO3} and some of their binary and ternary mixtures) from the melting point up to the thermodynamic conditions prevailing in the Earth's upper mantle ($\sim$ 1100--2100~K, 0--15~GPa). Our results are in very good agreement with the data available in the literature. To our knowledge a reliable molecular model for molten alkali carbonates covering such a large domain of thermodynamic conditions, chemical compositions and physicochemical properties has never been published yet.
\end{abstract}

\maketitle

\section{Introduction}
Carbonate melts receive an increasing interest in both fundamental and applied fields either in the view of developing technological devices (e.g. fuel cell technology)\cite{Chery2015,Chery2015rev,Cassir2012,Cassir2013,Lair2012} or for providing a better understanding of a number of geochemical processes (e.g. role of molten carbonates in the geodynamics of the Earth's mantle).\cite{Gaillard2008,Dasgupta2010,Dasgupta2013,Hammouda2015} The study of the physicochemical properties (thermodynamics, structure and transport) of, mainly alkali, molten carbonates, under atmospheric pressure has been initiated by the electrochemistry community a few decades ago.\cite{Janz1988} These systems were depicted as highly conducting liquids of low viscosity. From a geosciences viewpoint the focus is more recent and mainly set on alkaline-earth compositions under high pressures, whose properties are so far poorly known.\cite{Gaillard2008,Sifre20142,Kono2014} Indeed studies of these liquids at high $T-P$ require very specific devices adapted to the measurement a single observable. On the contrary, in a molecular dynamics simulation a number of properties can be obtained at once, even at extreme thermodynamic conditions. However molecular dynamics simulations require the knowledge of atomic interactions in the system under consideration. These interactions (the force field) can be deduced from electronic structure calculations (\emph{ab initio}) or based on a more empirical approach (classical molecular dynamics or simply referred to as MD in the following). The former method, although \emph{a priori} more accurate than the latter one is computationally much more demanding (by several orders of magnitude) and is thus inappropriate when the aim of the study is to span a large range of chemical composition and thermodynamic conditions. On the other hand, the relevance of a MD calculation relies on the quality of the implemented force field (FF) whose adjustment is system-specific. Since the first attempts by  \citeauthor{Janssen1990}\cite{Janssen1990,Tissen1990,Tissen1994} to model molten carbonates by two-body interactions, several modifications have been suggested. \citep{Habasaki1990,Koishi2000,Costa2008,Ottochian2016,Wilding2016} However the FF of \citet{Janssen1990} has some inherent defects. For instance the calculated pressure is shifted by $\sim$15~kbar,\cite{Ottochian2016} at the 1-bar experimental density. A decisive improvement towards an accurate description of the properties of molten carbonates in quantitative agreement with the experiments came with the force fields (FF) of \citet{Vuilleumier2014} and Corradini \emph{et al.}\cite{Corradini2016}. However these models are restricted in composition, to pure \ce{CaCO3} and to the \ce{Li2CO3}--\ce{K2CO3} (62:38 mol\%) eutectic mixture respectively. For the sake of conciseness and clarity of the comparisons with the experimental literature, alkaline-earth compositions will be investigated in a companion paper.\cite{moi} Here we present MD simulation results that sketch a systematic and consistent model of molten carbonates in the \ce{Li2CO3}--\ce{Na2CO3}--\ce{K2CO3} system over a large $T-P$ range ($\sim$ 1100--2100~K, 0--15~GPa). After presenting the details of the simulations, we discuss the thermodynamic properties (equation of state and surface tension), the liquid structure and the transport properties (diffusion coefficients, electrical conductivity and viscosity).

\section{Computational details}
\subsection{Ab initio molecular dynamics (AIMD)} 
The \emph{ab initio} molecular dynamics (AIMD) simulations were based on the density functional theory (DFT) within the Born-Oppenheimer framework. They were performed thanks to the free QUICKSTEP/CP2K software\cite{VandeVondele2005a} that applies a hybrid Gaussian plane-wave method.\cite{Lippert1997} For valence electrons, we used a double-zeta valence plus polarization (DZVP) basis set optimized for molecules,\cite{VdVHutter2007} while core electrons were replaced by the Goedecker-Teter-Hutter (GTH) norm-conserving pseudo-potentials.\cite{Goedecker1996,Hartwigsen1998,Krack2005} The cutoff for the electronic density was set to 700 Ry. Exchange and correlation interactions were accounted for by the gradient corrected BLYP functional\cite{Becke1988,Lee1988} using a semi-empirical D3 dispersion correction scheme with a 30~\AA\, cutoff.\cite{Grimme2010} All DFT calculations were run in the $NVT$ ensemble with the temperature set constant by a Nos\'e-Hoover thermostat.\cite{Nose1984a,Nose1984b} Timestep was fixed to 0.5 fs, and simulation run was of the order of 13--48 ps. \\
The starting configurations were issued from classical MD runs using a guess force field. \ce{Na2CO3} and \ce{K2CO3} simulations were performed at state points corresponding to their atmospheric melting point,\cite{vangroos1990,Klement1975} that is respectively (1140~K, 1.97~g/cm$^3$) and (1190~K, 1.90~g/cm$^3$), where densities were set to their experimental values.\cite{Liu2003} The pressure calculated along both runs is 0~$\pm$~0.5~GPa, as expected. To investigate the high pressure behavior, a third simulation was performed for \ce{K2CO3} at 1773~K and 2.41~g/cm$^3$ leading to a calculated pressure of 6~$\pm$~1.5~GPa. Table~S1 summarizes the simulation details of the three AIMD simulations. Note that the AIMD simulation of Corradini \emph{et al.}\cite{Corradini2016} for the \ce{Li2CO3}--\ce{K2CO3} eutectic mixture uses similar methodological features, and hence it will be considered as a benchmark for Li-bearing compositions investigated here by MD. 
\subsection{Classical molecular dynamics (MD)}
    Classical MD simulations were carried out using the free DL\_Poly\_2 software. \cite{Smith1996}
    The timestep was set to 1 fs and the studied systems were composed of $N \simeq$ 2000 atoms. To evaluate the equation of state (EoS) as discussed in section \ref{seos}, calculations in the $NPT$ ensemble were performed. The total simulation time was 0.9~ns, including a 0.5~ns equilibration with a Nos\'e-Hoover thermostat, to reach a statistical accuracy on the density of about $\Delta n/n \leqslant \pm 1~\%$. For transport coefficients, production runs of 10 to 30~ns were performed in the $NVE$ ensemble. Structure data were extracted from the same simulations. Note that for a better comparison with the AIMD-generated pair distribution functions (see section~\ref{sff}), some MD simulations were also specifically run on systems having the same density and number of atoms.

\section{Force field}
\label{sff}
\begin{table}[th]
\begin{center}
\begin{tabular}{|l|l|r|l|r|r|} 
\hline
 $i$			  &	$j$	& $A_{ij}$ 	(kJ/mol)	& $\rho_{ij}$ (\AA) &  $C_{ij}$ (\AA$^6$/mol) & $q_i$ (e)	\\ \hline 
Li            & O   & 3.0	$\times 10^5$	& 0.2228 	& 1210		& +0.82101\\\hline
Na           & O	& 11	$\times 10^5$ 	& 0.2228 	& 3000	    & +0.82101	\\\hline
K             & O	& 9.0 $\times 10^5$ 	& 0.2570 	& 7000     & +0.82101		\\\hline
O 	       & O 	& 5.0	 $\times 10^5$	& 0.252525	& 2300	    & $-$0.89429 	\\\hline
C             & O   & 0                             & 1           & 0        &  +1.04085 \\
\hline
\end{tabular}
\end{center}
\caption{Intermolecular potential parameters and partial charges. Note that the Born repulsion parameters between oxygen atoms of a same carbonate ion are $A_{\rm{OO}}^{intra-\rm{CO}_3}= 2 611 707.2$~kJ/mol and $\rho_{\rm{OO}}^{intra-\rm{CO}_3}= 0.22$~\AA.}
\label{tffsh}
\end{table}
The empirical force field is derived from an interaction potential summing the pair interactions between the atoms of a same carbonate molecule (intramolecular potentials $V_{ij}^{intra}$) and those between atom pairs belonging to different ions (intermolecular potentials $V_{ij}^{inter}$).  \\
The carbonate molecule is featured as flexible and non-dissociative by introducing the following pair potential between C and O atoms:
\begin{equation}
V_\text{CO}^{intra-\rm{CO}_3}(r_\text{CO})=\frac{1}{2}{k_\text{CO}}(r_\text{CO}-{r_{0,\text{CO}}})^2+{{q_\text{O} q_\text{C}}}/{4 \pi \epsilon_0 r_\text{CO}} \enspace .
\end{equation}
The force constant ${k_\text{CO}}=6118.17$~kJ/mol and the harmonic equilibrium distance ${r_{0,\text{CO}}}=$ 1.30~\AA \, are adjusted so as to obtain a mean C-O distance of 1.29~\AA.\cite{Antao2010,Vuilleumier2014} Moreover the coulombic interactions are implemented in order to prevent the atomization of the molecule at high pressure. The oxygen atoms of a same molecule repel one another via the pair potential:
\begin{equation}
V_\text{OO}^{intra-\rm{CO}_3}({r}_\text{OO})={A_
\text{OO}^{intra-\rm{CO}_3}}\exp(-{r}_\text{OO}/{{\rho_\text{OO}^{intra-\rm{CO}_3}}})	\enspace .
\end{equation}
The latter one ensures on average the planarity of the carbonate ion, where ${A_\text{OO}^{intra-\rm{CO}_3}}=2 611 707.2$~kJ/mol and ${\rho_\text{OO}^{intra-\rm{CO}_3}}=$ 0.22~\AA.
The intermolecular interactions are ruled by the pair potential:
\begin{equation}
V_{ij}^{inter}({r}_{ij})={A_{ij}}\exp(-{r}_{ij}/{\rho_{ij}})-{{C_{ij}}}/{{r}_{ij}^6}+{{q_iq_j}}/{4\pi\epsilon_0{r}_{ij}} \enspace ,
\end{equation}
with $i,j$= O, C, Li, Na, K. The van der Waals part (first two terms) of this potential is set to zero between cations, thus enabling a straightforward transferability of the force field to any carbonate mixture in the Li--Na--K system.
Parameters of the FF are summarized in Table~\ref{tffsh}. While the charges were constrained by the FF of \citet{Vuilleumier2014}, the van der Waals interactions have been benchmarked on experimental volumetric studies and on AIMD structure data (note that dispersion interactions have little effect on the shape of the pair distribution functions, but affects strongly the density). Thus atmospheric density and compressibility\cite{Liu2003} are accurately reproduced by our model as discussed in section~\ref{sthermo}.
The suitability of the force field was further constrained by fitting at best the pair distribution functions of AIMD simulations, obtained either by Corradini \emph{et al.}\cite{Corradini2016} for the \ce{Li2CO3}--\ce{K2CO3} eutectic mixture of ratio 0.62:0.38 mol\% (LKe), or specifically calculated by ourselves for \ce{Na2CO3} and \ce{K2CO3}. The agreement between MD simulations and AIMD structure data is excellent (Figure~S1). Notice that for LKe in particular, the fitting of the AIMD curves by MD is at least as good, as when using Corradini \emph{et al.}'s FF (Figure 1 of reference \onlinecite{Corradini2016}). The microscopic structure is thoroughly described in section~\ref{sstruct}.\\
To be complete, and although the present FF is not aimed at studying the crystalline phase, we checked its ability to reproduce the experimentally determined structures of the monoclinic crystals of \ce{Li2CO3}, \ce{Na2CO3} and \ce{K2CO3} at room conditions. The densities calculated by anisotropic relaxation ($NST$ ensemble) of these structures are 2.12 g/cm$^3$ for \ce{Li2CO3}, 2.50 g/cm$^3$ for \ce{Na2CO3} and 2.45 g/cm$^3$ for \ce{K2CO3} to be compared to 2.10, 2.54 and 2.44~g/cm$^3$, respectively\cite{Idemoto1998,Arakcheeva2010,Gatehouse1973} (see also Table~S2 for lattice parameters).

\section{Thermodynamics}\label{sthermo}
\subsection{Equation of state}
\label{seos}
\begin{table}
\begin{center} 
\begin{tabular}{|l|c|c|c|c|} 
\hline
 & \multicolumn{2}{c|}{$n_{1100\,K}$ (g/cm$^{3}$)} & \multicolumn{2}{c|}{$\alpha$ (10$^{-4}$g/cm$^{3}$/K)}\tabularnewline
\cline{2-5} 
 & MD  & Exp  & MD  & Exp \tabularnewline
\hline 
\ce{Li2CO3}		    			& 1.796		& 1.792 $^\blacklozenge$    & 	4.251		&   3.729 $^\blacklozenge$      \\
                                    &           &  1.793 $^\blacktriangle$  &               &   3.385 $^\blacktriangle$     \\  
                                    &           & 1.795 $^\bigstar$		    &               & 3.459 $^\bigstar$             \\ \hline
\ce{Na2CO3}			    			& 1.991		& 1.986 $^\blacklozenge$    & 5.515			&   4.487$^\blacklozenge$       \\
                                    &           &   1.980 $^\blacktriangle$ &               & 4.696 $^\blacktriangle$       \\
                                    &           &  1.988 $^\bigstar$	    &               & 4.511 $^\bigstar$             \\ \hline
\ce{K2CO3}		    		       	& 1.931		& 1.928 $^\blacklozenge$    & 	4.222	    &   4.421 $^\blacklozenge$      \\
                                    &           &  1.925 $^\blacktriangle$  &               & 3.563 $^\blacktriangle$      \\
                                    &           &  1.939 $^\bigstar$		&               &  4.615 $^\bigstar$	        \\ \hline
\end{tabular}
\end{center}
\caption{Density and thermal expansion coefficient ($n=n_{1100\, K}-\alpha(T-1100)$) from MD simulations and comparison with the experimental literature: references \onlinecite{Janz1988} ($^\blacklozenge$), \onlinecite{Kojima2009} ($^\blacktriangle$) and \onlinecite{Liu2003} ($^\bigstar$).} 
\label{t_compn}
\end{table}
\begingroup
\squeezetable
\begin{table*}
\begin{tabular}{|l | r | c | c | c | c | c | c | c | c | c |}  
\hline
		&$T_{ref}$ 	& $n_{ref}^0$ 		& $\alpha_0$	&	$\alpha_1$	&  $K_{ref}^0$&  $b_1$ 		&$b_2$		&	 $K^{\prime 0}$		& $T$	&	$P$ \\
		&(K) 	& ($\text{g/cm}^3$)	& (K$^{-1}$)	& 	(K$^{-2}$)	& (GPa) 	  & (K$^{-1}$)	& (K$^{-2}$)& 							& (K)	&	(GPa) 
\\  \hline
Li$_2$CO$_{3}$	& 1100	& 1.80	&	-2.57$\times$10$^{-4}$ & 2.41$\times$10$^{-8}$	& 12.56	& 9.4$\times$10$^{-4}$	&	8.5$\times$10$^{-7}$ &	8.0 	& 1100-1923		& 0-6	\\		\hline	 	
Na$_2$CO$_{3}$	& 1140	& 1.97	&	-2.64$\times$10$^{-4}$ & -4.61$\times$10$^{-8}$	& 9.01	& 11.4$\times$10$^{-4}$	&	12.7$\times$10$^{-7}$ &	8.5 & 1100-2073	&	0-15	\\	 \hline	 	
K$_2$CO$_{3}$  	& 1190	& 1.89	&	-2.56$\times$10$^{-4}$ & -10.05$\times$10$^{-8}$& 5.96	& 12.0$\times$10$^{-4}$	&	12.2$\times$10$^{-7}$ &	7.6 & 1100-2073		&	0-15		\\ \hline
\end{tabular}
\caption{Parameters of the third-order Birch Murnaghan equation of state and $T-P$ domain of validity for each pure end-member composition. See text and equations~(\ref{eBMEOS}), (\ref{enT}) and (\ref{ekT}) for details.}
\label{teosparam}
\end{table*}
\endgroup
\begin{table}
\begin{center} 
\begin{tabular}{|l|rrr|} 
\hline
        &                  LKe    & &    LNKe \\ \hline
        
 $n^0_{Exp}$ (g/cm$^3$) & 1.88    &  &  1.91\\
 $n^0_{MD}$ (g/cm$^3$) & 1.88 &(1.85$^\dagger$) & 1.91\\
 $n^0_{MD, \, mix}$ (g/cm$^3$) & 1.87  & & 1.91  \\ \hline
 $K^0_{Exp}$ (GPa)	  & 8.9 &  & 9.1\\
 $K^0_{MD}$ (GPa) & 8.6	& (6.81$^\dagger$) & 8.9 \\
 $K^0_{MD, \, mix}$ (GPa)  &  8.6 & & 8.8 \\ \hline
\end{tabular}
\end{center}
\caption{Density and bulk modulus of the LKe and LNKe mixtures at $T=$1100~K calculated either by MD simulation ($n^0_{MD}$, $K^0_{MD}$) or by using ideal mixing rules ($n^0_{MD, \, mix}$, $K^0_{MD, \, mix}$). Experimental data $n^0_{Exp}$, $K^0_{Exp}$ are given for comparison.\cite{Liu2003} Results issued from the MD model of Corradini \emph{et al.}\cite{Corradini2016} ($^\dagger$) are given in parenthesis.}
\label{tid} 
\end{table}
\begin{figure}
\centering
{\includegraphics{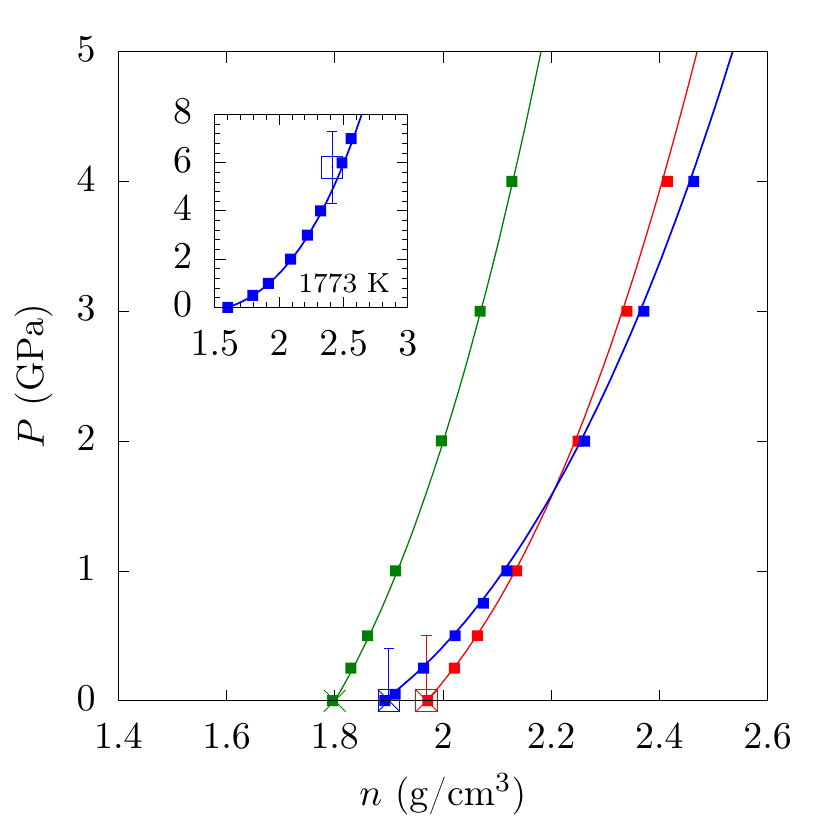}}				                     			
\caption{Compression isotherms of molten \ce{Li2CO3} (green), \ce{Na2CO3} (red) and \ce{K2CO3} (blue) at their respective melting temperature: plain squares are MD simulations, open squares are AIMD simulations and crosses are experimental data of \citeauthor{Liu2003}\cite{Liu2003}. The isotherms (full curves) were obtained by interpolating the MD points with the third-order Birch-Murnaghan equation (see text). In the inset are shown the 1773~K MD isotherm for K$_2$CO$_3$ and the AIMD result at the same temperature and 2.41~g/cm$^{3}$.}
\label{feos}
\end{figure}
The density of molten alkali carbonates at atmospheric pressure has been reported by several experimental studies \cite{Janz1961,Spedding1970,Janz1988,Zhu1991,Kojima2009,Liu2003,Kojima2003,Kojima2008,Kojima2009}. For the pure end-members, \ce{Li2CO3}, \ce{Na2CO3} and \ce{K2CO3} a global consistency between these studies is found. Using the present FF, MD fits very well the above studies considering the small experimental uncertainties and the very limited temperature range ($\sim$ 100~K) investigated at 1 bar (Table~\ref{t_compn}). In this context notice that the model of Corradini \emph{et al.}\cite{Corradini2016} recovers reasonably the 1-bar density of \ce{K2CO3}, but it doesn't perform as well for pure \ce{Li2CO3}, the deviations reaching $-$5~\%.\\
Despite the geochemical interest for assessing the buoyancy of carbonated fluids in the mantle, no systematic measurements of liquid density at high pressure have been reported yet, except for the study of \citeauthor{Dobson1996}\cite{Dobson1996}, for \ce{K2CO3} at 4~GPa. Besides the study of \citeauthor{Dobson1996} has since been challenged by Liu \emph{et al.}\cite{Liu2007}. These latter authors  provided estimates of high pressure volumetric behavior for \ce{K2CO3},\cite{Liu2003,Liu2007} by using the third-order Birch Murnaghan equation of state (BMEoS)\cite{Birch1947} that is well-known for its accuracy in the modeling of the compressibility of molten and crystalline silicates \cite{Rigden1989,Guillot2007}. It writes:
\begin{equation}
\begin{split}
       P=&\frac{3}{2}K^{0}(T)\Bigg(\Big(\frac{n}{n_T^{0}}\Big)^{7/3}-\Big(\frac{n}{n_T^{0}}\Big)^{5/3}\Bigg)\\
  &\times\left(1-\frac{3}{4}(4-K^{\prime 0})\Bigg(\Big(\frac{n}{n_T^{0}}\Big)^{2/3}-1\Bigg)\right) \enspace ,
\end{split}
\label{eBMEOS}\end{equation}
where $n_T^{0}$ is the atmospheric density at temperature T, $K^{0}_T$ the bulk modulus and  $K^{\prime 0}$ its pressure derivative at 1~bar. Such an equation expressing the density of the carbonated liquid at any thermodynamic condition is appealing, although $K^{\prime 0}$ has so far only been derived by indirect thermodynamic considerations (see below).\\
To determine the BMEoS of liquid \ce{Li2CO3}, \ce{Na2CO3} and \ce{K2CO3} from MD, their density was calculated along several isotherms at different pressures. The values of $n_T^{0}$ and $K^{0}(T)$ were fitted by using empirical temperature dependencies: 
\begin{equation} n_T^{0}=n_{ref}^0 e^{\int_{T_{ref}}^{T} - (\alpha_0+\alpha_1 T) \, \mathrm dT} \,, \label{enT}\end{equation}
and 
\begin{equation}
K^0_T=\frac{K_{ref}^0}{1+b_1(T-T_{ref})+b_2(T-T_{ref})^2}  \enspace .
\label{ekT}
\end{equation}
The reference temperature $T_{ref}$ is defined for numerical purposes and set to the atmospheric melting point for each composition (Table \ref{teosparam}), $n_{ref}^0$ and $K_{ref}^0$ are the density and bulk modulus at 1-bar and $T_{ref}$. Values of the parameters $\alpha_0$, $\alpha_1$, $b_1$ and $b_2$ are given in Table~\ref{teosparam}. \\
The excellent agreement between calculated points and the fitted isotherms (Figures~\ref{feos} and~S2) illustrates the ability of the BMEoS to describe carbonate melts, at least on the investigated pressure-temperature range. Furthermore, the obtained BMEoS for \ce{Li2CO3}, \ce{Na2CO3} and \ce{K2CO3} match not only the experimental values at 1~bar, but also those of our AIMD calculations. However, AIMD simulations are performed at constant density with a relatively small system size, and display large pressure fluctuations which yield uncertainties, typically of the order of $\sim$~1~GPa  (inset of Figure~\ref{feos}). \\
The 1-bar bulk modulus $K^0_T$ (inverse of the isothermal compressibility) has been  experimentally determined by sound-speed measurements for a number of mixtures.\cite{OLeary2015} It is found in excellent accordance with the ones resulting from the parameterization of the BMEoS using MD (e.g. in Table \ref{tid}). $K^0_T$ consistently decreases with the cationic radius (Table \ref{teosparam}), unlike its P-dependence, $K^{\prime 0}$, that is  almost invariant with chemical composition. Using a fusion curve analysis, \citet{Wang2016} have constrained $K^{\prime 0}$ for liquid K$_2$CO$_3$, at pressure below 5 GPa, to 14.4  instead of 7.6 by MD. This disagreement can be explained by the indirect method used by the authors to obtain $K^{\prime 0}$ from the Clapeyron curve. In this method (i) many thermodynamic data (which introduce uncertainties) are required, and (ii) the melting curve is not accurately known in the light of the differences between the studies of \citeauthor{Klement1975}\cite{Klement1975}, \citet{Wang2016} and \citeauthor{Li2015}.\cite{Li2015}\\
For carbonate mixtures, discrepancies on atmospheric density measurements can reach up to 5~\%.\cite{Janz1961,Spedding1970,Janz1988,Liu2003,Kojima2003,Kojima2008} The values from the study of Liu \emph{et al.}\cite{Liu2003} seem to be of first relevance since the authors have proven their reproducibility, have accounted for surface tension corrections and have tested their apparatus on standard liquids.
The comparison of our MD results with their study shows the excellent transferability of our force field to the determination of the density of mixtures, as illustrated in Table~\ref{tid} on some selected compositions, namely the eutectic ternary mixture (\ce{Li2CO3}, \ce{Na2CO3} and \ce{K2CO3} in proportions 43.5:31.5:25 mol\%,\citep{Maru1984} LNKe) and a binary eutectic mixture (\ce{Li2CO3} and \ce{K2CO3} in proportions 62:38 mol\%, LKe). Note that the density of mixtures follows an ideal mixing rule as it was reported by Liu \emph{et al.}\cite{Liu2003}:
\begin{equation}
n_{mix}= \cfrac{\sum_i x_i {M}_i}{\sum_i x_i \bar{V}_i} \quad ,
\label{emelid}
\end{equation} 
where $x_i$ is the molar fraction of species $i$ of molar mass ${M}_i$ and molar volume $\bar{V}_i$.\\
For the bulk modulus, an ideal mixing rule,
\begin{equation}
K_{mix}= \frac{\sum_i x_i \bar{V}_i}{\sum_i x_i \bar{V}_i/K_i} \enspace , 
\label{emelidbeta}
\end{equation} 
is also verified (Table \ref{tid}). \\
Therefore the use of the above mixing rules for the density and the bulk modulus, combined with the fact that $K^{\prime 0}$ varies very little for the three end-members, leads to an accurate estimation of the density of mixtures in the system \ce{Li2CO3}--\ce{Na2CO3}--\ce{K2CO3} over a large range of $T-P$ conditions. For instance, considering the eutectic ternary mixture at 1500~K and 3~GPa, direct simulation gives a density of 2.15 g/cm$^3$, while the use of equations~(\ref{emelid}) and (\ref{emelidbeta}), with $K^{\prime 0}$ in the range $7.6-8.5$ gives 2.16 g/cm$^3$ after using equation~(\ref{eBMEOS}).

\subsection{Evolution of the liquid structure with temperature and pressure}\label{sstruct}
\begin{figure}
\centering
{\includegraphics{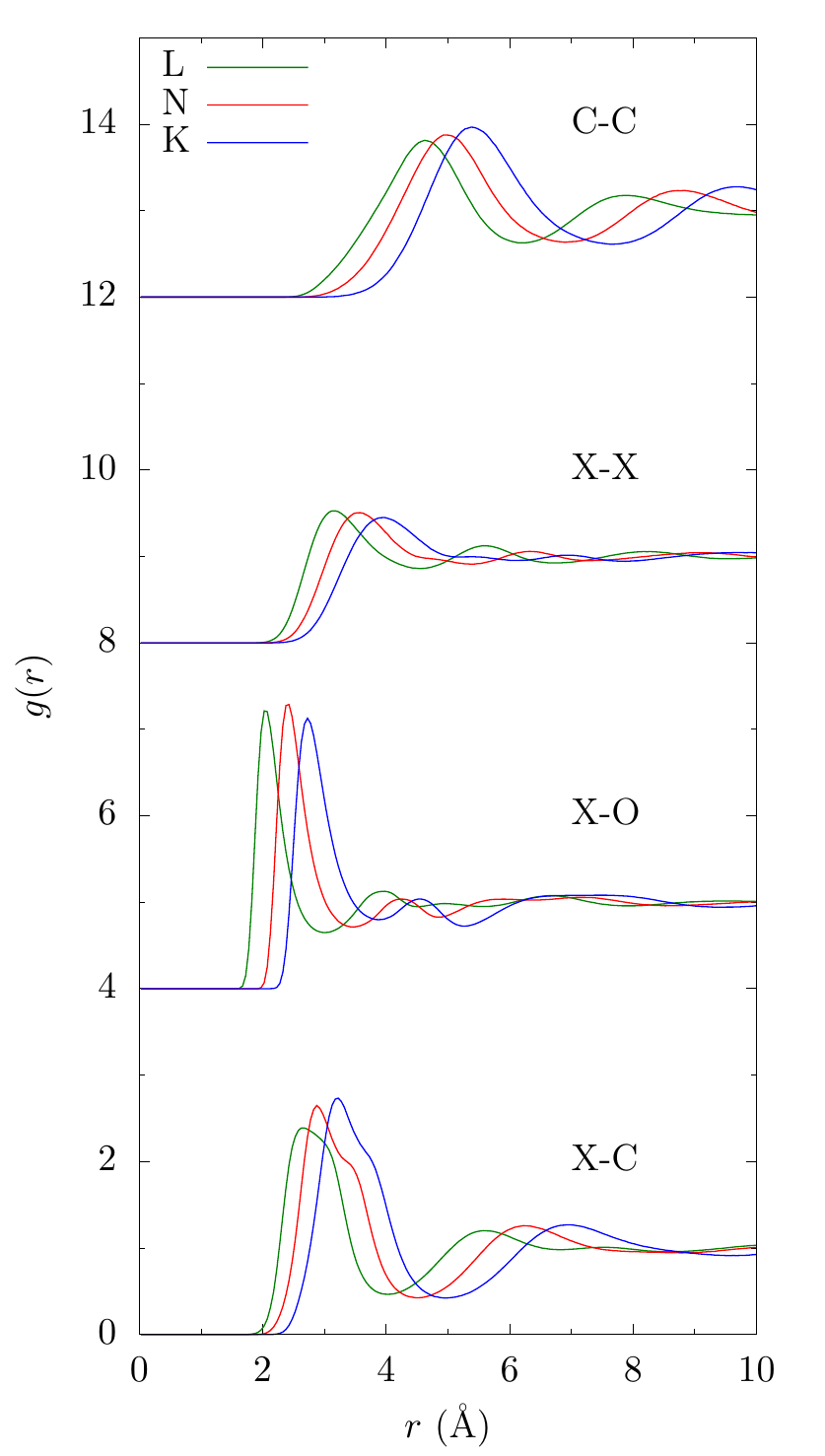}}                 
\caption{Pair distribution functions C$-$C, X$-$X, X$-$O and X$-$C where C and O are carbon and oxygen atoms of the \ce{CO3} groups and X$=$Li, Na or K, in \ce{Li2CO3} (L), \ce{Na2CO3} (N) and \ce{K2CO3} (K) at 1~bar and 1100~K. To help visualization the different PDFs were shifted vertically.}
\label{fstrcut}
\end{figure}
\begin{figure}
\centering
\includegraphics{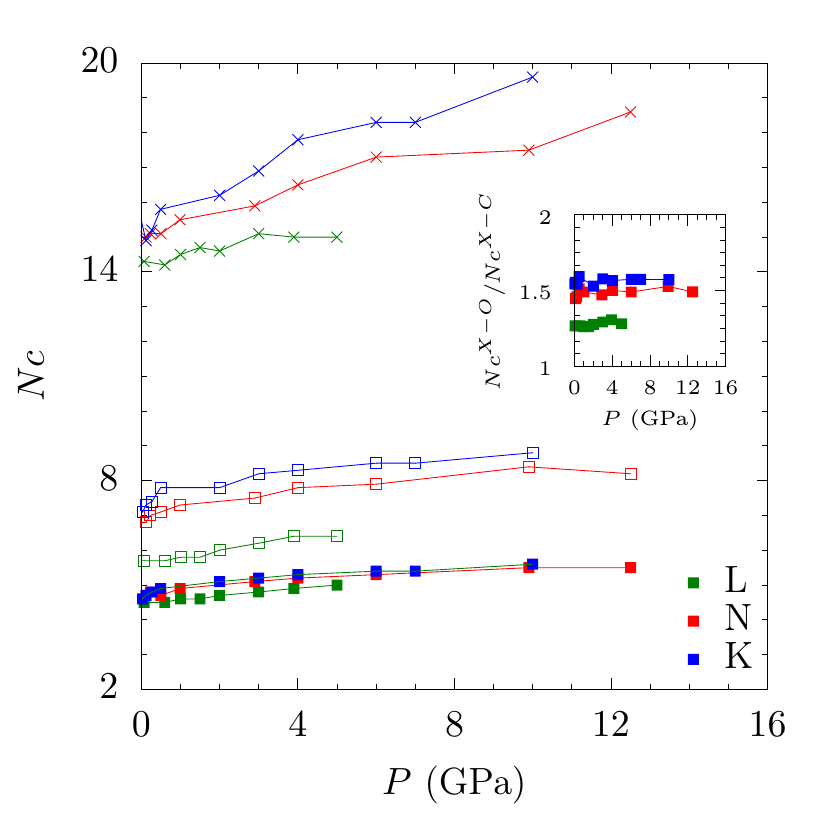}               
\caption{Evolution with pressure of the coordination number of the C$-$C (crosses), X$-$C (plain squares) and X$-$O (empty square) pairs where X=Li, Na or K, in: \protect\\
        \ce{Li2CO3} (L) at 1140~K (0~GPa), 1200~K (0.1~GPa), 1350~K (0.5--1~GPa) and 1500~K (1.5--5~GPa),\\
         \ce{Na2CO3} (N) at 1200~K (0--1~GPa), 1650 K (3--6 GPa) and 2073 K (10--12.5~GPa),\\
         \ce{K2CO3} (K) at 1200~K (0--0.5~GPa), 1500~K (2--4~GPa), 1773~K (6~GPa) and 2073~K (7--10~GPa).}
\label{fnc}
\end{figure}
\begingroup
\squeezetable
\begin{table*}
\begin{center}
\begin{tabular}{|l | c c c|  c c | c c | c c c| c c c|} \hline
				&$T$ 	&	$P$			& $n$ 				& $r^\text{X-X}$		&  	$N_c^\text{X-X}$	  	&  $r^\text{C-C}$ 	& $N_c^\text{C-C}$	&  $r^\text{X-C}$ 	& $N_c^\text{X-C}$  & $N_c^\text{C-X}$ &  $r^\text{X-O}$ 	& $N_c^\text{X-O}$ & $N_c^\text{O-X}$ \\ 
				&(K) 	&	(GPa)		& (g/cm$^{3}$) 	& (\AA)	        	& 			      	& (\AA)				& 				& (\AA)				& 	    & & (\AA)				& 		&	 \\ \hline
\ce{Li2CO3}	
				& 1100 & 0		& 1.80				&	3.15			& 	2.5			& 4.63 & 5.2	& 2.65			& 1.1			& 1.9	&		2.04		&	1.8	& 1.0 \\ 
				& 	   & 				& 					&	4.75			& 	12.4			& 6.22 & 14.3	& 4.03			& 4.5			& 9.0	&		3.03		&	5.9	 & 3.9\\ \hline
\ce{Na2CO3}	
				& 1100 & 0				& 1.94				&	3.56			& 	2.8				& 5.05 & 5.0	& 	2.89		 & 1.0			& 1.9	&		2.40		&	1.6	& 0.9 \\
				& 	   & 				& 					&	5.38			& 	13.7				& 6.90 & 15.0	& 	4.48		& 4.7 			& 9.4	&		3.48		&	6.8	& 4.6\\ \hline
\ce{K2CO3}
				& 1100 & 0				& 1.89				&	3.94			& 	2.9				& 5.40 & 4.8	& 	3.25		& 1.1			& 2.3	&		2.73		&	1.5	& 1.2	\\	
				&	   &				& 					&	6.20		& 	21.6				& 7.70 &	15.5 & 	5.05		& 4.8 			& 9.5	&		3.82		&	7.2	& 5.0	\\ \hline
\end{tabular}
\end{center}
\caption{First neighbors in molten carbonates at 1100 K: distance $r$ and coordination numbers $N_c$ (with X=Li, Na or K and O and C are the oxygen and the carbon atom of \ce{CO3}, respectively). The first line reports values corresponding to the first maximum of the pair distribution function, the second line values evaluated at its first minimum.}
\label{tnc}
\end{table*}
\endgroup
Despite their liquid nature, related to a long-range disorder, molten carbonates show a residual local ordering coming from the competition between ionic, dispersive and steric forces. Among these, the coulombic forces are essential as they impose an alternation of charges of opposite signs, as qualitatively confirmed by X-ray and neutron diffraction studies of molten carbonates.\cite{Zarzycki1961,Kohara1998}  Thus the pair distribution functions (PDF) of X$-$C and C$-$C (where C is the carbon atom of \ce{CO3} and X=Li, Na or K) are nearly in opposition of phase (Figure~S1). However the structure of molten carbonates is complicated further, in comparison to simple ionic liquids, by the peculiar shape of the carbonate ion and by the anion-cation asymmetry of charge. The anion is surrounded by cations that subsequently attract carbonate anions into a second coordination shell. In turn the anions encircle the cation by pointing one or two oxygen atoms in its direction (see Table~\ref{tnc}). Actually the number of pointing O of each carbonate is barely sensitive to the size of the cation ($\sim$1.5 in each case, Table~\ref{tnc}). The number density progresses as follows at 1~bar and 1100~K:  $\rho_{\ce{Li2CO3}}=$ 24.4 mol/L $>$ $\rho_{\ce{Na2CO3}}=$ 18.7 mol/L $>$ $\rho_{\ce{K2CO3}}=$ 14.0 mol/L. Therefore some structural features can be described as a simple homothetic transformation upon volumetric shrinkage from \ce{K2CO3} to \ce{Li2CO3} (\emph{e.g.} C$-$C pair on Figure \ref{fstrcut}). Interestingly, this simple behavior leads to a pseudo-ideal mixing rule of the cation-carbonate pair distribution functions. As a matter of fact the X$-$C PDFs associated with the pure end-members are quasi-identical to the ones observed in the corresponding mixtures (not shown).\\
According to the parametrization procedure we used, the PDFs from MD compare very well with the ones from AIMD (Figures~S1a, S1b and S1c). Moreover the quality of the MD description is preserved at high $T-P$ as illustrated by the example of \ce{K2CO3} (Figure~S1d). Actually the agreement between MD and AIMD is even better under pressure, except for the cation-cation (X$-$X) pair. With regard to the structural changes occurring with increasing pressure and temperature, it is noteworthy that on the $T-P$ range under study the local structure in the first shell depends very little on the temperature. For example, for \ce{K2CO3} at 4~GPa, the cation-oxygen $N_c^\text{X-O}$ coordination number is 8.3 at 1500~K and 8.0 at 2073~K. Consequently the evolution of the coordination numbers $N_c^\text{C-C}$, $N_c^\text{X-C}$, $N_c^\text{X-O}$ at increasing pressure and temperature can thus be assigned essentially to pressure (Figure \ref{fnc}). Surprisingly for all compositions, $N_c^\text{X-O}$ evolves very little with pressure, it increases somewhat before levelling off. The ratio $N_c^\text{X-O}$/$N_c^\text{X-C}$ plotted in the inset of Figure \ref{fnc} is indicative of the orientation of the carbonate ions around the cations (there are 1.3 oxygen atoms pointing towards Li, $\sim$1.5-1.6 towards Na and K) and is virtually independent of pressure. Although the pressure evolution of the coordination number of the C$-$C pair is a bit steeper, there is nothing remarkable in this evolution. This is at variance with the hypothesized reminiscence in the liquid structure of crystalline polymorphs at the corresponding pressure.\cite{Hudspeth2018} The absence of a significant structural rearrangement of molten carbonates upon pressure should translate in smoothly varying transport coefficients over this pressure range.\\
\citet{Wilding2016} have studied the structure of \ce{Na2CO3} at 1 bar and from 1100 to  1750~K, using MD and a FF adapted from \citet{Tissen1990}. The authors assumed the formation of a low-dimensional network made of short carbonate-cation-carbonate chains argued by the presence of a shouldering in the first peak of  $g_{\text{C-C}}(r)$ at 2.3~\AA. Such a shouldering does not appear in our MD study, neither in the pair distribution function generated by AIMD. We believe that the aforementioned structural feature is due to the inaccuracy of the \citet{Janssen1990} force field, even when supplemented by an harmonic intramolecular potential for \ce{CO3}, as used by \citet{Wilding2016}

\subsection{Surface Tension}
\begin{figure}
\centering
\includegraphics{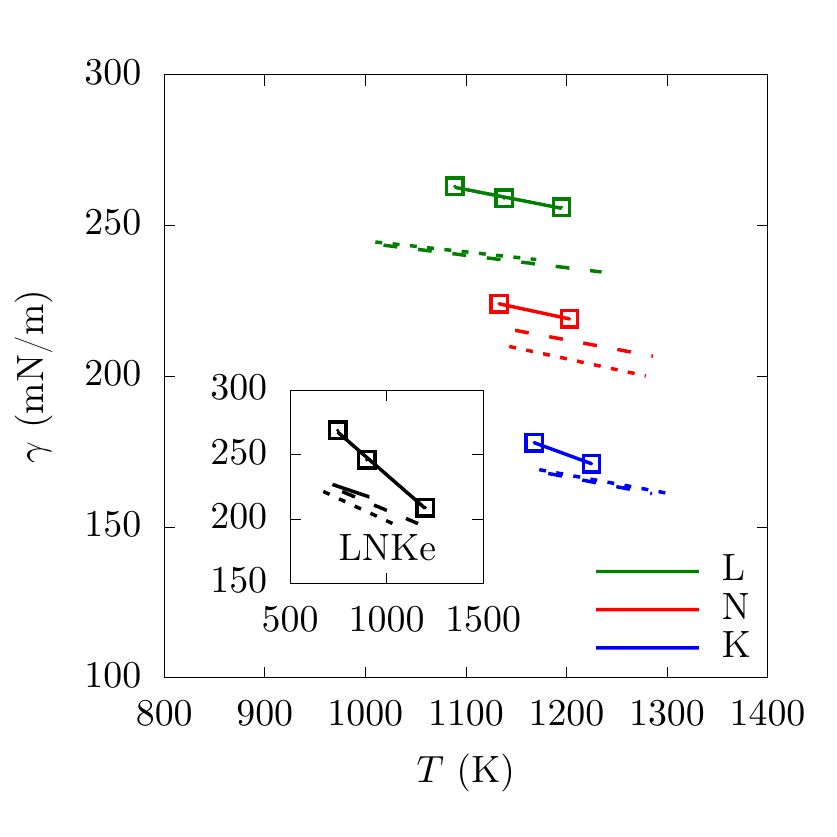}                                  \caption{Surface tension of the end-members \ce{Li2CO3} (L), \ce{Na2CO3} (N) and \ce{K2CO3} (K) calculated in MD (squares and plain lines as a guide to the eye) and measured experimentally (dashed line\cite{Kojima2009} and dotted line~\cite{Janz1988}). In the inset are the values for the ternary eutectic mixture (LNKe) calculated in MD (squares and plain lines) and measured experimentally  (dashed line\cite{Kojima2009}, long dash\cite{Igarashi1992} and dotted line\cite{Ward1965}).}
\label{fcompgamma}  
\end{figure}
Surface tension can be evaluated by explicitly simulating the coexistence of the liquid with its vapor along a plane interface,\cite{Alejandre1995,Aguado2002,Salanne2007} and by calculating the long time limit of the average of some components of the stress tensor. More precisely, assuming that the $z$ axis is perpendicular to the interface, the surface tension is given by:\cite{Kirkwood1949}
\begin{equation}
\centering
\gamma=\underset{t \to \infty}{\lim} \gamma(t)\,,\, \text{with}\,\gamma(t) = \frac{L_z}{2}\Big\langle\Pi_{zz}-\frac{1}{2}\big(\Pi_{xx}+\Pi_{yy}\big)\Big\rangle \enspace ,
\label{est}
\end{equation}
where $L_z$ is the length of the simulation box and $\Pi_{xx}$, $\Pi_{yy}$ and $\Pi_{zz}$ are the diagonal components of the stress tensor (see equation~(\ref{estresst}) in next section). We studied several compositions at different temperatures, but for the sake of clarity, we chose to present the result of only a few of them to picture the main trends (Figure~\ref{fcompgamma}). Our model recovers the right order of magnitude of experimental measurements of the surface tension, but seems to overestimate it systematically by 10--20~\%. Moreover the temperature dependencies are satisfying although one notices an overestimation of the activation energy in the case of the LNKe mixture. The hierarchy between the different compositions is respected: the bigger the cation, the smaller the surface tension. Thus the surface tension decreases when the asymmetry of size between the cation and the anion decreases.\cite{Gonz2005}\\
As mentioned above, the overestimation of the MD-calculated surface tension seems to be systematic (of the order of +10--15~\% compared with \citet{Kojima2009}). As our FF describes with accuracy the structure and the density as well as other properties (EoS, viscosity) which depend also strongly on cohesive forces\cite{Marchand2011} (see next section), this observation is somehow a surprise and raises concerns on possible systematic errors due to the simulation parameters. Some authors have pointed out that away from the interface the bulk, instead of showing isotropic properties, could wrongly contribute a non-zero amount to equation (\ref{est}).\cite{SPensado} We tested the possible effect of this anomalous contribution in the case of the \ce{Li2CO3}--\ce{Na2CO3}--\ce{K2CO3} equimolar mixture at 900~K, by calculating the surface tension with a liquid depth increased by a factor 7 ($z$ dimension from $\sim$ 29 to 200~\AA), all else being equal. This resulted in a value of $\gamma$ that is almost unchanged or maybe slightly decreased (from 239 $\pm$~3 to 232 $\pm$~6), which suggest a weak contribution of the bulk. We have tested several other parameters of the simulation which could affect the calculation of the surface tension namely, the size of the simulation box, the truncation cutoff of the van der Waals potential, and the parameters of the Ewald summation.\cite{Gonz2005} Although the calculated value changes a little with these parameters, the variations seem too weak to explain the discrepancy with the experimental data (calculated values differ by less than 5~\%). Larger scaling tests with much greater simulations boxes would be needed to completely rule out the role of finite size effect. However, it could be that our FF is less suitable to describe interfacial (rather than bulk) properties as it possibly fails to render some polarization effects. In any case, the discrepancies with the experiments we obtained are similar in magnitude with previous MD studies of ionic liquids.\cite{Aguado2002,Salanne2007} and this first attempt of simulating the surface tension of molten carbonates by MD leads to good qualitative predictions.

\section{Transport properties}
Aiming at a systematic description of transport properties as a function of chemical composition, temperature and pressure, a large number of simulations were performed at different thermodynamic conditions for pure  \ce{Li2CO3}, \ce{Na2CO3} and \ce{K2CO3}, as well as for some binary and ternary compositions. Each run was carried out in the $NVE$ ensemble for a duration of 10--20~ns, which is sufficient to investigate the diffusive regime and to calculate accurately the self-diffusion coefficients $D_s$ of each chemical species ($s=$~Li, Na, K and CO$_3^{2-}$), the  electrical conductivity $\sigma$ and the viscosity $\eta$, defined respectively by:\cite{AllenTild,Hess2002}
\begin{eqnarray} 
D_s&=& \lim_{t\to \infty} \frac{1}{6t} \,\frac{1}{N_s}\sum_{i=1}^{N_s}
\Big \langle \vert \mathbf{r}_i(t)-\mathbf{r}_i(0)\vert ^{2}
\Big\rangle \enspace, \label{eD}\\
\sigma&=& \lim_{t\to \infty} \frac{1}{6t} \, \frac{e^2}{k_{\rm B}TV}
\Big\langle \Big\vert\sum_{i=1}^N z_i
\big(\mathbf{r}_i(t)-\mathbf{r}_i(0)\big)\Big\vert ^{2}
\Big\rangle \enspace, \label{econd} \\ \eta&=& \lim_{t\to \infty}
\frac{V}{k_{\rm B}T}\int_{0}^{t} {\rm d}\tau \Big\langle \Pi_{\alpha
  \beta}(\tau)\cdot \Pi_{\alpha \beta}(0) \Big\rangle \enspace,
\end{eqnarray} 
where the off-diagonal pressure tensor components $\Pi_{\alpha \beta}(t)$, with $\alpha, \beta=x,y,z$ is given by:
\begin{equation}
\begin{split}
  \Pi_{\alpha \beta}(t)= \frac{1}{V}\sum_{i=1}^N &\Big(m_iv_{i\alpha}(t)v_{i\beta}(t)\\
  &+\sum_{j>i} r_{ij\alpha}(t).F_{ij\beta}(t) \Big)\enspace,    
\end{split}
\label{estresst}
\end{equation}
where $N$ is the total number of ions in the simulation box of volume $V$, $N_s$ the number of ions of species $s$, $k_{\rm B}$ the Boltzmann constant, $e$ the elementary charge, $m_i$ the mass of ion $i$ and $\mathbf{r}_i(t)$ its position, $v_{i\alpha}(t)$ the component $\alpha$ of its velocity and $F_{ij\beta}(t)$ is the component $\beta$ of the force exerted by ion $j$ on ion $i$, separated by distance $r_{ij}(t)$ at time $t$.  The ensemble average $\big\langle \dots \big\rangle$ is taken over many time origins. The parameter $z_i$ is the conduction charge of ion $i$ (i.e. $\sigma$ is effectively an ionic conductivity), chosen here as the formal charge, as it is usual for simple ionic liquids. \cite{Adams1977} Note that the optimal choice for the conduction charges, either formal or partial charges, is still an open question.\citep{Vuilleumier2014,Corradini2016}
The estimation of the self-diffusion coefficient $D_s$ relies on an average over the $N_s$ ions of a specific species $s$ (sum $\sum_{i=1}^{N_s}$ in eq (\ref{eD})), which ensures a small uncertainty of $\sim$1~\%. On the contrary, conductivity and viscosity are collective quantities that necessarily require a large number of atoms (and long time runs) to be correctly estimated with an error bar of 5--10~\% in this study. For this reason, AIMD estimations are today only crude and suffer from high uncertainties. As a matter of fact, until now only two classical MD studies have provided estimations of the viscosity and electrical conductivity of carbonate melts, namely \ce{CaCO3}\cite{Vuilleumier2014} and the \ce{Li2CO3}--\ce{K2CO3} eutectic mixture  (62:38 mol\%) .\cite{Corradini2016}\\
As usually observed for molten salts, the temperature dependence of the transport properties calculated by MD is well reproduced by an Arrhenius activation law. An activation volume $V_{a}^{X}$ is introduced to take into account the pressure dependence of the physical quantity $X=D_s$, $\sigma$ and $\eta$, as in references \onlinecite{Vuilleumier2014,Corradini2016}. The following expressions are used:
\begin{eqnarray}
D_s (P,T)&=&D_s^0 \, e^{-( E_{a}^{D_s} + P V_{a}^{D_s})
  /RT} \enspace, \label{eqDArrh}\\
\sigma (P,T)&=&\sigma^0 \, e^{- ( E_{a}^{\sigma} + P V_{a}^{\sigma} )
  /RT} \enspace, \label{eqSigmaArrh}\\
\eta (P,T)&=&\eta^0 \, e^{ ( E_{a}^{\eta} + P V_{a}^{\eta} )
  /RT} \enspace,\label{eqViscoArrh}
\end{eqnarray}
where $E_{a}^{X}$ is the activation energy associated to the physical quantity $X$. The values of $X^0$, $E_{a}^{X}$ and $V_{a}^{X}$ were determined by fitting the MD data for the three end-members \ce{Li2CO3}, \ce{Na2CO3} and \ce{K2CO3} (see Figures~S3, S4, S5, S6 and S7). These values are given in Tables \ref{tfitdiff} and \ref{tfittransp}, for $D_s$, $\sigma$ and $\eta$ and for the pressure and temperature range mentioned in Table~\ref{teosparam}. \\
A crucial advantage of MD simulations is that all these transport coefficients are calculated in the same theoretical framework, which allows one to check the validity of approximate relations between electrical conductivity and self-diffusion coefficients (Nernst-Einstein equation) and between self-diffusion coefficients and viscosity  (Stokes-Einstein equation). Furthermore, the role played by each chemical species and the correlations between their  contributions to the transport properties can be easily investigated, as we shall see below.

\subsection{Self-diffusion coefficients}
\begin{figure*}
\centering
\subfloat[Li$^+$\label{fcompLi}]{\includegraphics{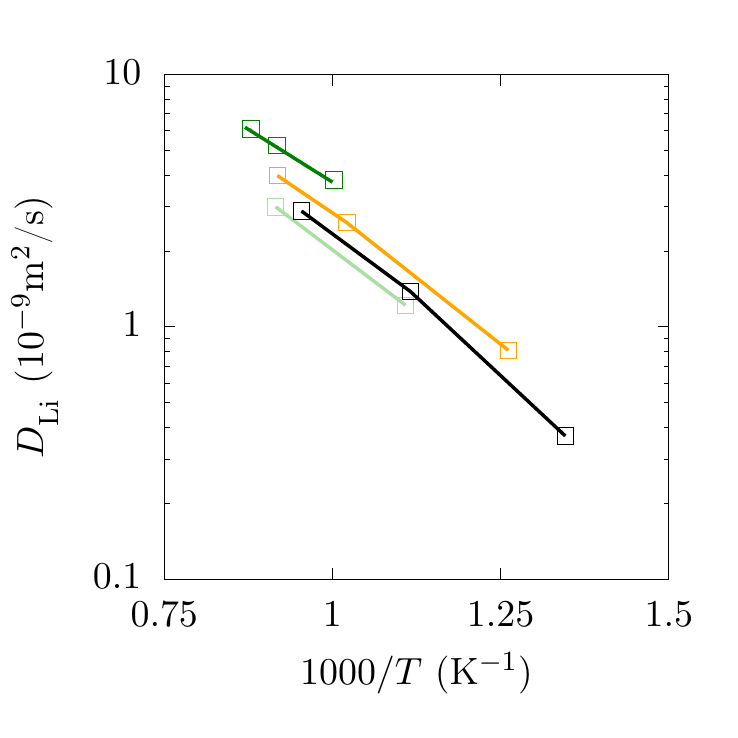}}
\subfloat[Na$^+$\label{fcompDNa}]{\includegraphics{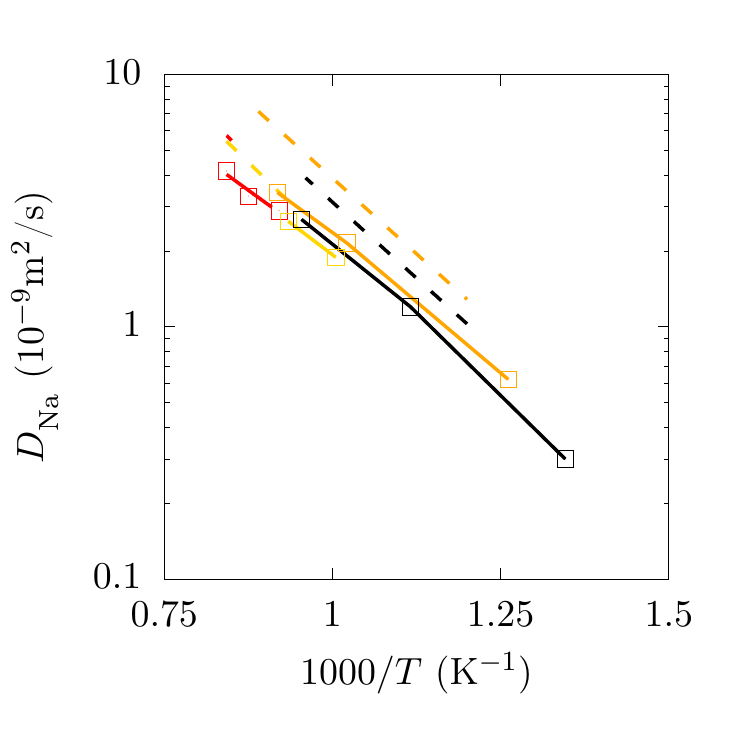}}\\ 
\subfloat[K$^+$\label{fcompDK}]{\includegraphics{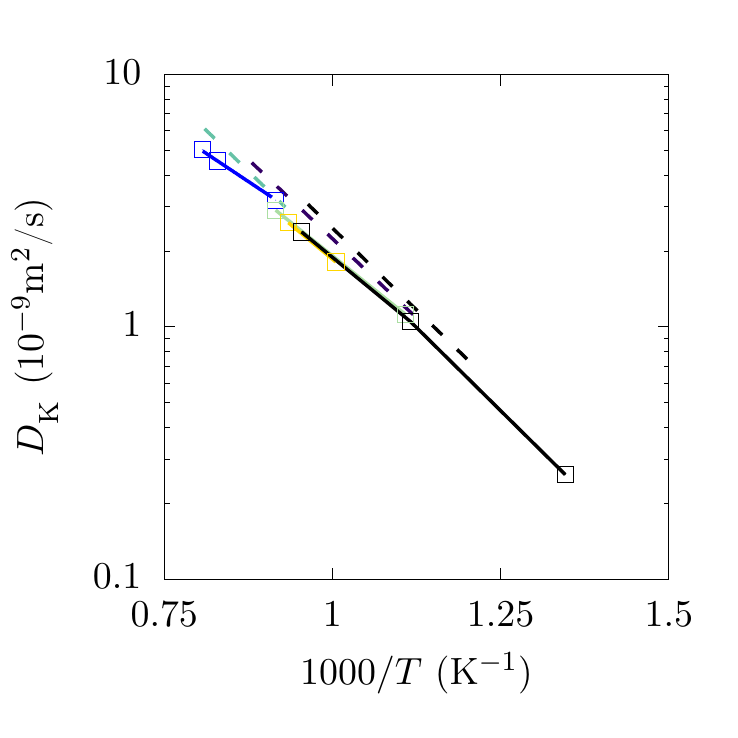}}
\subfloat[CO$_3^{2-}$\label{fcompDC}]{\includegraphics{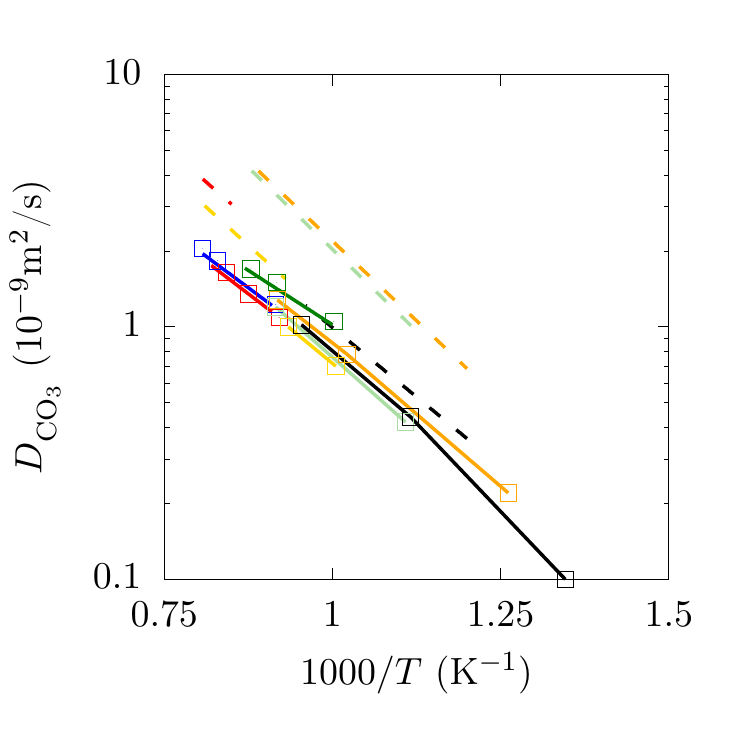}}\\\vspace{-0.5cm}
\subfloat{ {\includegraphics{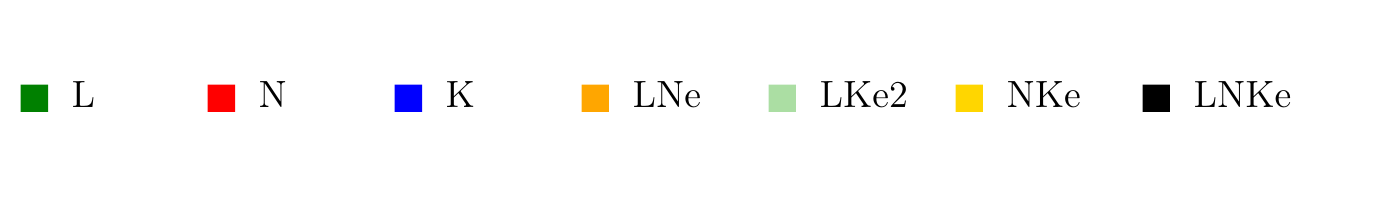}}}\vspace{-0.5cm}
\caption{Diffusion coefficient of ions in \ce{Li2CO3} (L), \ce{Na2CO3} (N), \ce{K2CO3} (K) and the eutectic mixtures \ce{Li2CO3}--\ce{Na2CO3} (LNe), \ce{Li2CO3}--\ce{K2CO3} (LKe2), \ce{Na2CO3}--\ce{K2CO3} (NKe) and \ce{Li2CO3}--\ce{Na2CO3}--\ce{K2CO3} (LNKe) calculated by MD (squares and plain lines as a guide to the eye) and experimentally measured by \citet{Spedding1965,Spedding1966} and \citet{Mills1966} (dashed lines).}
\label{fcompD}
\end{figure*}
\begingroup
\squeezetable
\begin{table*}
\begin{center}
\begin{tabular}{|l | c |c |c | c| c |c|} 
\hline
		&   ${D}_{0,{X}}$ & $E_{a}^{D_X}$ & $V_{a}^{D_X}$ &${D}_{0,{CO_3}}$ & $E_{a}^{D_{CO_3}}$ & $V_{a}^{D_{CO_3}}$\\ 
		& ($10^{-9}$m$^2$/s) & (kJ/mol) & (cm$^3$/mol) & ($10^{-9}$m$^2$/s) & (kJ/mol) & (cm$^3$/mol)		\\ \hline
Li$_2$CO$_{3}$	&  176 & 32 & 2.59-0.12P+0.0037P$^2$ & 54 & 33 & 2.4-0.18P-0.0145P$^2$			 \\   \hline 
Na$_2$CO$_{3}$	& 171 & 37 & 4.2-0.25P+0.0081P$^2$ & 82 & 39 & 4.0-0.24P+0.0079P$^2$ 	
 \\    \hline 
K$_2$CO$_{3}$  	&  135 & 34 & 5.2-0.43P+0.0179P$^2$ & 78 & 38 & 5.3-0.48P+0.0210P$^2$\\ \hline
\end{tabular}
\end{center}
\caption{Parameters of the Arrhenius activation law (\ref{eqDArrh}) obtained by the interpolation of all MD simulation points (with X = Li, Na or K). $P$ is in GPa.}
\label{tfitdiff}
\end{table*}
\endgroup
\begin{figure}
\centering
    \subfloat[\,Li$_2$CO$_3$-K$_2$CO$_3$\label{fchemlak}]{{\textsc{\includegraphics{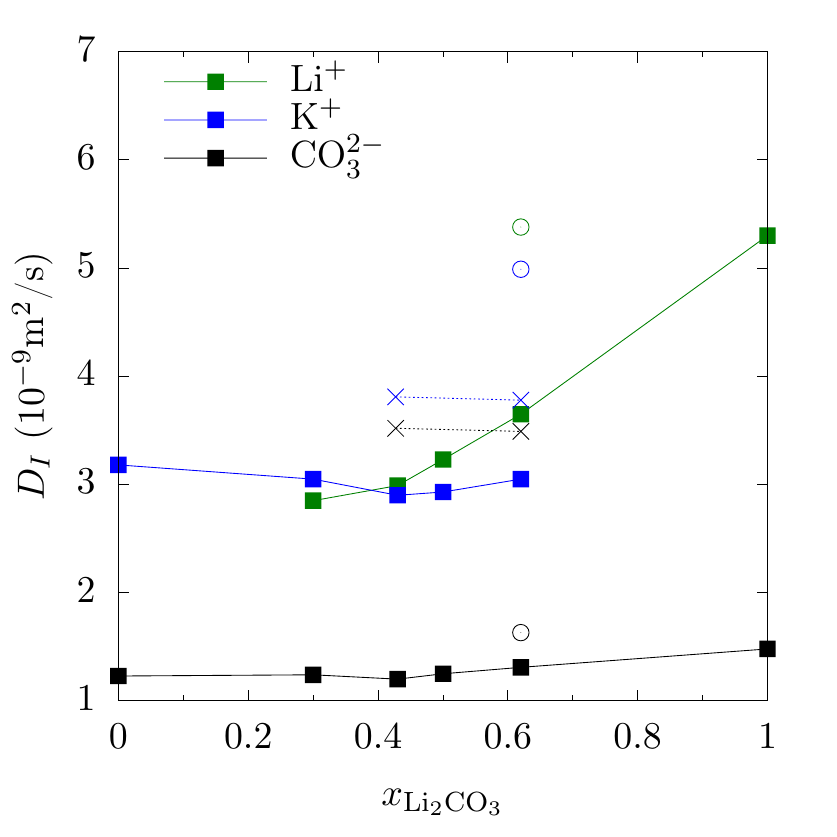}}}} \\
        \subfloat[\,Li$_2$CO$_3$-Na$_2$CO$_3$\label{fchemlana}]{{\includegraphics{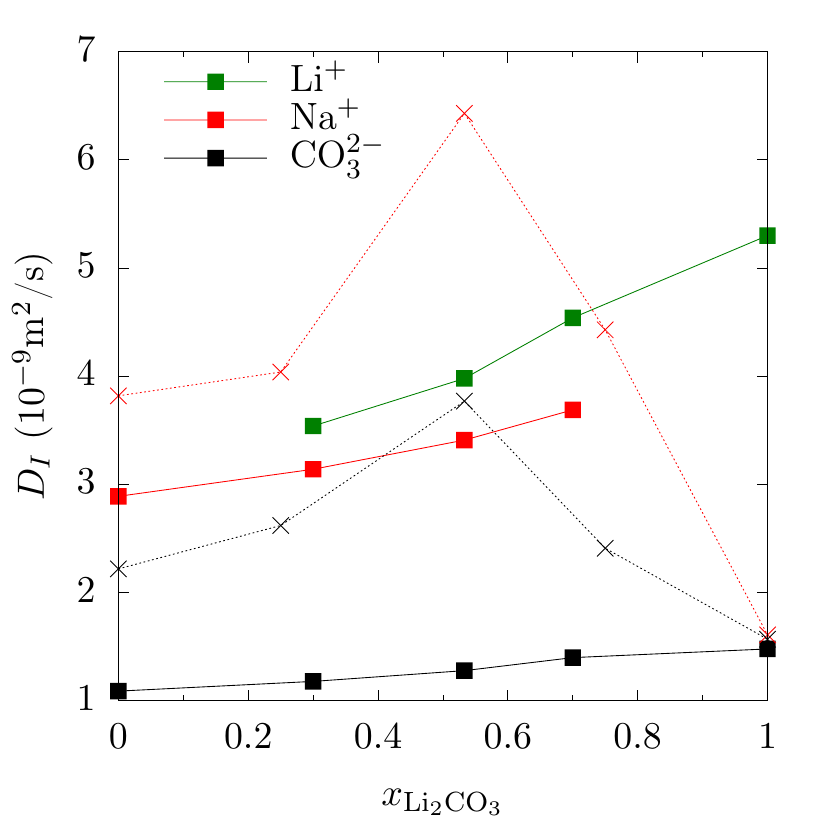}}}
\caption{Diffusion coefficients of ions in Li$_2$CO$_3$--K$_2$CO$_3$ and Li$_2$CO$_3$--Na$_2$CO$_3$ mixtures as a function of chemical composition ($x_{\rm{Li}_2\rm{CO}_3}$) at 1100~K and 1~bar: MD calculations of this study (plain squares and plain lines), experimental data\cite{ Spedding1965,Spedding1966,Mills1966} (crosses and dashed lines) and MD data of Corradini \emph{et al.}\cite{Corradini2016}  for the LKe mixture (empty circles). Note that no experimental data for the diffusion coefficient of \ce{Li+} are available yet.}
\label{fchemla}
\end{figure}
We have first estimated the ion self-diffusion coefficients from our AIMD simulations, which were long enough to reach the diffusive regime. Thus at 1~bar we found $D_{\rm Na} =$ 3.4~$\pm$~0.3 $10^{-9}$ m$^2$/s and $D_{\rm CO_3} =$ 1.5~$\pm$~0.3 $10^{-9}$ m$^2$/s in \ce{Na2CO3} at 1140~K, and $D_{\rm K} =$ 4.0~$\pm$~0.2 $10^{-9}$ m$^2$/s and $D_{\rm CO_3} =$ 1.8$~\pm$~0.3 $10^{-9}$ m$^2$/s in \ce{K2CO3} at 1190~K, values which compare well with those obtained from MD (3.5 and 1.4, and 4.6 and 1.8, respectively) a finding which emphasizes the reliability of our empirical force field. \\
The only diffusivity data in molten alkali carbonates have been provided by Mills and Spedding\cite{Spedding1965,Spedding1966,Mills1966} who measured the self-diffusion coefficients of Na$^+$, K$^+$ and CO$_3^{2-}$ ions in pure, binary and ternary alkali carbonate melts, 
at atmospheric pressure only. Note that the self-diffusion coefficient of Li$^+$ has not been measured yet. \\
As illustrated in Figure \ref{fcompD}, the agreement between MD and experimental results is satisfactory for $D_{\rm K}$ and $D_{\rm Na}$, at least in the pure \ce{Na2CO3} melt and in the \ce{Na2CO3}--\ce{K2CO3} eutectic mixture (NKe), but the diffusivity of the CO$_3^{2-}$ anion measured in various melt compositions is systematically higher by a factor of 1.5--3 than the one calculated by MD or by AIMD.\\
The activation energy (32--39~kJ/mol) depends slightly on the ion species and on the melt composition (Table~\ref{tfitdiff} and Figure~\ref{fcompD}), which is frequent in molten salts,\citep{Spedding1965} whereas the experimental values are between $\sim$ 40 and 53~kJ/mol, with an error reportedly below 1~kJ/mol. Note that the activation energies ($\sim$~36~kJ/mol) determined by the MD simulations of Corradini \emph{et al.}\cite{Corradini2016} for the molten LKe mixture are close to the ones we found in the same mixture (41~kJ/mol) as well as in the pure end-members (Table~\ref{tfitdiff}). However their values of $D_{\rm Li}$ and $D_{\rm K}$ are higher than ours, by a factor of $\sim$1.5.\\
General trends can be inferred from our MD simulations. By comparing the cation diffusivity in the three end-member compositions at a given temperature, the following hierarchy is found: $D_{\rm Li} > D_{\rm Na} > D_{\rm K} > D_{\rm CO_3}$, a behavior also found in experiments. As expected the largest ion CO$_3^{2-}$ has the slowest diffusivity. This hierarchy of the self-diffusion coefficients with the ionic radius is also found in binary and ternary mixtures. \\
However upon mixing, the diffusion coefficient of an ion species may change with the melt composition. Hence, when diffusivity data are plotted as functions of the \ce{Li2CO3} molar content in mixtures $x_{\rm Li_2CO_3}$, $D_{\rm Na}$ and $D_{\rm CO_3}$ display a maximum at the eutectic composition,\cite{Spedding1965,Spedding1966} as shown in Figure~\ref{fchemla}. According to \citeauthor{Spedding1965}\cite{Spedding1965,Spedding1966}, this feature could be due to a high level of structural disorder at the eutectic point, which could also explain the low melting temperature for this composition. At variance, our MD simulations do not display such a composition effect. Furthermore the microscopic structure, as expressed by the PDFs, does not seem more disordered at the eutectic point (compare Figure~S1a to Figures S1b and S1c). In fact the calculated self-diffusion coefficients of Na$^+$, CO$_3^{2-}$ (and also Li$^+$) increase as the \ce{Li2CO3} content increases in the binary mixture (Figure \ref{fchemla}). A different situation is observed in the \ce{Li2CO3}--\ce{K2CO3} mixture where $D_{\rm K}$ and $D_{\rm CO_3}$ are almost constant and exhibit a very slight minimum at 0.42, whereas $D_{\rm Li}$ is increasing with $x_{\rm  Li_2CO_3}$. A consequence is that $D_{\rm Li} < D_{\rm K}$ when $x_{\rm  Li_2CO_3}<0.42$ (see Figure ~\ref{fchemlak}). Thus, the major effect of mixing is that Na$^+$ and K$^+$ hinder the motion of the light Li$^+$ ion when $x_{\rm  Li_2CO_3}$ is low. Notice that the mobility measurements of \citet{Yang1987} support the above findings. Those trends are also in accordance with previous numerical studies \citep{Tissen1990,Tissen1994,Koishi2000,Costa2008,Ottochian2016} based on the force field by \citeauthor{Janssen1990}\cite{Janssen1990}.\\
To our knowledge no diffusivity data are available under pressure. Our simulations predict an Arrhenius behavior according to equation~(\ref{eqDArrh}). Thus the self-diffusion coefficients strongly decrease with pressure (see Figures~S3, S4 and S5), a behavior also observed in the \ce{CaCO3} melt.\citealp{Vuilleumier2014} Concerning the comparison between MD and AIMD calculations we notice that at 6~GPa and 1773~K in the \ce{K2CO3} melt, the self-diffusion of K$^+$ and CO$_3^{2-}$ calculated by MD are slightly lower than that obtained from AIMD: $D_{\rm K}=$ 3.6 and 5.0~$\pm$~0.5  $10^{-9}$~m$^2$/s and $D_{\rm CO_3}=$ 1.7 and 3.5 $\pm$ 0.5 $10^{-9}$~m$^2$/s for MD and AIMD respectively, because densities are a little bit different in the two calculations (2.49 and 2.41~g/cm$^3$ for MD and AIMD, respectively).

\subsection{Electrical conductivity}
The knowledge of the electrical conductivity of alkali carbonates is requisite for their industrial applications as electrolytes in fuel cell devices. Several experimental and numerical studies have be devoted to this issue. The measurements by Kojima \emph{et al.} are broadly consistent with those reported by \citet{Janz1988} for the end-members and various binary\cite{Kojima2007} and ternary\cite{Kojima2008} mixtures. \citet{Gaillard2008} also measured the electrical conductivity of the equimolar mixtures \ce{Na2CO3}--\ce{K2CO3} (NK) and \ce{Li2CO3}--\ce{Na2CO3}--\ce{K2CO3} (LNK) at atmospheric pressure. Moreover the electrical conductivity of the LKe eutectic mixture was measured by \citet{Lair2012}. The electrical conductivity data at atmospheric pressure are usually fitted by an Arrhenius law (see equation (\ref{eqSigmaArrh})). However some deviations were noticed at low temperature ($<$ 973~K) by Kojima \emph{et al.}, and for this reason they introduced a quadratic function of temperature (in log$T$) to fit their data. \\
\begin{figure}
\centering
{\includegraphics{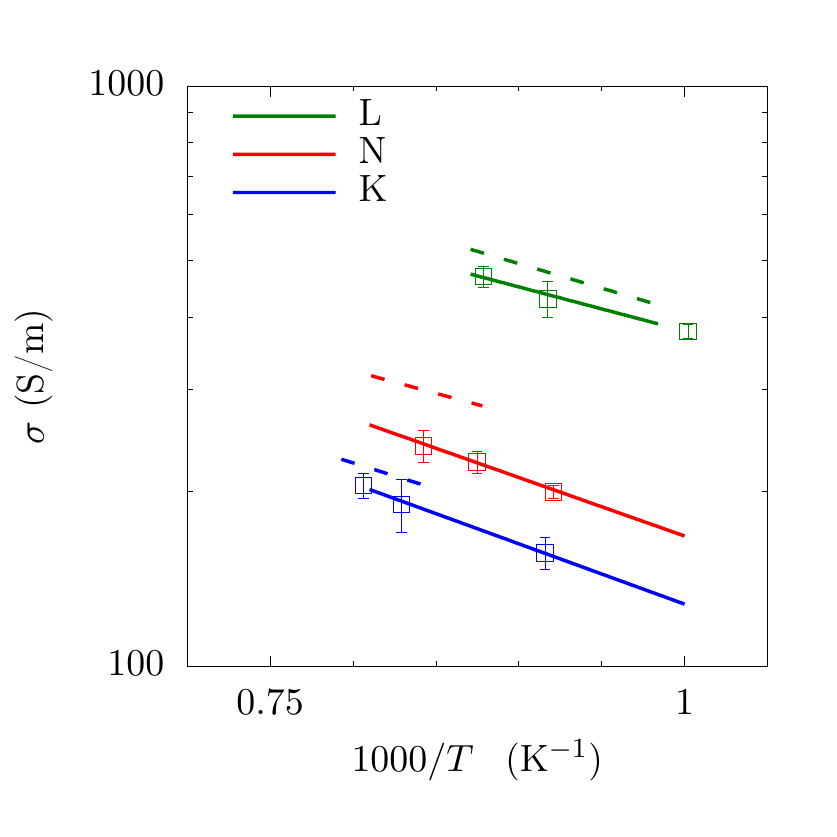}}       \caption{Electrical conductivity at 1 bar from MD (squares and plain lines) and from experiments of \citet{Kojima2009} for \ce{Li2CO3} (L), \ce{Na2CO3} (N) and \ce{K2CO3} (K). The plain line is the Arrhenius activation law adjusted on the 1~bar MD values. It differs somewhat from the ($P$-dependent) law adjusted on all simulation points (Table~\ref{tfittransp}), the parameters being: $\sigma_0$(L)=2170~S/m, $E_{a}^{\sigma}$(L)=15~kJ/mol,  $\sigma_0$(N)=1700~S/m, $E_{a}^{\sigma}$(N)=19~kJ/mol,  $\sigma_0$(K)=1400~S/m, $E_{a}^{\eta}$(K)=20~kJ/mol  for \ce{Li2CO3}~(L), \ce{Na2CO3}~(N) and \ce{K2CO3}~(K).}
\label{fcompcond}
\end{figure}
\begin{figure}
\centering
{\includegraphics{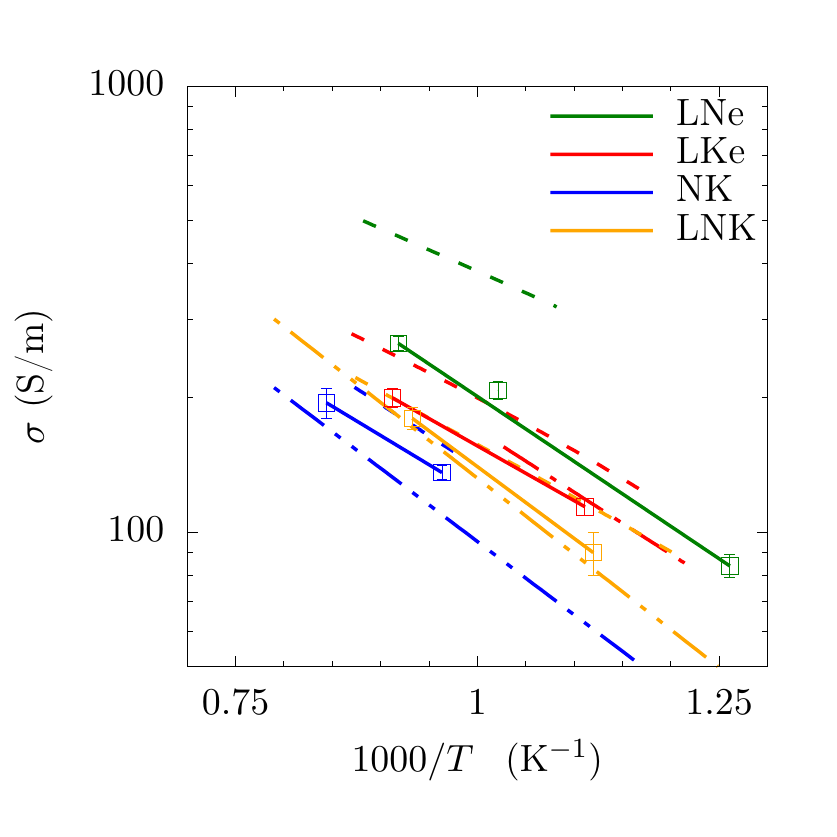}}     \caption{Electrical conductivity at 1 bar from MD (squares and plain lines) and from experiments of \citet{Kojima2009} (dashed line), \citet{Lair2012} (long dashed and dotted lines) and \citet{Gaillard2008} (long dashed and double dotted lines) for several mixtures: eutectic \ce{Li2CO3}--\ce{Na2CO3} (56:44 mol.\%, LNe), \ce{Li2CO3}--\ce{K2CO3} (62:38 mol.\%, LKe) and equimolar \ce{Na2CO3}--\ce{K2CO3} (NK) and \ce{Li2CO3}--\ce{Na2CO3}--\ce{K2CO3} (LNK).}
\label{fcompcond2}
\end{figure}
As shown in Figure \ref{fcompcond}, the electrical conductivity at 1~bar is negatively correlated to the cation size, as mentioned by \citet{Kojima2008} and Sifr\'e \emph{et al.}\cite{Sifre20142} The agreement between MD calculations and the available 1-bar experimental data\cite{Kojima2009} is good for \ce{Li2CO3} and \ce{K2CO3} (Figure \ref{fcompcond}), within the error bars of both the simulations and the measurements (expected to be of the order of 10--15~\%). Nevertheless the simulations underestimate somewhat the conductivities with respect to the experimental values of Kojima\cite{Kojima2009}. This discrepancy could be due to experimental inaccuracy. Indeed, the MD results are in very good agreement with the measurements of \citet{Gaillard2008} for the equimolar NK and LK systems and with those of \citeauthor{Lair2012}\cite{Lair2012} for the LKe mixture, whose values are systematically below that of \citeauthor{Kojima2009}\cite{Kojima2009} (Figure~\ref{fcompcond2}).\\
\begingroup
\squeezetable
\begin{table*}
\begin{center}
\begin{tabular}{|l|c|c|c|c|c|c|} 
\hline
		& $\sigma_0$ 	& $E_{a}^{\sigma}$	& $V_{a}^{\sigma}$	& $\eta_0$	& $E_{a}^{\eta}$ & $V_{a}^{\eta}$	\\ 
		&(S/m) & (kJ/mol)	& ($\text{cm}^3$/mol) & ($\text{Pa}\cdot\text{s}$)& (kJ/mol) & ($\text{cm}^3$/mol) \\ \hline
Li$_2$CO$_{3}$	& 3295.3 & 18 & 1.1+0.166P-0.0148P$^2$ & 2.7$\times$10$^{-4}$ & 24 & 2.2+0.01P  
 \\  \hline 
Na$_2$CO$_{3}$	& 2143.7 & 22 & 3.6-0.208P+0.0072P$^2$ &  2.5$\times$10$^{-4}$ & 27 & 3.5-0.10P 
 \\  \hline 
K$_2$CO$_{3}$  	& 1101.6 & 17 & 3.9-0.241P+0.0076P$^2$ & 2.2$\times$10$^{-4}$ & 27 & 4.7-0.20P \\ \hline
\end{tabular}
\end{center}
\caption{Parameters of the Arrhenius activation laws (\ref{eqSigmaArrh}) and (\ref{eqViscoArrh}) for electrical conductivity and viscosity, obtained by the interpolation of all MD simulation points. $P$ is in GPa.}
\label{tfittransp}
\end{table*}
\endgroup
The activation energy of calculated electrical conductivity is between 17 and 22~kJ/mol for the three end-members, which is slightly higher than the values found by \citet{Kojima2009} (15--20~kJ/mol). In the LKe mixture at atmospheric pressure, $E_{a}^{\sigma}$ calculated by Corradini \emph{et al.}\cite{Corradini2016} ($E_a=20.22$~kJ/mol) is close to the ones measured by \citet{Kojima2007} and \citet{Lair2012} ($E_a=21.91$~kJ/mol for Kojima). As for the activation energies determined by \citet{Gaillard2008} in the equimolar \ce{Na2CO3}--\ce{K2CO3} and \ce{Li2CO3}--\ce{Na2CO3}--\ce{K2CO3} mixtures, they are higher, around 31~kJ/mol. Notice that the electrical conduction is a collective process including ion-ion correlations (see below). Therefore the activation energy associated to the electrical conductivity differs from that associated to the diffusion process (compare Table~\ref{tfittransp} and Table~\ref{tfitdiff})).\\
Up to a few kbars, the electrical conductivity depends weakly on pressure (see activation volumes in Table~\ref{tfittransp}), whereas at higher pressures it significantly decreases with pressure (see Figure~S6).\\
Assuming that the ions move independently from each other (the ion-ion cross correlations are then neglected), the "ideal" electrical conductivity, $\sigma^{NE}$ is given by the Nernst-Einstein equation: 
\begin{equation}
\begin{split}
\sigma^{NE}=&\frac{e^2}{k_{\rm B}TV}\sum_{s} N_s z_s^{2}D_s \\
 =&\frac{e^2}{k_{\rm B} TV}\lim_{t\to \infty} \frac{1}{6 t} 
\sum_{s} z_{s}^2 \sum_{i=1}^{N_s} \langle
\big(\vec{\Delta}_i^{(s)}(t) \big)^2\rangle
\enspace,
\label{eNE}
\end{split}
\end{equation}
where $e$ is the elementary charge, $k_{\rm B}$ the Boltzmann constant and $N_s$, $z_s$ and $D_s$ are the number of ions, the conduction charge, equal to the formal charge here, and the self-diffusion coefficient of the chemical species $s=$Li, Na, K, and \ce{CO3} respectively. In the right-hand side of equation~(\ref{eNE}), we made use of the definition (\ref{eD}) of the self-diffusion coefficient by writing $\vec{\Delta}_i^{(s)}(t)=\vec{r}_i^{(s)}(t)-\vec{r}_i^{(s)}(0)$.\\
The Haven ratio $H=\sigma/\sigma^{NE}$ is defined as the ratio of the exact electrical conductivity $\sigma$, here calculated using the Einstein relation (\ref{econd}) to the Nernst-Einstein conductivity (\ref{eNE}). Note that the same electric charges are used (here, the formal ones) for evaluating $\sigma$ and $\sigma^{NE}$, therefore the value of $H$ is independent of the charges assigned to the ions. \\
The value of $H$ is usually less than 1 for molten salts, which is explained by the large contribution of the anticorrelations between ions of the same charge that tends to decrease the ideal conductivity $\sigma^{NE}$\cite{Kashyap2011}. Simulations studies showed that $H \simeq 0.81$ for CaCO$_3$ melt,\cite{Vuilleumier2014} $H \sim 0.7$ for \ce{Li2CO3}--\ce{Na2CO3} mixtures\cite{Ottochian2016} and $H=$0.5-1.0 for \ce{Li2CO3}--\ce{K2CO3} mixtures\cite{Costa2008}. However, the Haven ratio was found greater than 1 ($H=1.8$) for LKe.\cite{Corradini2016} Note that $H$ may depend on pressure and temperature. In this work, we find a Haven ratio which is slightly greater than 1 for \ce{Li2CO3} (e.g. $H=1.04$ at the melting point) and lower than 1 for \ce{Na2CO3} and \ce{K2CO3} (e.g. $H=0.99$ and 0.94, respectively, at the melting point). This ratio seems to depend slightly on temperature, and more appreciably on pressure, especially for \ce{Na2CO3} and \ce{K2CO3}. For instance, at 4~GPa and 1500~K (near the melting point) the $H$~ratio of \ce{K2CO3} drops to 0.47 (Table~\ref{tGKvsNE}). The amplitude of $H$ is often considered to be indicative of the degree of dynamic correlations between ions. So a value close to 1 should confirm the independence hypothesis of the Nernst-Einstein approximation. Nevertheless, we show below that such correlations are present in the melt even when $H\simeq 1$ and that the variations of $H$ with chemical composition and thermodynamic conditions are due to subtle changes in dynamic correlations between ionic motions. \\
To better understand the role of ion-ion correlations it is suitable to develop the exact Einstein expression of the conductivity given by equation~(\ref{econd}). Note that in this expression the charge displacement is squared as opposed to equation~(\ref{eD}) that is the sum of each squared ionic displacement. After a few algebra equation~(\ref{econd}) leads to the following expression
\begin{equation}
\begin{split}
\sigma=&\frac{e^2}{k_{\rm B} TV} \lim_{t\to \infty} \frac{1}{6t} \langle
\Big(\sum_{s} \sum_{i=1}^{N_s} z_s \vec{\Delta}_i^{(s)}(t)
\Big)^2 \rangle \\ 
=& \frac{e^2}{k_{\rm B} TV} \lim_{t\to \infty} \frac{1}{6t} \Big\{ \sum_{s}
z_s^2 \Big( \sum_{i=1}^{N_s} \langle
\big(\vec{\Delta}_i^{(s)}(t)\big)^2 \rangle \\
& + \sum_{i=1}^{N_s}
\sum_{j\ne i}^{N_s} \langle \vec{\Delta}_i^{(s)}(t).
\vec{\Delta}_j^{(s)}(t)\rangle \Big) \\
& + 
\sum_{s} \sum_{s'\ne s} z_s z_{s'}
\Big(\sum_{i=1}^{N_s} \sum_{j=1}^{N_{s'}} \langle
\vec{\Delta}_i^{(s)}(t).\vec{\Delta}_j^{(s')}(t)\rangle \Big)
\Big\} \enspace ,
\end{split}
\end{equation}
where $N_s$ is the number of ions of species $s$, $z_s$ is their charge and $\Delta_i^{(s)}(t)$ is the displacement of the ion $i$ of species $s$. The first term in the RHS of the last equation is the Nernst-Einstein conductivity (\ref{eNE}). We then get the expression of the Haven ratio in terms of the correlations between ions:
\begin{equation}
H=\frac{\sigma}{\sigma^{NE}}=1+ \sum_{s} H_s+ \sum_{s}
\sum_{s'\ne s} H_{ss'}
\end{equation}
where $H_s$ expresses the average cross correlations (through a scalar product) between the displacements of ions of species $s$:
\begin{equation}
H_s=\lim_{t\to \infty} \frac{1}{6t} \frac{z_{s}^{2}}{\sum_{s}N_s
  z_{s}^2 D_s}\sum_{i=1}^{N_s} \sum_{j\ne i}^{N_s} \langle
\vec{\Delta}_i^{(s)}(t).\vec{\Delta}_j^{(s)}(t)\rangle \enspace,
\end{equation}
and $H_{ss'}$ expresses the average cross correlations between the displacement of an ion $i$ of species $s$ and that of an ion $j$ of another species $s'$:
\begin{equation}
H_{ss'}= \lim_{t\to \infty} \frac{1}{6t}
\frac{z_s z_{s'}}{\sum_{s}N_s z_{s}^2 D_s}
\sum_{i=1}^{N_s} \sum_{j=1}^{N_{s'}} \langle
\vec{\Delta}_i^{(s)}(t).\vec{\Delta}_j^{(s')}(t)\rangle \enspace.
\label{eness}
\end{equation}
If the trajectories of the ions are independent from each other, $H_{s}=H_{ss'}=0$ and the Nernst-Einstein approximation is recovered (\emph{i.e.} $H=1$). However, $H=1$ doesn't imply that $H_{s}$ and $H_{ss'}$~$\sim 0$. For instance in \ce{Li2CO3} (see Table~\ref{tGKvsNE}), $H$ is very close to 1 ($H=1.04$) although $H_{\rm LiLi}=0.04$, $H_{\rm CO_3}=-0.32$ and $H_{\rm Li CO_3}=0.32$. These results mean that the displacements of Li cations are mostly not correlated to each other, whereas those of \ce{CO3} are negatively correlated to each other, \emph{i.e.} two \ce{CO3} have a high probability to move in opposite directions. At variance, the cation-anion correlation term $H_{\rm Li_2CO_3}=0.32$ is positive, meaning that cations and anions are also moving in the opposite direction (notice that in equation~(\ref{eness}) $z_{s}z_{s'}<0$). The net result is that $H_{\rm CO_3}$ and $H_{\rm Li CO_3}$ are canceling each other, leading to $H=1+H_{Li}=1.04$. A different situation is observed with \ce{Na2CO3} and \ce{K2CO3}, for which the cancellation effect implies the three contributions $H_{\rm X}$, $H_{\rm CO_3}$ and $H_{\rm X CO_3}$. As a result the anion-cation correlations $H_{\rm X CO_3}$ contribute positively (0.48 for \ce{Na2CO3} and 0.46 for \ce{K2CO3}), whereas the correlations between ions of the same species decrease $H$ by a value of $H_{{\rm X}}+H_{{\rm CO}_{3}}$ ($-$0.49 for \ce{Na2CO3} and $-$0.52 for \ce{K2CO3}), but these two contributions almost balance each other providing a Haven ratio $H=1+H_{\rm X}+H_{\rm CO_3}+H_{\rm XCO_3}$ equal to~$\sim~1$ (0.99 for \ce{Na2CO3} and 0.94 for \ce{K2CO3}). \\This approach can be carried out on other compositions, including binary and ternary mixtures. This issue will be more thoroughly discussed in a future paper. The conclusion we want to emphasize here is that the Nernst-Einstein relation leads to a reasonable estimation of the electrical conductivity, not because the underlying assumption (the ions move independently from one another) is valid, but because of a cancellation effect between ion-ion correlations. Therefore the Nernst-Einstein relation must be carried with great caution, in particular when it is used to extract the ion diffusion coefficients.
\begin{table}
\begin{center}
\begin{tabular}{|c|c|c|c|c|c|c|c|} \hline
       & $T$(K)  &  $P$ (GPa)& $\sigma$ (S/m) & $H$  & $H_{\rm X}$ & $H_{\rm CO_3}$ & $H_{\rm XCO_3}$ \\ \hline
\ce{Li2CO3} &  1100   &  0        &   430          & 1.04 &  0.04 &  -0.32   & 0.32    \\ 
            &  1500   & 4.0       &   470          & 1.06 &  0.08 &  -0.34   & 0.32    \\   \hline
\ce{Na2CO3} &  1140   &  0        &   224          & 0.99 & -0.24 &  -0.25   & 0.48    \\ 
            &  1500   & 4.0       &   154          & 0.80 & -0.28 &  -0.31   & 0.39    \\    \hline
\ce{K2CO3}  &  1200   &  0        &   196          & 0.94 & -0.38 &  -0.14   & 0.46    \\ 
            &  1500   & 4.0       &   97           & 0.47 & -0.38 &  -0.24   & 0.35    \\  \hline
\end{tabular}
\caption{Haven ratio and its ionic contributions. Note that the error bar on these quantities is of the order of the error bar on the electrical conductivity: $\sim$5--10~\%.}
\label{tGKvsNE}
\end{center}
\end{table} 

\subsection{Viscosity}
\begin{figure}
\centering
{\includegraphics{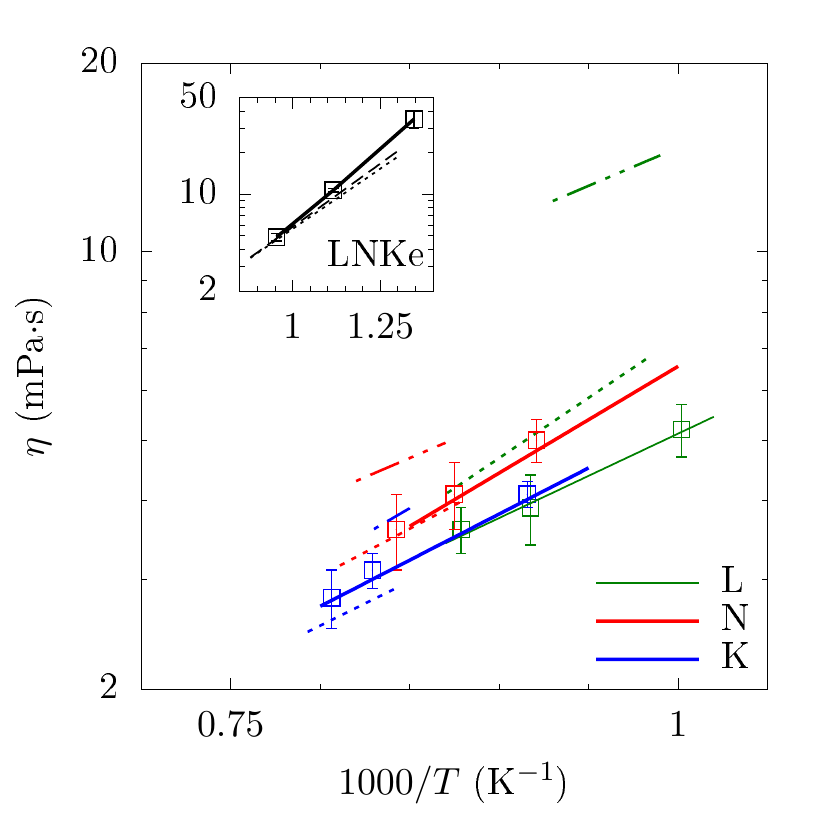}}                 
\caption{Viscosity  of the end-members \ce{Li2CO3} (L), \ce{Na2CO3} (N) and \ce{K2CO3} (K) at 1 bar, calculated by MD (squares and plain lines) and measured in references~\onlinecite{Janz1988,Sato1999} (dotted line) and \onlinecite{DiGe16} (long dashed and double dotted lines). The plain line is the Arrhenius activation law adjusted on the 1-bar MD values. It differs somewhat from the law adjusted on all simulation points (Table~\ref{tfittransp}), the parameters being: $\eta_0$(L)= 1.8~$\times$10$^{-4}$~Pa.s, $E_{a}^{\eta}$(L)=26~kJ/mol,  $\eta_0$(N)= 1.3~$\times$10$^{-4}$~Pa.s, $E_{a}^{\eta}$(N)=33~kJ/mol,  $\eta_0$(K)= 1.8~$\times$10$^{-4}$~Pa.s, $E_{a}^{\eta}$(K)=28~kJ/mol for \ce{Li2CO3} (L), \ce{Na2CO3} (N) and \ce{K2CO3} (K).\\
In the inset is plotted the viscosity of the eutectic ternary mixture (LNKe) calculated by MD (squares and plain line) and measured by \citet{An2016} (dotted line), indistinguishable from the data of \citeauthor{Janz1988}\cite{Janz1988}, and by \citet{Ejima1987} (dashed line).}
\label{fcompvisco}
\end{figure}
\begin{figure}
\centering
{\includegraphics{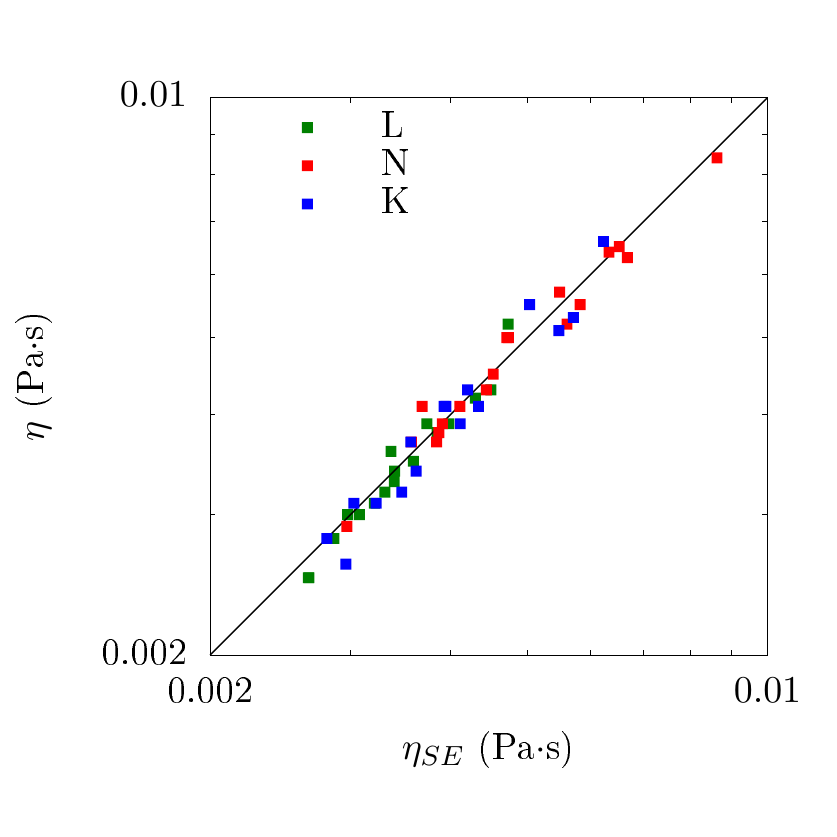}}
\caption{Comparison of the viscosity calculated using the Green-Kubo formula, $\eta$, and the Stokes-Einstein equation, $\eta_{SE}$, for all $T$ and $P$. The $d$ parameter, figuring in equation~(\ref{eSE}) has been adjusted for \ce{Li2CO3} (L), \ce{Na2CO3} (N) and \ce{K2CO3} (K) to 1.6, 2.2 and 2.4 \AA \, respectively, so as to align the data on the $\eta=\eta_{SE}$ bisector (black line).}
\label{fSE}
\end{figure}
Viscosity of molten alkali carbonates at ambient pressure has been measured since the sixties and critically reviewed by \citeauthor{Janz1988}\cite{Janz1988}. Its value stands between 3 and 30~mPa.s, depending on composition and temperature, as compared to 1~mPa.s, the typical viscosity of alkali chloride melts (NaCl at 1100~K) and liquid water (0.8 mPa.s at 300~K) at atmospheric pressure. \\
Afterwards, Sato \emph{et al.} performed accurate measurements (errors within 3~\%) of the viscosity of \ce{Li2CO3}, \ce{Na2CO3} and \ce{K2CO3} (as well as \ce{Rb2CO3} and \ce{Cs2CO3}) \cite{Sato1999}, and of \ce{Li2CO3}--\ce{Na2CO3} and \ce{Li2CO3}--\ce{K2CO3} binary melts\cite{Sato19992}, by using the oscillating method. However this technique may present some drawbacks (dependence on the apparatus, long time analysis\dots). On the other hand, the measurement of low viscosities at high temperature with the rotation method is tricky, but do not suffer from these disadvantages.\cite{Kim2015} Using this approach with a high-temperature rheometer system, \citet{Kim2015} obtained results that confirm those of \citeauthor{Sato19992} for \ce{Li2CO3}, the \ce{Li2CO3}--\ce{K2CO3} eutectic mixture of ratio 0.43:0.57 mol\% (LKe2) and for the LNKe mixture, which has the lowest melting temperature (668~K).\cite{Sato1999,Sato19992} However \citet{DiGe16} have recently applied a similar but improved method, as they claimed, and found viscosities systematically higher, by 30 to 100~\%, compared to previous studies. \\
Concerning our simulations, they satisfactorily reproduce the experimental data of \citet{Sato1999} for \ce{Na2CO3} and \ce{K2CO3} and in particular the activation energy (see Figure~\ref{fcompvisco}). As mentioned above, there is a large disagreement between the experimental study of \citet{DiGe16} and the one of \citeauthor{Sato1999}.\cite{Sato1999} In particular the viscosity measured by \citeauthor{DiGe16} for \ce{Li2CO3} is twice the one measured by \citeauthor{Sato1999}, and leads to a surprisingly large difference between the viscosity of \ce{Li2CO3} and those of \ce{Na2CO3} and \ce{K2CO3}. Indeed, such a contrasting behavior between the small Li$^+$ cation and the bigger Na$^+$ and K$^+$ cations, is not observed for other transport coefficients. As for our simulations of \ce{Li2CO3}, the calculated activation energy is compatible with the data but the magnitude of the viscosity is $\sim$~15~\% lower than that measured by \citet{Sato19992} and \citeauthor{Kim2015}\cite{Kim2015} This result could be explained by the neglecting of some electronic effects (\emph{e.g.} polarizability) in our model. Still, with regard to the viscosity of \ce{Na2CO3} and \ce{K2CO3} the calculated values are very close to the experimental ones (within the experimental uncertainties).\\
On the inset of Figure \ref{fcompvisco} the measured and calculated viscosity of the \ce{Li2CO3}--\ce{Na2CO3}--\ce{K2CO3} eutectic mixture is also shown. A good agreement is found with both the reference data of \citet{Janz1988} and \citet{Ejima1987} and the recent measurements of \citeauthor{An2016}\cite{An2016}, emphasizing the ability of our force field to model the contribution of all the ions in mixtures. With regard to the LKe mixture, the viscosity estimated by our MD simulations is 10--30~\% higher than the one measured by \citeauthor{Sato19992},\cite{Sato19992} with an activation energy equal to  42.6~kJ/mol (as compared to 34~kJ/mol),  whereas the one calculated by Corradini \emph{et al.}\cite{Corradini2016} is by 20--30~\% lower (with a lower activation energy of 29.6~kJ/mol ).\\
Like the other transport coefficients, the evolution of viscosity with temperature at high pressure is conveniently reproduced by an Arrhenius activation law given by equation~(\ref{eqViscoArrh}), the parameters of which are given in Table~\ref{tfittransp}.  As shown in Figure~S7 for low and moderate pressures ($P <$~1~GPa), the viscosity depends almost exclusively on temperature. However beyond a few GPa, the weight of the activation volume $V_{a}^{\eta}$ is no longer negligible. Very recently, \citet{Stagno2018} reported viscosity measurements for \ce{Na2CO3} in the 1.7--4.6~GPa pressure range, that are in agreement with our calculations (deviation within 0--50~\%, Figure~S7b).\\
Interestingly, the activation energy differs from that of electrical conductivity (Table \ref{tfittransp}). The viscosity is indeed dominated by the motion of slow ions, as opposed to the electrical conductivity which is controlled by the fastest species. Still the viscosity can be described fairly well by the phenomenological Stokes-Einstein equation:
\begin{equation}
\eta_{SE}=\frac{k_{\rm B}T}{2\pi Dd} \enspace ,
\label{eSE} 
\end{equation}
using the arithmetic mean of the diffusion coefficients, $D=\sum_s x_s D_S/\sum_s x_s$ where $x_s$ is the molar fraction of ion of species $s$ in the melt, and $d$ is the hydrodynamic diameter of the diffusing ions. By choosing $d=1.6$, 2.2 and 2.4~\AA \, for \ce{Li2CO3}, \ce{Na2CO3} and \ce{K2CO3} respectively, we found that the behavior of the MD-calculated viscosity with pressure and temperature is correctly reproduced by equation~(\ref{eSE}) as demonstrated by Figure \ref{fSE}.\\ 
Finally, we find that the electrical conductivity $\sigma$ (in S/m) and the viscosity $\eta$ (in Pa.s) are related by the following empirical formula: 
\begin{equation}
\sigma=\frac{A}{\eta^{0.8}} \enspace,
\end{equation}
where $A=$ 5.375, 2.615 and 1.825 for \ce{Li2CO3}, \ce{Na2CO3} and \ce{K2CO3} respectively (Figure~S8). A comparable relation between these two transport properties has been also highlighted from experimental data of various melt compositions.\cite{Sifre20142,Vuilleumier2014} The simple relations between $D$, $\sigma$ and $\eta$ can be very useful to infer one of these quantities from the others. They also show that these transport properties are essentially ruled by the diffusion of all ionic species in carbonate melts. This is in contrast with silicate melts whose viscosity is governed by the relaxation time of the Si-O bonds network, whereas the conductivity is mainly controlled by the mobility of network modifier elements, at least at low temperature near the melting point.

\section{Conclusion}
We have studied the thermodynamics (equation of state and surface tension), the microscopic structure and the transport properties (diffusion coefficients, electrical conductivity and viscosity) of molten alkali carbonates (\ce{Li2CO3}, \ce{Na2CO3}, \ce{K2CO3} and some of their binary and ternary mixtures) in the $T-P$ range: $\sim$ 1100--2100~K, 0--15~GPa. For this we have developed an empirical force field using benchmark data from experiments (density and compressibility of the end-members at 1~bar) and from AIMD simulations (microscopic structure of four liquids). The density and compressibility data for binary and ternary mixtures that were not used in the FF parameterization procedure are also very well reproduced. We also showed for the end-members that the density of the crystalline phase at room conditions could be obtained with a good accuracy. Furthermore, the global agreement between the MD results and the full set of experimental data (thermodynamics ans transport coefficients) is very satisfying. This gives consistent evidence for the ability of the FF to be transferable in terms of composition and thermodynamic conditions. In this respect we believe it is a significant improvement compared to the previously existing force fields for molten alkali carbonates. \\
Using this force field, the equation of state of the end-members was first evaluated and modeled by a third-order Birch-Murnaghan formula. Next we have shown that mixtures behave ideally regarding the density and the compressibility. The analysis of the PDFs showed that the microscopic structure is that of an ionic liquid and the only modification we could identify upon increasing pressure was a compaction effect. As for the transport properties (diffusion coefficients, electrical conductivity and viscosity), they evolve smoothly (Arrhenius-like) over the studied $T-P$ domain. \\
Moreover, for the first time the surface tension of molten carbonates was calculated using MD. Although the calculated values are systemically higher than the experimental values published in the literature, the discrepancy is moderate ($\sim +15$~\%), as observed in previous numerical studies dealing with ionic liquids, and could be due to computational uncertainties (size effects). \\
Finally we discussed the significance of the Nernst-Einstein and the Stokes-Einstein equations, that relate the diffusion coefficients to the electrical conductivity and to the viscosity, respectively. We showed that, by using \emph{ad hoc} parameters, both formula lead to reasonable values. However they are not always representative of the transport mechanism itself. In particular we have demonstrated that the fairly good predictions provided by the Nernst-Einstein equation result from a partial canceling of interionic dynamic correlations.

\section*{Supplementary Material}
See supplementary material for comparisons of the PDFs issued by MD and by AIMD simulations, a summary of all calculated properties with their uncertainties, plots of the evolution of these properties with $T$ and $P$, convergence plots of the correlation functions and details on the AIMD simulations including an example input file.
 
\begin{acknowledgments}
The research leading to these results has received funding from the R\'egion Ile-de-France and the European Community’s Seventh Framework Program (FP7/2007-2013) under Grant agreement (ERC, N$^\circ$ 279790). The authors acknowledge GENCI for HPC resources (Grant No. 2015-082309).
\end{acknowledgments}

\bibliography{M}
 
\end{document}


\title{Supplementary Material for "Atomistic simulations of molten carbonates: thermodynamic and transport properties of the \ce{Li2CO3}--\ce{Na2CO3}--\ce{K2CO3} system"}

\author{Elsa Desmaele}\email{elsa.desmaele@gmail.com}
\affiliation{Sorbonne Universit\'e, CNRS, Laboratoire de Physique Th\'eorique de la Mati\`{e}re Condens\'ee, LPTMC, F75005, Paris, France}
\author{Nicolas Sator}
\affiliation{Sorbonne Universit\'e, CNRS, Laboratoire de Physique Th\'eorique de la Mati\`{e}re Condens\'ee, LPTMC, F75005, Paris, France}
\author{Rodolphe Vuilleumier}
\affiliation{PASTEUR, D\'epartement de chimie, \'Ecole normale sup\'erieure, PSL University, Sorbonne Universit\'e, CNRS, 75005 Paris, France}
\author{Bertrand Guillot}\email{guillot@lptmc.jussieu.fr}
\affiliation{Sorbonne Universit\'e, CNRS, Laboratoire de Physique Th\'eorique de la Mati\`{e}re Condens\'ee, LPTMC, F75005, Paris, France}

\date{\today}

\maketitle

\onecolumngrid

 \tableofcontents
 \listoffigures
 \listoftables
 
\clearpage 
\begin{table*}
    \begin{tabular}{|l |r | c | c |  c | c|} 
    \hline
		&	$N$	& $t_{sim}$ (ps) & $n$ (g/cm$^{3}$)			& $T$  (K) 	& $\langle P \rangle$	   	(GPa)		\\ \hline				
Na$_2$CO$_{3}$  & 	768	& 	48.0			&		1.97		&	1140		&	0 $\pm$ 0.5		\\ \hline	
K$_2$CO$_{3}$ & 204	& 	13.2		& 1.90				&	1190		& 0 $\pm$ 0.5				\\
	 & 	204	& 	24.0			&	2.41			&	1773		&	6 	$\pm$ 1.5	\\ \hline
\end{tabular}
\caption{Details of the AIMD simulations.}
\label{tdft}
\end{table*}

\begin{table*}
    \begin{tabular}{|l |c | c |  c | c| c |} 
    \hline
		&	$n$ (g/cm$^{3}$)			& a (\AA) 	& 	b (\AA)	& c (\AA) & $\beta$ ($^\circ$)\\ \hline				
\ce{Li2CO3} & 	2.12	& 8.77	& 5.09 & 6.16 & 119.7	\\ \hline	
Exp.\cite{Idemoto1998} & 2.10 & 8.36 & 4.97 & 6.19 & 114.7 \\ \hline
\ce{Na2CO3}& 2.50	& 9.08 & 5.31 & 5.97 & 	101.9	\\ \hline
Exp.\cite{Arakcheeva2010} & 2.54 & 8.91 & 5.24 & 6.05 & 101.4 \\ \hline
\ce{K2CO3}	 & 2.45	& 5.67 & 9.98 & 6.83 & 100.2 \\ \hline
Exp.\cite{Gatehouse1973}
 & 2.47 & 5.64 & 9.84 & 6.87 & 98.7 \\ \hline
\end{tabular}
\caption[Crystal structure parameters]{Crystal structure parameters for monoclinic \ce{Li2CO3}, \ce{Na2CO3} and \ce{K2CO3} (300~K, 1~bar) from the experimental literature (Exp.) and calculated by anisotropic MD simulations ($NST$ ensemble) starting from the experimentally determined structures.}
\label{tsupcryst}
\end{table*}

\begin{figure*}
\subfloat[LKe ($n$=1.90~g/cm$^3$, $T$=1000~K)\label{frdfmdvsdftli}]{\includegraphics{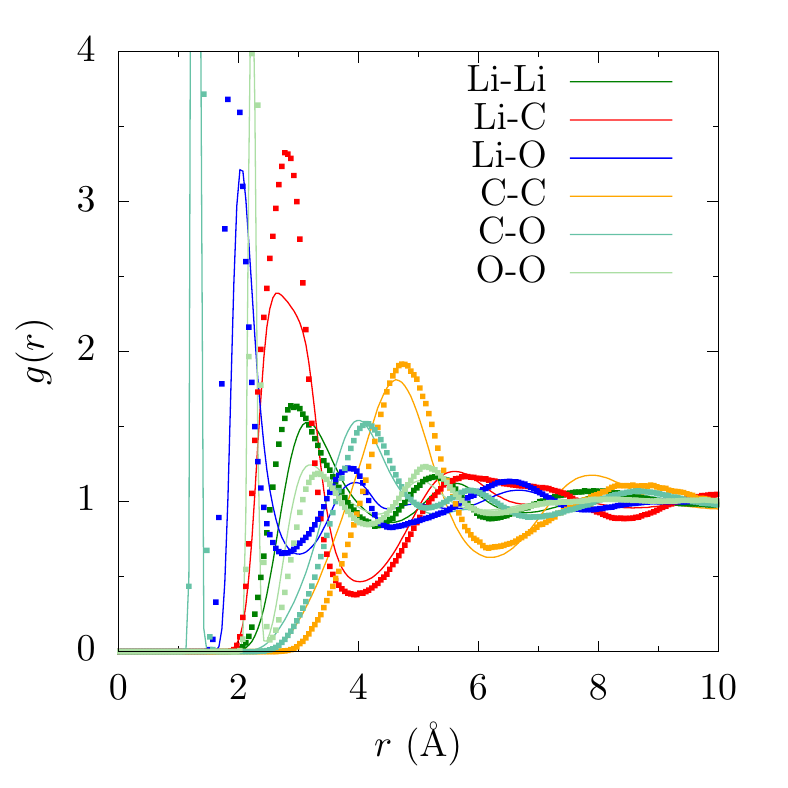}}
\subfloat[\ce{Na2CO3} ($n$=1.97~g/cm$^3$, $T$=1140~K)\label{frdfmdvsdftna}]{\includegraphics{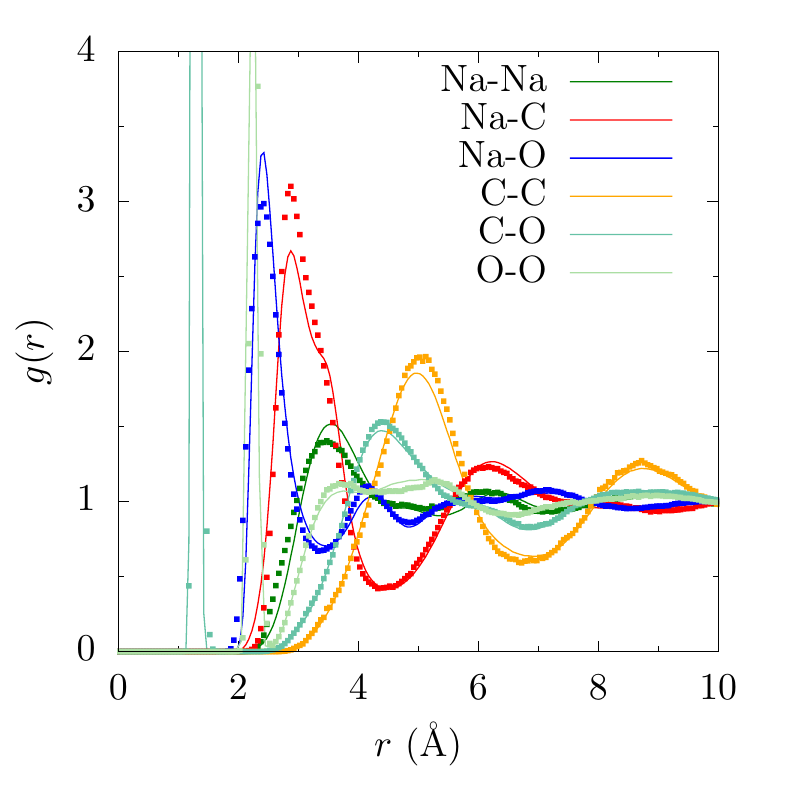}}\\ 
\subfloat[\ce{K2CO3} ($n$=1.90~g/cm$^3$, $T$=1190~K)\label{frdfmdvsdftk1}]{\includegraphics{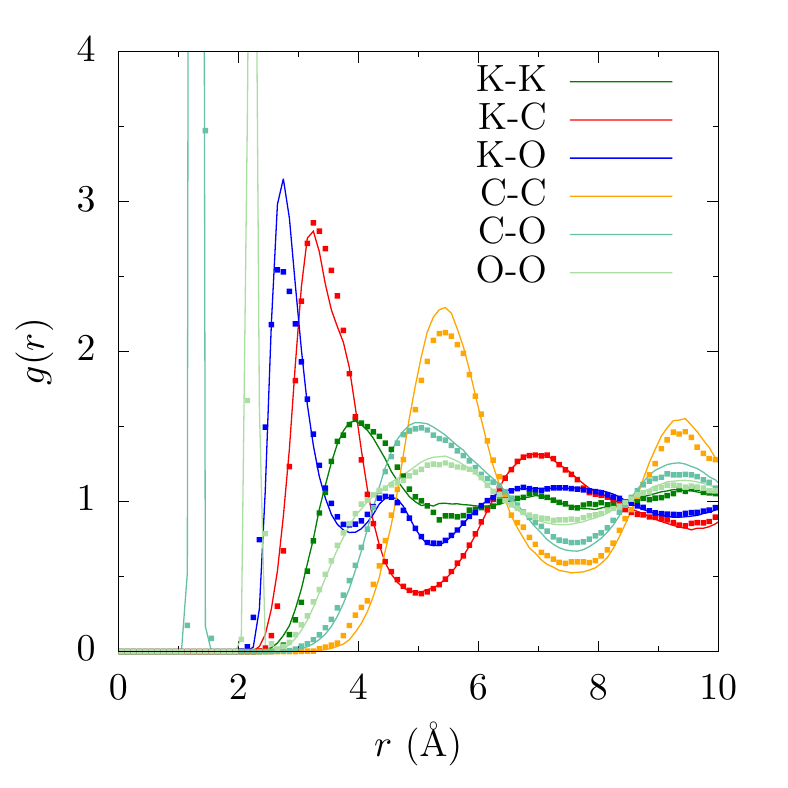}}
\subfloat[\ce{K2CO3} ($n$=2.41~g/cm$^3$, $T$=1773~K)\label{frdfmdvsdftk2}]{\includegraphics{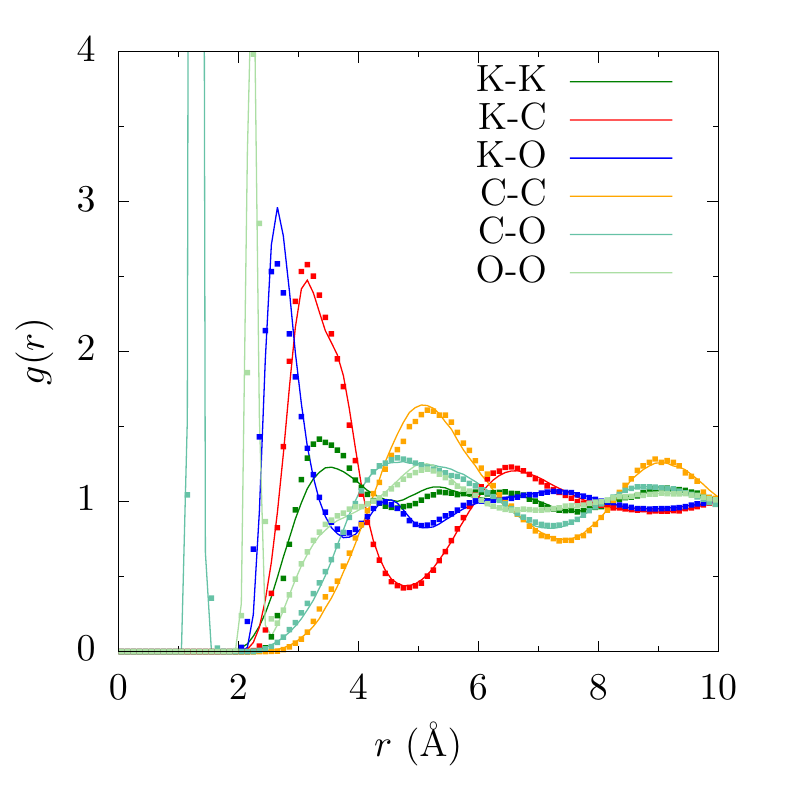}}				
\caption[Pair distribution functions from MD and AIMD]{Pair distribution functions obtained by MD (full line) and AIMD (squares). Chemical compositions, temperatures and densities are mentioned above the figures. In the case of panel~(a), only $g(r)$ associated with the \ce{Li2CO3} component of the \ce{Li2CO3}--\ce{K2CO3} eutectic mixture (LKe) are plotted.}
\label{frdfmdvsdft1}    
\end{figure*}

\begin{figure*}
\subfloat[\ce{Li2CO3}\label{feos1li}]{\includegraphics{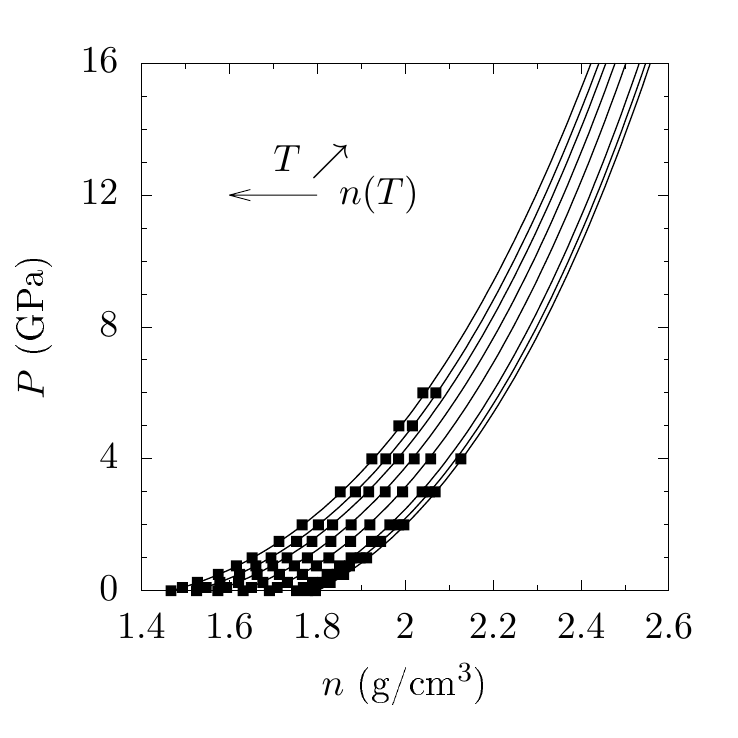}}
\subfloat[\ce{Na2CO3}\label{feos1na}]{\includegraphics{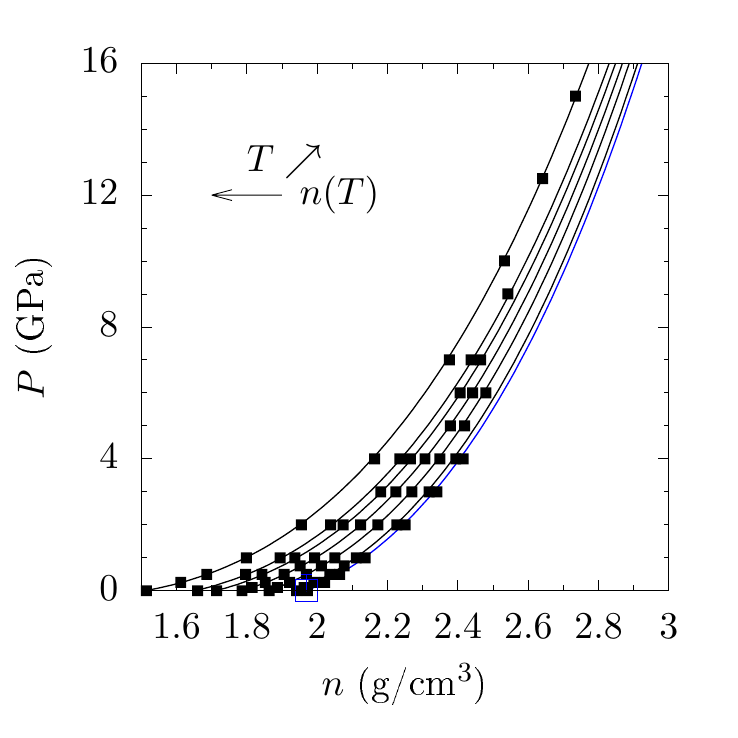}}\\
\subfloat[\ce{K2CO3}\label{feos1k}]{\includegraphics{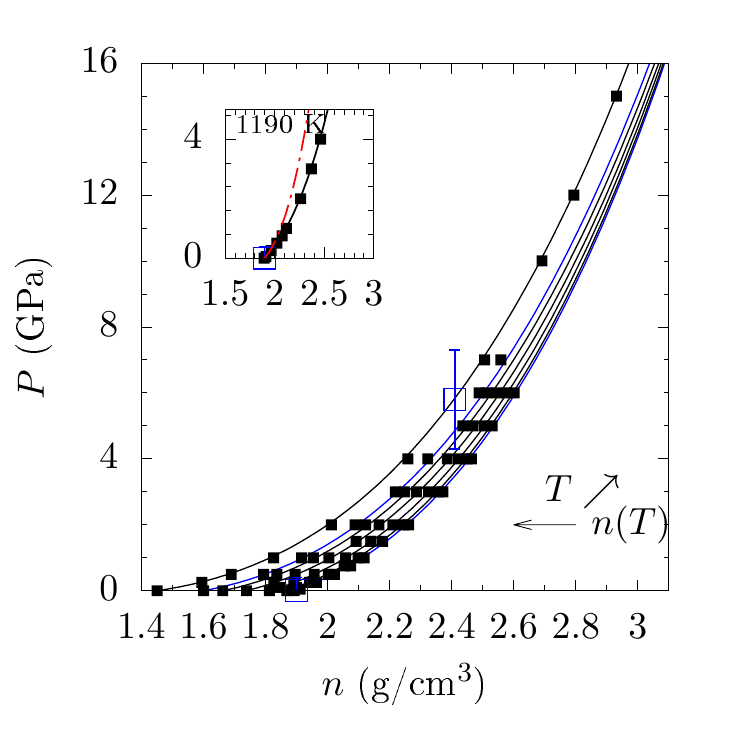}}
\caption[Compression isotherms for \ce{Li2CO3}, \ce{Na2CO3} and \ce{K2CO3}]{MD (black) and AIMD (blue) calculated density-pressure points. Compression isotherms (black curves) are interpolated from the MD points by the Birch-Murnaghan equation of state (BMEoS). The temperature of the MD isotherms are \{1100~K, 1140~K, 1200~K, 1350~K, 1500~K, 1650~K, 1773~K, 1923~K\} for $\ce{Li2CO3}$, \{1140~K, 1200~K, 1350~K, 1500~K, 1650~K, 1773~K, 2073~K\} for $\ce{Na2CO3}$  and \{1190~K, 1240~K, 1350~K, 1500 K, 1650~K, 1773~K, 2073~K\} for $\ce{K2CO3}$. Blue squares are AIMD points (1140~K for \ce{Na2CO3}, 1190~K and 1773~K for \ce{K2CO3}). For K$_2$CO$_3$ a BMEoS isotherm issued from the experimental thermodynamics data and fusion curve analysis at 1190~K is also plotted in the inset (red curve).\cite{Liu2003,OLeary2015,Wang2016}}
\label{feosall}
\end{figure*}

\clearpage

\begin{figure*}[t]
    \subfloat[\,Li in \ce{Li2CO3}]{\includegraphics{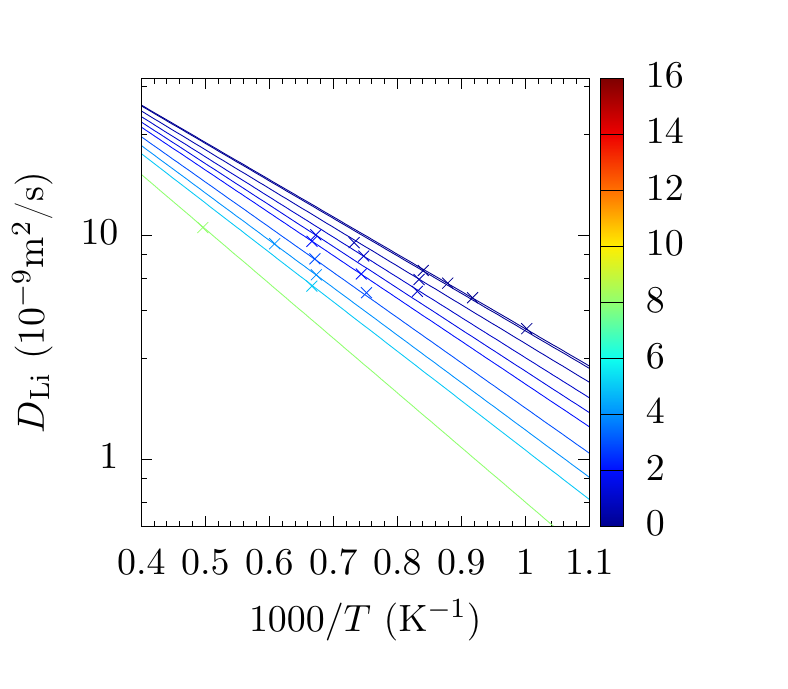}} 
    \subfloat[\,CO$_3$ in \ce{Li2CO3} as a function of $T$ and $P$]{\includegraphics{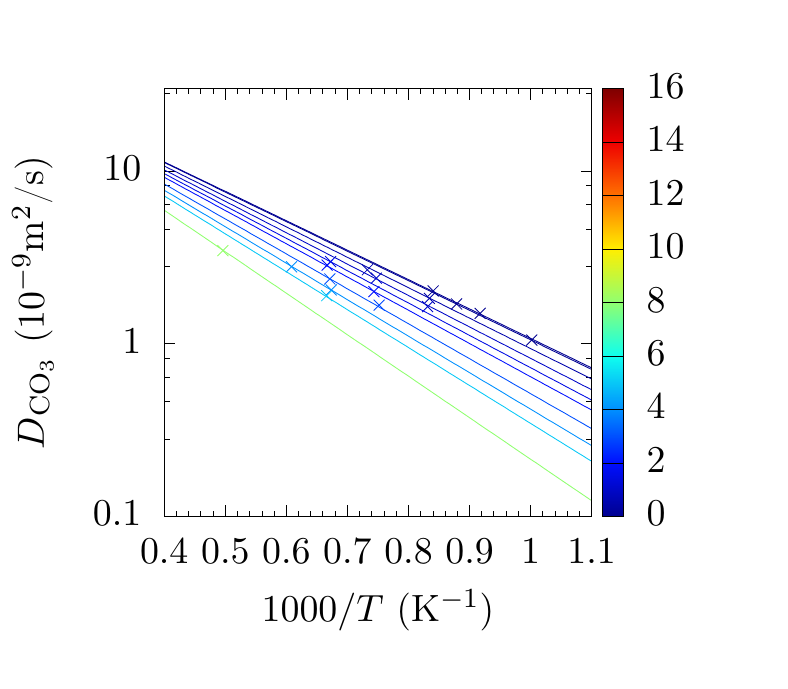}}
\caption[Diffusion coefficients in \ce{Li2CO3} as a function of $T$ and $P$]{Diffusion coefficients in $\ce{Li2CO3}$ calculated by MD (crosses) and isobaric activation laws obtained by fitting all the simulation points. The pressures of the isobars are \{0.001, 0.05, 0.5, 1, 1.5, 2, 3, 4, 5, 8\} in GPa and are referred to with a color code (vertical scale).}
\label{fdiff1}  
\end{figure*}

\begin{figure*}
    \subfloat[\,Na in \ce{Na2CO3}]{\includegraphics{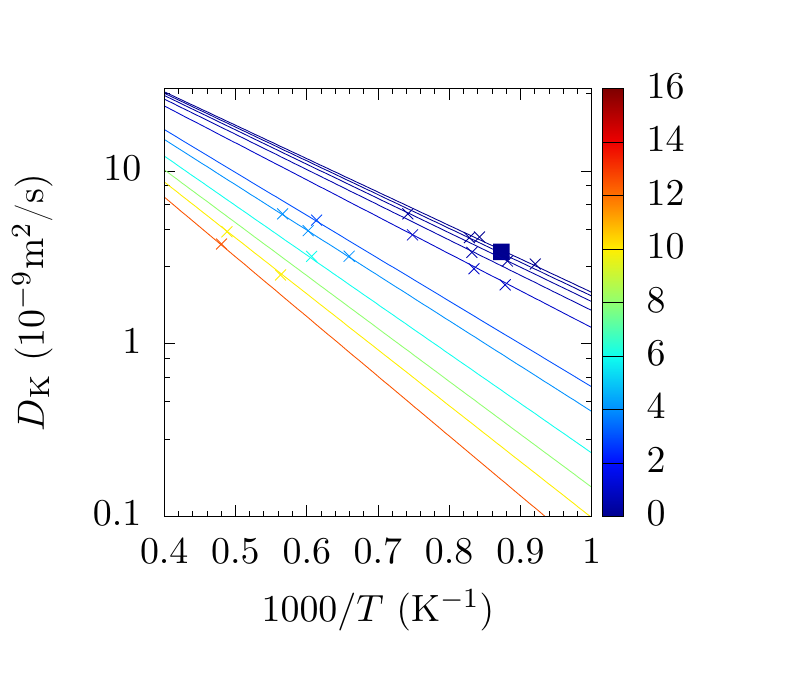}} 
    \subfloat[\,CO$_3$ in \ce{Na2CO3}]{\includegraphics{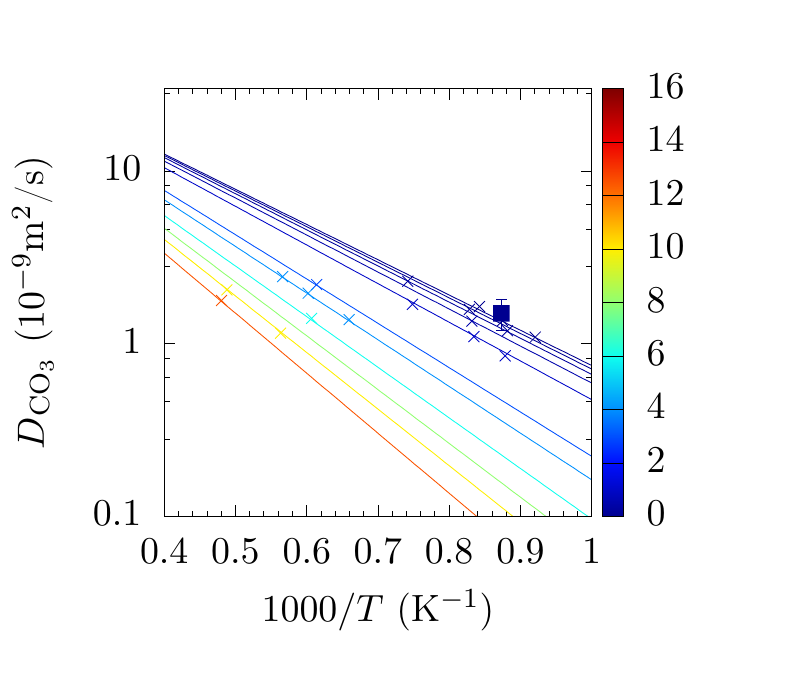}}
\caption[Diffusion coefficients in \ce{Na2CO3} as a function of $T$ and $P$]{As Figure~\ref{fdiff1}, but for $\ce{Na2CO3}$. The values calculated from the AIMD simulations are represented by plain squares (see Table~\ref{tdft}). The pressures of the isobars are \{0.001, 0.1, 0.25, 0.5, 1, 3, 4, 6, 8, 10, 12.5\} in GPa and are referred to with a color code (vertical scale).}
\label{fdiff2}  
\end{figure*}

\begin{figure*}
    \subfloat[\,K in \ce{K2CO3}]{\includegraphics{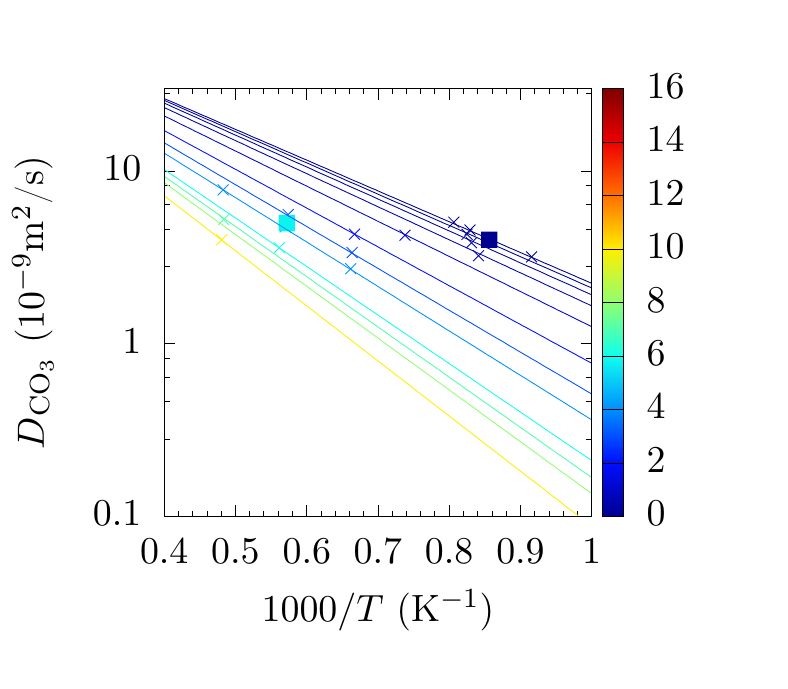}}
    \subfloat[\,CO$_3$ in \ce{K2CO3}]{\includegraphics{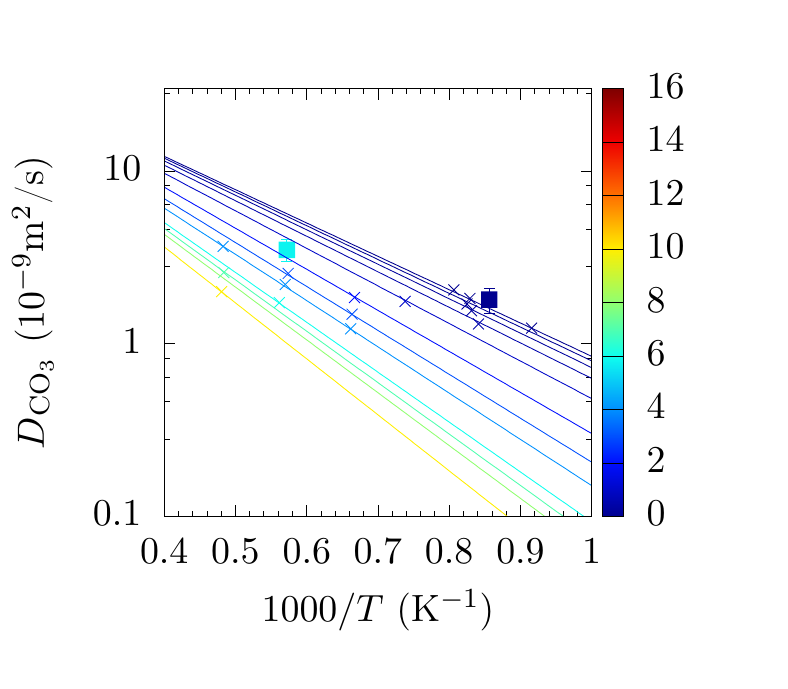}}
\caption[Diffusion coefficients in \ce{K2CO3}  as a function of $T$ and $P$]{As Figure~\ref{fdiff1}, but for \ce{K2CO3}. The values calculated from the AIMD simulations are represented by plain squares (see Table~\ref{tdft}).  The pressures of the isobars are \{0.001, 0.25, 0.5, 1, 2, 3, 4, 6, 8, 10\} in GPa and are referred to with a color code (vertical scale).}
\label{fdiff3}  
\end{figure*}

\begin{figure*}
\centering
    \subfloat[ \,\ce{Li2CO3}]{\includegraphics{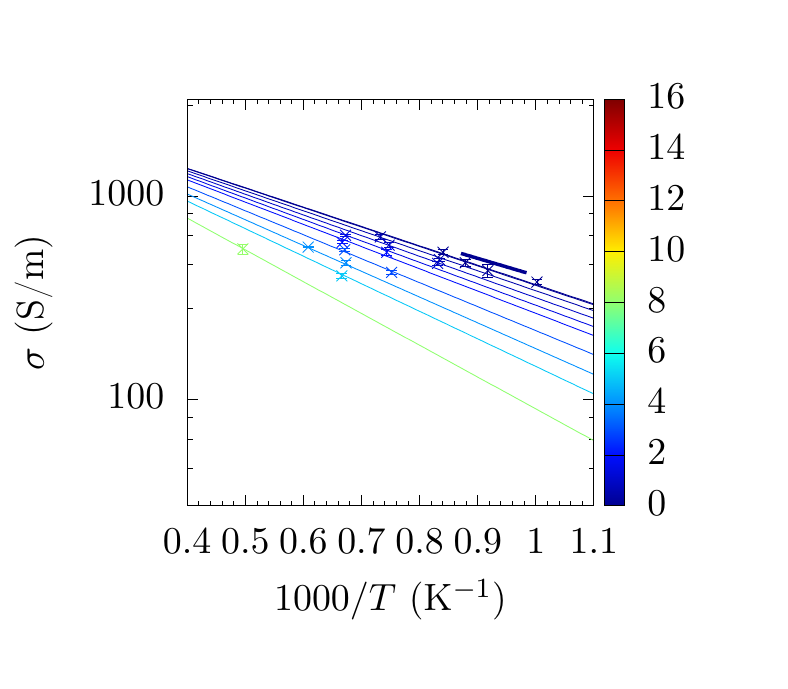}}
    \subfloat[ \,\ce{Na2CO3}]{\includegraphics{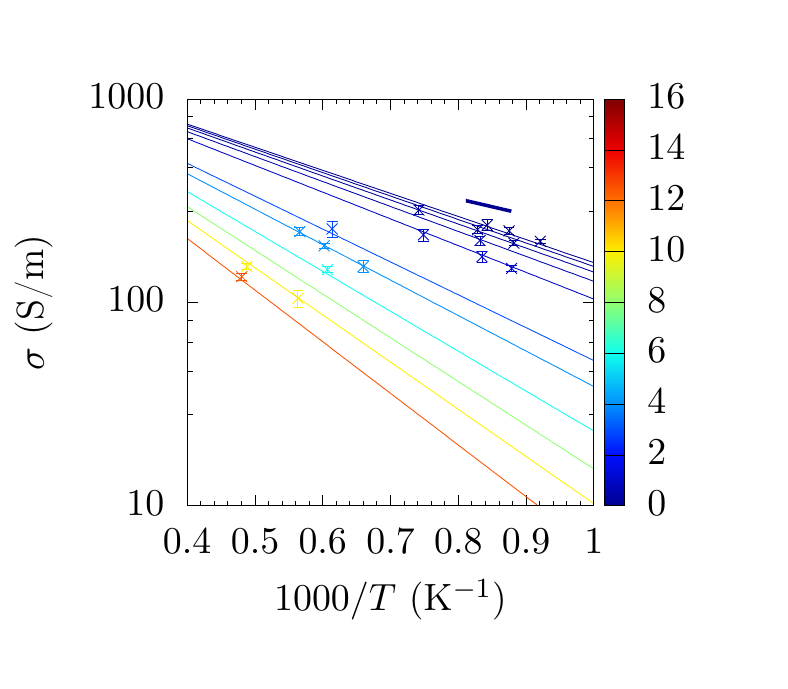}}\\
    \subfloat[\, \ce{K2CO3}]{\includegraphics{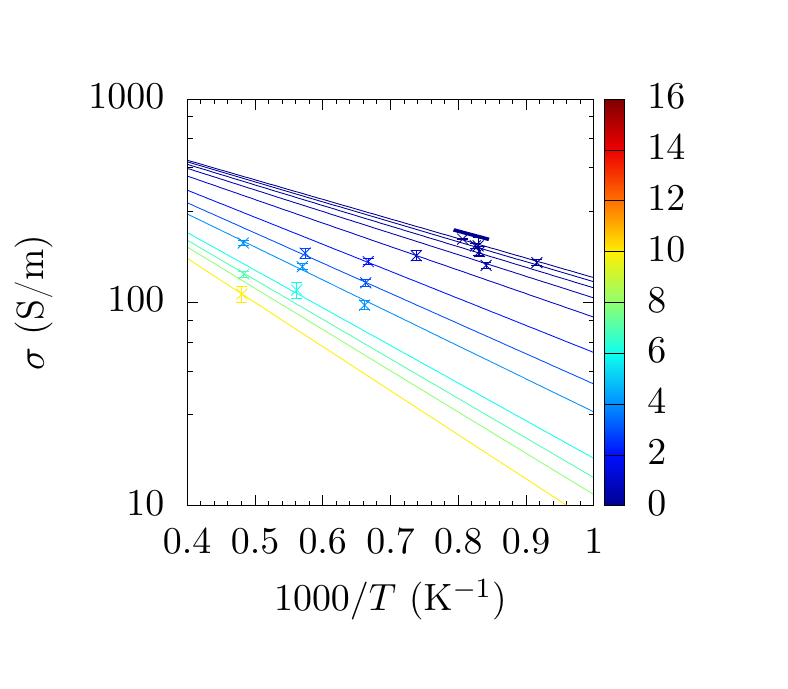}}
\caption[Electrical conductivity as a function of $T$ and $P$]{Electrical conductivity from MD (crosses) and isobaric activation laws (plain lines). The pressures (in GPa) of the isobars are \{0.001, 0.05, 0.5, 1, 1.5, 2, 3, 4, 5, 8\} for \ce{Li2CO3}, \{0.001, 0.1, 0.25, 0.5, 1, 3, 4, 6, 8, 10, 12.5\} for \ce{Na2CO3}  and \{0.001, 0.25, 0.5, 1, 2, 3, 4, 6, 8, 10\} for \ce{K2CO3} and are referred to with a color code (vertical scale). Bold segments are experimental values of \citeauthor{Kojima2009}\cite{Kojima2009} at $P=0$.}
\label{fcond} 
\end{figure*}

\begin{figure*}
\centering
\subfloat[\,\ce{Li2CO3}]{{\includegraphics{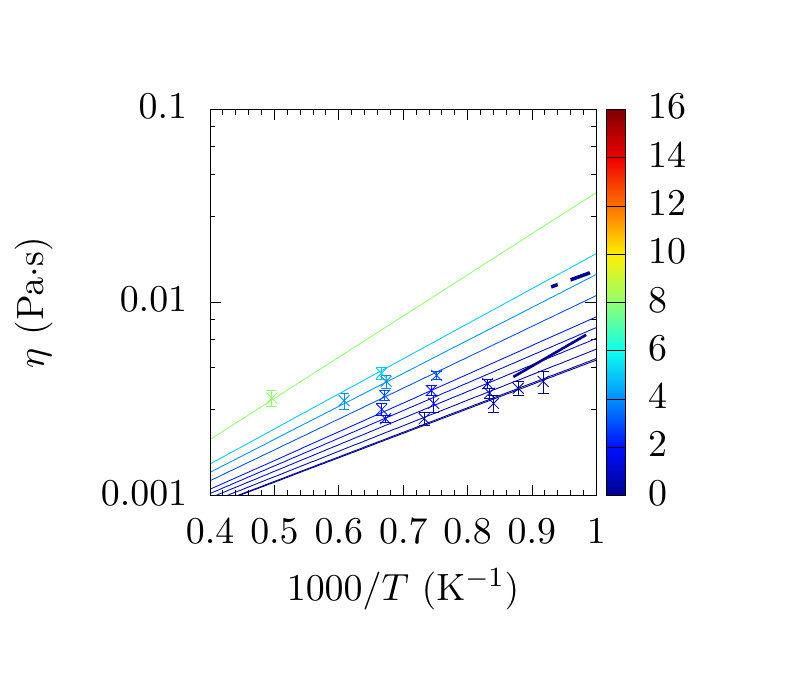}}\label{fviscoli}}
\subfloat[\,\ce{Na2CO3}]{{\includegraphics{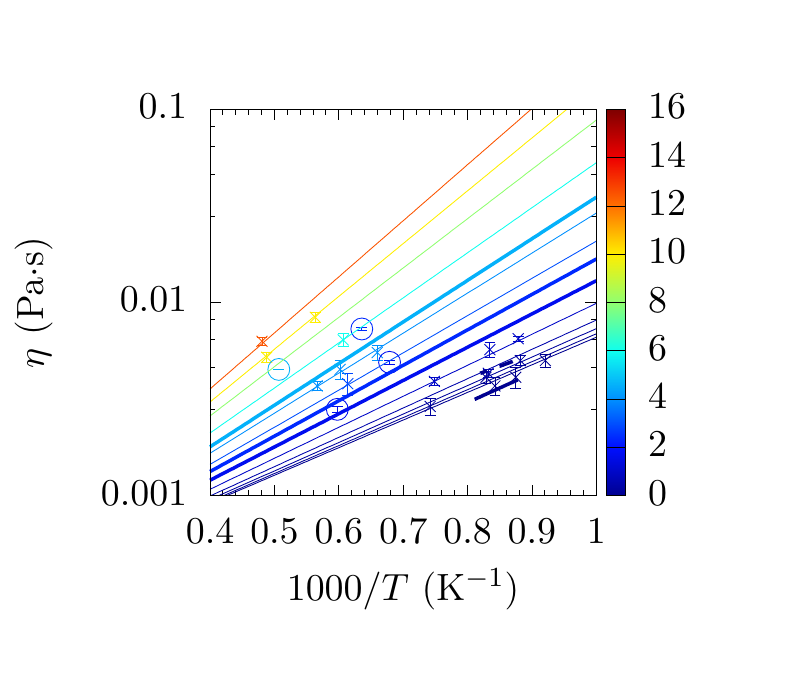}}\label{fviscona}}\\
\subfloat[\,\ce{K2CO3}]{{\includegraphics{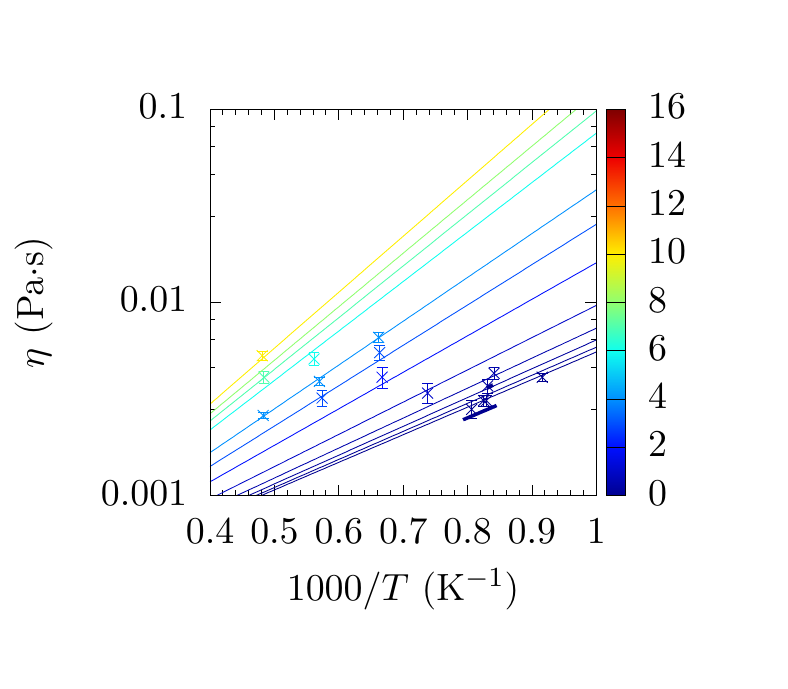}}\label{fviscok}}
\caption[Viscosity as a function of $T$ and $P$]{Viscosity from MD (crosses) and isobaric activation laws obtained for each composition by interpolation of all simulation points. The pressures (in GPa) of the isobars are \{0.001, 0.05, 0.5, 1, 1.5, 2, 3, 4, 5, 8\} for $\ce{Li2CO3}$, \{0.001, 0.1, 0.25, 0.5, 1, 3, 4, 6, 8, 10, 12.5\} for $\ce{Na2CO3}$  and \{0.001, 0.25, 0.5, 1, 2, 3, 4, 6, 8, 10\} for $\ce{K2CO3}$ and are referred to with a color code (vertical scale). Bold segments are experimental data, those from \citet{Janz1988} and \citet{Sato1999} are almost indistinguishable and are thus plotted together as plain lines, whereas the data of \citet{DiGe16} are plotted as dotted line. For \ce{Na2CO3} the high pressure data recently reported by \citet{Stagno2018} are plotted as circles, the corresponding isobars (1.7, 2.4 and 4.6~GPa) obtained from the present MD study are reported for comparison (bold blue lines).}
\label{fvisco}
\end{figure*}

\begin{figure}[ht]
\centering
\includegraphics{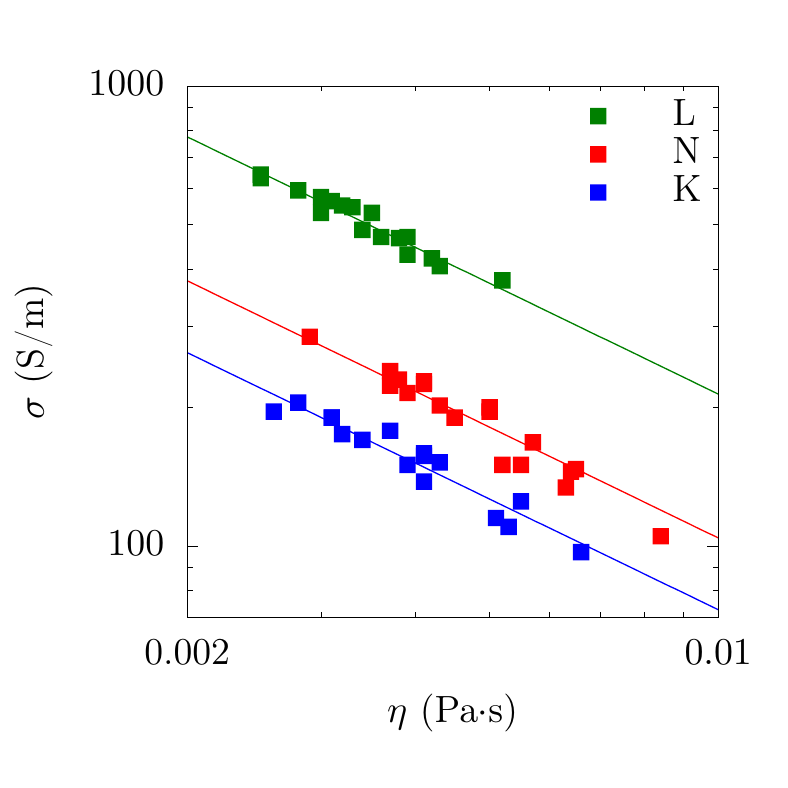}  \caption[Electrical conductivity as a function of viscosity]{Electrical conductivity as a function of viscosity. The squares are the MD data and the lines are empirical fits of the MD data by equation $\sigma=A/\eta^{0.8}$ (where $\sigma$ is in S/m and $\eta$ in Pa$\cdot$s) with $A=$ 5.375, 2.615 and 1.825 for \ce{Li2CO3} (L), \ce{Na2CO3} (N) and \ce{K2CO3} (K), respectively.}
\label{sigmaeta}
\end{figure}

\clearpage

\begin{figure*}
\subfloat[]{\includegraphics{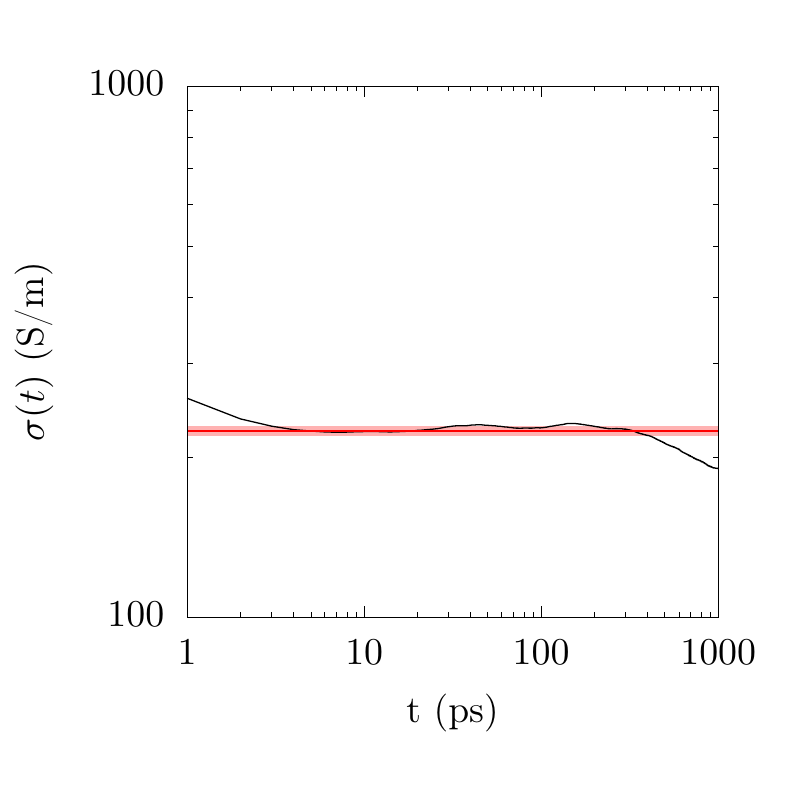}
 \label{fcondt1}	}	
\subfloat[]{\includegraphics{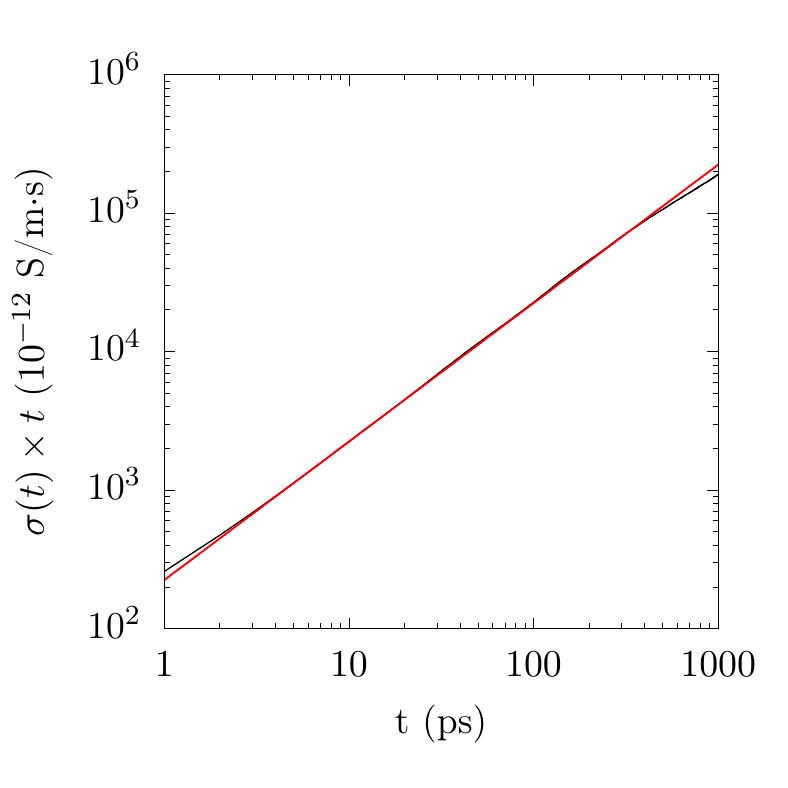}}
  \label{fcondt2}    	   	                          
\caption[Method for the calculation of the electrical conductivity]{Example of the method used to estimate the electrical conductivity from the MD simulations. (a) For the conductivity to be calculated the running average $\sigma (t) = \frac{e^2}{k_BTV} \cdot \frac{1}{6t}\Big\langle \Big\vert\sum_{i} z_i \big(\mathbf{r}_i(t)-\mathbf{r}_i(0)\big)\Big\vert ^{2} \Big\rangle$ has to reach a plateau (black line). (b) Equivalently convergence is attained when $\sigma(t) \times t$ is linear. In red we plot (a) the calculated conductivity, $\sigma$, or (b) $\sigma \times t$, bracketed within uncertainty (red shaded area).}
\label{fcondt}
\end{figure*}

\begin{figure*}
\subfloat[]{\includegraphics{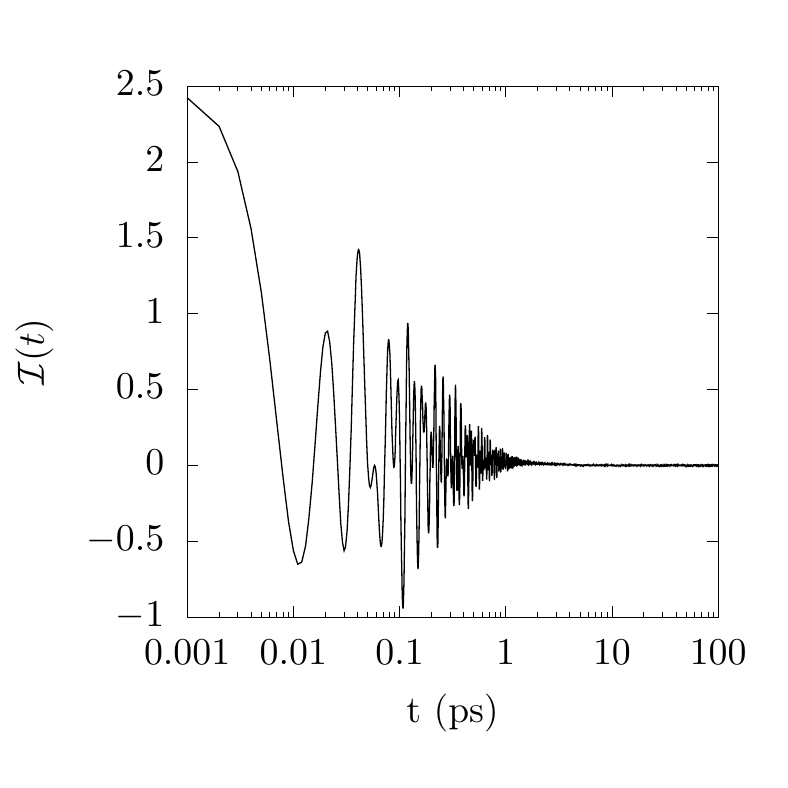}
 \label{fstress}	}	
\subfloat[]{\includegraphics{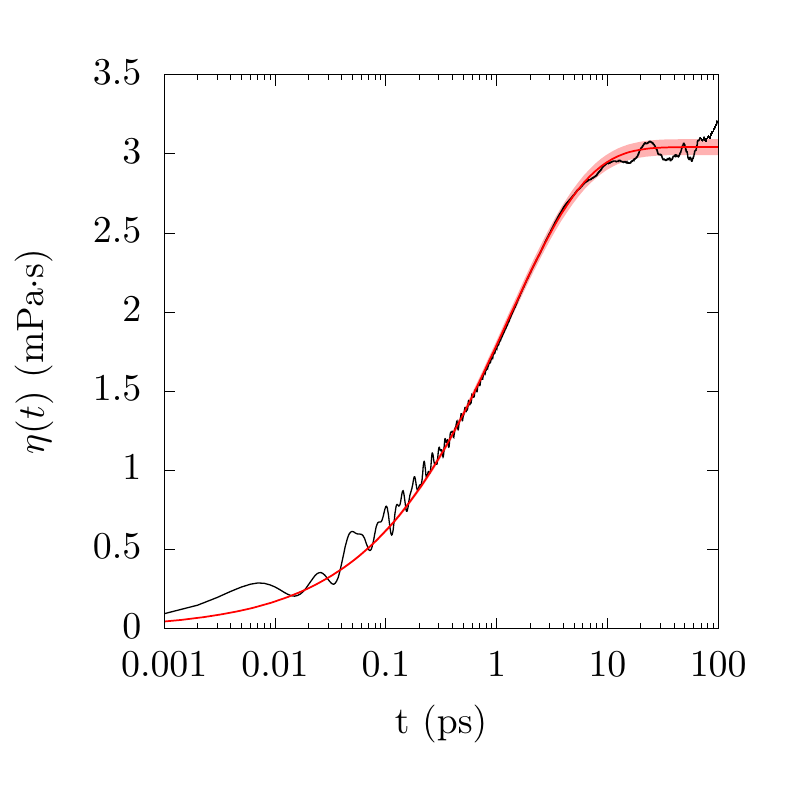}}
  \label{fviscot}    	   	                          
\caption[Method for the calculation of the viscosity]{Example of the method used to estimate the viscosity from the MD simulations. (a) For the viscosity to be calculated the running average $\mathcal{I}(t)=\frac{V}{RT}\langle \Pi_{\alpha \beta}(t)\cdot \Pi_{\alpha \beta}(0) \rangle$, where $\Pi_{\alpha \beta}$ stand for the off-diagonal stress tensor components, has to converge to zero (see equation (12) of main text). (b) To calculate the viscosity, $\eta(t)=\int_{0}^{t}\text{d} t^\prime \mathcal{I}(t^\prime)$ is fitted by a stretched exponential: $ \eta_\infty\Big(1- \mathrm{e}^{(-t/\tau)^\beta}\Big)$ (red), where $\eta_\infty$ is the viscosity calculated for the considered MD run (i.e. $\eta=\eta_\infty$), $\tau$ is a relaxation time and $\beta \in \mathopen{]}0\,;1\mathclose{]}$ is a stretching factor, typically $\beta \in \mathopen{]}0.5\,;0.65\mathclose{]}$.}
\end{figure*}

\begin{figure*}
\subfloat[]{\includegraphics{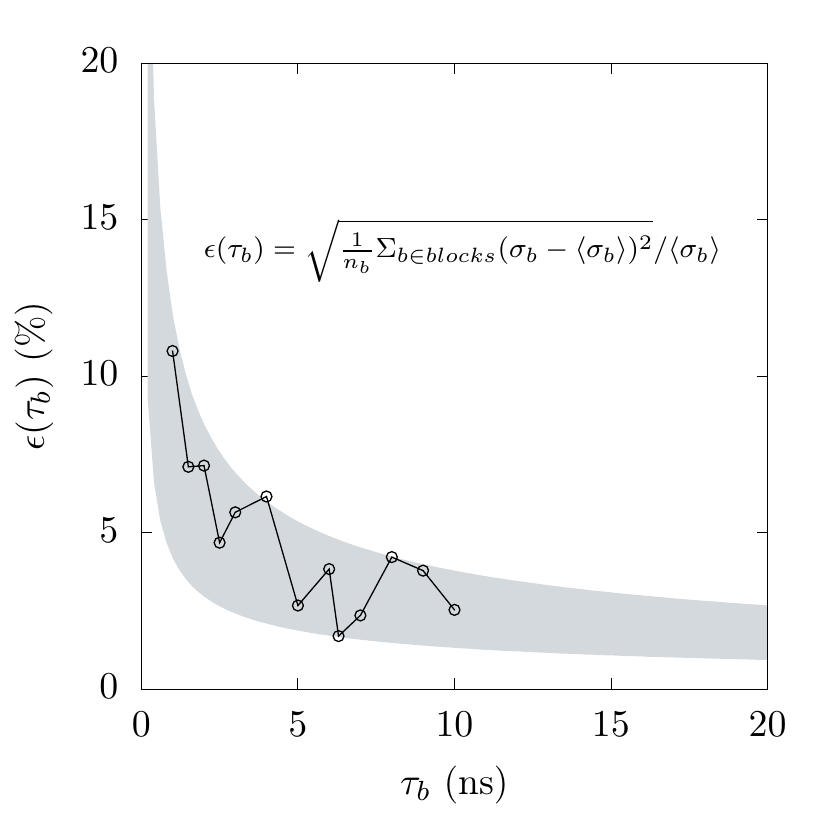}}
\subfloat[]{\includegraphics{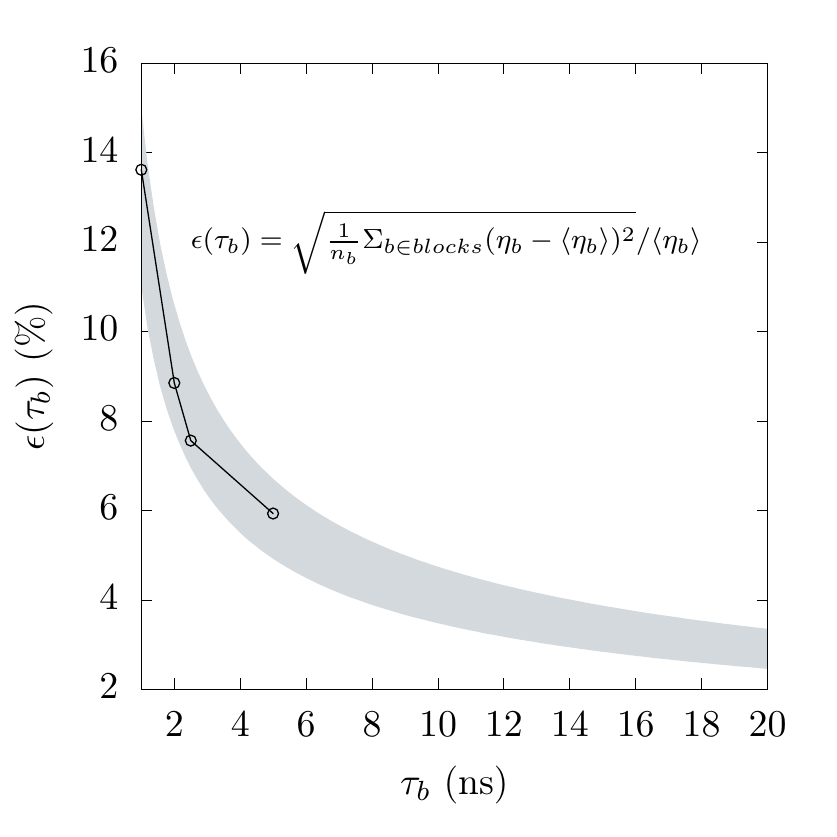}}
 \label{ferrcond}		   	                          
\caption[Uncertainties on the electrical conductivity and on the viscosity as standard deviations using block averaging]{Estimation of the standard deviation associated to the calculation of the conductivity and of the viscosity as correlation functions (equation (11) and (12) of main text), using a block averaging method. For blocks of different length $\tau_b$, we calculate (a) $\sigma (t) = \frac{e^2}{k_BTV} \cdot \frac{1}{6t}\Big\langle \Big\vert\sum_{i} z_i \big(\mathbf{r}_i(t)-\mathbf{r}_i(0)\big)\Big\vert ^{2} \Big\rangle$ or (b) $\eta(t)=\int_{0}^{t}\text{d} t^\prime \mathcal{I}(t^\prime)$, where $t$ is a time at which convergence is attained (see Figure~\ref{fcondt} and S10). We arbitrarily chose $t=20$~ps. Standard deviation as a function of block length, $\epsilon(\tau_b)$, decreases as $1/\tau_b$, as represented by the gray shade. For a typical run length ($10 - 30$~ns), $\epsilon(\tau_b)$ is below 5~\% for the conductivity and for the viscosity. This example is based on a 20~ns long simulation of the \ce{Li2CO3} melt at 1100~K and 1~bar.}
\end{figure*}

\clearpage

\begingroup
\squeezetable
\begin{table}
\begin{center}
\begin{tabular}{l c c c c c c c c c} 
		&$T$ 	& $\langle T \rangle$ & 	$n$ 		& $\langle P\rangle$ 	&	$\sigma$	&	$\eta$						   &  ${D}_{\text{Li}}$ 	&${D}_{\text{CO}_3}$	\\ 
		&(K) 	&			(K)	      &(g/cm$^{3}$)	& (kbar)	    &  (S/m) 			& (mPa$\cdot$s) & ($10^{-9}$m$^2$/s) & ($10^{-9}$m$^2$/s)\\ \hline
\\
\ce{Li2CO3}\\			
	& 1000 & 998  & 1.84  & 0 & 378 $\pm$ 10	& 5.2 $\pm$ 0.5 & 3.82 & 1.05 \\				        
	& 1100 & 1090 & 1.80 & 0 &	430 $\pm$ 30	& 3.9 $\pm$ 0.5	&	5.25	 & 1.50	\\ 
	& 1140 & 1138 & 1.78 & 0 &	470 $\pm$ 20	& 3.6 $\pm$ 0.3	& 6.10		& 1.70	&  	\\ 
\\	
	& 1200 & 1190 & 1.77 & 0.7 & 530 $\pm$ 15		& 3.0 $\pm$ 0.3	& 6.95 		& 2.02\\
\\
	& 1200 & 1199 & 1.82 & 5   & 487 $\pm$ 10		& 3.4 $\pm$ 0.2	& 6.33 & 1.83	 \\	
	& 1350 & 1365 & 1.77 & 6   & 630 $\pm$ 20	& 2.5 $\pm$ 0.2 	& 9.24		& 2.68	\\
\\
	& 1200 & 1203 & 1.88 & 10  & 467 $\pm$ 10		& 3.8 $\pm$ 0.2	& 5.61 &	1.64		\\	
	& 1350 & 1338 & 1.83 & 10  & 574 $\pm$ 15		& 3.0 $\pm$ 0.3	& 8.06 & 2.40 \\
\\
	& 1500 &	1487	& 1.83  & 15	& 642 $\pm$ 10	& 2.5 $\pm$ 0.1 & 10.01 & 2.99 	 \\
\\	
	& 1350 & 1345 & 1.92 & 20  & 530 $\pm$ 20 	& 3.5 $\pm$ 0.2 & 6.70 & 2.00\\
	& 1500 & 1500 & 1.88 & 20  & 593 $\pm$ 10		& 2.8 $\pm$ 0.2 & 9.37 & 2.85	\\	
\\
	& 1350 & 1330 & 1.99 & 29  & 422 $\pm$ 10		& 4.2 $\pm$ 0.2 & 5.53 & 1.67  \\
	& 1500 & 1490 & 1.95 & 30  & 545 $\pm$ 10	 	& 3.3 $\pm$ 0.2 & 7.83 & 2.37  	\\
\\	
	& 1500 & 1485 &	2.02 & 39 & 470 $\pm$ 15	&	3.9 $\pm$ 0.3	& 6.65 & 2.05	 \\
	& 1650 & 1644 & 1.99 & 40 &	562 $\pm$ 5	  	 	& 3.1 $\pm$ 0.3 & 9.14 & 2.79 		\\	
\\
	& 1500 & 1501 &	2.08 & 50 & 406 $\pm$ 10		& 4.3 $\pm$ 0.3 & 5.92 & 1.90\\	
\\
	& 2000 & 2016 & 2.12 & 81  & 550 $\pm$ 30 & 3.2 $\pm$ 0.3	 & 10.80 & 3.46 \\
\\	
\hline
\\
\end{tabular}
\end{center}
\caption[Summary of the calculated properties for \ce{Li2CO3}]{Thermodynamic averages and transport coefficients issued from all  MD simulations of $\mathbf{Li_2CO_3}$.}
\label{tantrans1}
\end{table}
\endgroup

\begingroup
\squeezetable
\begin{table}
\begin{center}
\begin{tabular}{l c c c c c c c c c} 
		&$T$ 	& $\langle T \rangle$ & 	$n$ 		& $\langle P\rangle$ 	&	$\sigma$	&	$\eta$						   &  ${D}_{\text{Na}}$ 	&${D}_{\text{CO}_3}$	\\ 
		&(K) 	&			(K)	      &(g/cm$^{3}$)	& (kbar)	    &  (S/m) 			& (mPa$\cdot$s) & ($10^{-9}$m$^2$/s) & ($10^{-9}$m$^2$/s)\\ \hline
\\
\ce{Na2CO3}\\
	& 1100	& 1086 & 2.00 & 0 & 200 $\pm$ 5 & 5.0 $\pm$ 0.4	& 2.89	& 1.09\\							        
	& 1140 & 1143 & 1.97 & 0 &	225 $\pm$ 5	& 4.1 $\pm$ 0.5	&	3.48 	& 1.36	 						\\ 
\texttt{ 
AIMD}  &  1140 & 1145 & 1.97 & 0 $\pm$ 5  & / & / & 3.4$\pm$0.3 & 1.5 $\pm$0.3   \\
	 & 1200 & 1187 & 1.94 & 0 & 242 $\pm$15	& 3.7 $\pm$0.5 			&	4.15 	&	 1.62		 			\\ 
	& 2073	& 2055	& 1.50	& 0		& 465 $\pm$ 10 	& 0.86 $\pm$ 0.03	& 21.42	& 9.78
\\
	& 1200	& 1207	& 1.93	& 1.2  	& 228 $\pm$ 10	& 4.1 $\pm$ 0.3	& 4.11 & 1.58\\
\\	
	& 1140 	& 1134	& 2.02 	& 2.3 & 196 $\pm$ 5 & 5.0 $\pm$ 0.3 &	3.01	& 1.19			\\
	& 1350	& 1348	& 1.92	& 2.5 &	285 $\pm$ 15 & 	2.9 $\pm$ 0.1	&	5.65	& 2.30	\\
\\	
	& 1200	& 1202	& 2.04	& 5.0 & 202 $\pm$ 10 	& 	4.3 $\pm$ 0.2  	 & 3.38	& 1.35 \\
\\
	& 1140 	& 1138	& 2.14	& 9.9 & 147 $\pm$ 5 	& 6.5 $\pm$ 0.2  & 2.19	& 0.85\\
	& 1200  & 1198	& 2.11	& 9.9 & 168 $\pm$ 10	& 5.7 $\pm$ 0.5	 & 2.72	& 1.10	 	\\
	& 1350	& 1336	& 2.05	& 9.7 & 215 $\pm$ 15 	& 3.9 $\pm$ 0.2  & 4.27 & 1.69		\\
\\	
	& 1650 & 1629 & 2.18 & 29 	 & 230$\pm$20	& 3.8$\pm$ 0.5 	& 5.20 & 2.20 				\\	
\\
	& 1500 & 1516 & 2.31 & 40 	 & 154 $\pm$ 10	& 5.5 $\pm$ 0.5 	& 3.21 	& 1.38	\\
	& 1650 & 1661 & 2.27 & 40	 & 190 $\pm$ 5	& 4.5 $\pm$ 0.5 	& 4.51 & 1.96	\\ 		
	& 1773 & 1766 & 2.23 & 40	 & 223 $\pm$ 10	& 3.7 $\pm$ 0.2 	& 5.65 &  2.45				\\	
\\
	& 1650 & 1648 & 2.41 & 60	 & 145$\pm$5	& 6.4 $\pm$ 0.5	& 3.20 & 1.40 				\\
	\\
	& 1773 & 1775 & 2.59 & 100	& 105 $\pm$ 10	& 8.4 $\pm$ 0.5	& 2.50 & 1.15				\\
	& 2073 & 2049 & 2.53 & 	99 	& 150 $\pm$ 5	& 5.2 $\pm$ 0.3	& 4.45 & 2.05  				\\			
\\
	& 2073 & 2081 & 2.64 & 125	 & 134 $\pm$ 5	& 6.3 $\pm$0.3 	& 3.78 & 1.78 				\\	
\\ 
\hline
\\
\end{tabular}
\end{center}
\caption[Summary of the calculated properties for \ce{Na2CO3}]{Thermodynamic averages and transport coefficients issued from all  MD simulations of $\mathbf{Na_2CO_3}$. Diffusion coefficients estimated from the AIMD simulation are also mentioned (\texttt{AIMD}).}
\label{tantrans2}
\end{table}
\endgroup

\begingroup
\squeezetable
\begin{table}
\begin{center}
\begin{tabular}{l c c c c c c c c c} 
		&$T$ 	& $\langle T \rangle$ & 	$n$ 		& $\langle P\rangle$ 	&	$\sigma$	&	$\eta$						   &  ${D}_{\text{K}}$ 	&${D}_{\text{CO}_3}$	\\ 
		&(K) 	&			(K)	      &(g/cm$^{3}$)	& (kbar)	    &  (S/m) 			& (mPa$\cdot$s) & ($10^{-9}$m$^2$/s) & ($10^{-9}$m$^2$/s)\\ \hline
\\
\ce{K2CO3}\\
	& 1100 & 1092 & 1.94 &	0 & 157 $\pm$ 5			&	4.1 $\pm$ 0.2	& 	3.18	& 1.23				\\  
	& 1200 & 1206 &	1.89 & 0 &	190 $\pm$20			& 	3.1 $\pm$ 0.2	&	 4.55	&	1.83 				\\ 
\texttt{AIMD}  & 1190 	& 1168	& 1.90 	& 0 $\pm$ 4 	&  	/				&		/			& 4.0 $\pm$ 0.2 & 1.8 $\pm$ 0.3   \\	
	& 1240 & 1240 &	1.87 & 0 & 205 $\pm$ 5 & 	2.8$\pm$ 0.3	& 5.05 	&	2.03 					\\ 		
\\	
	& 1200 & 1212 & 1.92 & 1.2 	 & 195 $\pm$ 10	& 3.1 $\pm$ 0.2	& 4.30 & 1.70 				\\ 	
\\
	& 1190 & 1203 & 1.96 & 2.7	 & 178 $\pm$ 10 	& 3.7 $\pm$ 0.3	& 3.85 & 1.55  	\\	
\\
	& 1190 & 1189 & 2.02 & 4.9	 & 152 $\pm$ 5  	& 4.3 $\pm$ 0.3 	& 3.23 & 1.30		\\
\\	
	& 1350 & 1355 & 2.06 & 10 	 & 170 $\pm$ 10 & 3.4 $\pm$ 0.4	& 4.25 & 1.76 \\		
	& 2073 & 2025 & 1.83 & 9.4	 & 318 $\pm$ 5	& 1.6 $\pm$ 0.1	& 14.05 & 6.35\\
\\
	& 1500 & 1499 & 2.17 & 20  	 & 159 $\pm$ 5	& 4.1 $\pm$ 0.5	& 4.31 & 1.85	  \\
\\	
	& 1500 & 1507 & 2.29 & 30 	 & 125 $\pm$ 5	& 5.5 $\pm$ 0.5	& 3.37 & 1.48 \\	
	& 1773 & 1742 & 2.22 & 29 	 & 175 $\pm$ 10	& 3.2 $\pm$ 0.3		&	5.60 	&	2.55 						\\ 	
\\
	& 1500 & 1511 & 2.39 & 40	& 97 $\pm$ 5		& 6.6 $\pm$ 0.4 	& 2.72 & 1.22	\\ 	
	& 1773 & 1755 & 2.32 & 40 	& 150 $\pm$ 5	& 3.9 $\pm$ 0.2 	& 4.75 &  2.20				\\	
	& 2073 & 2072 & 2.26 & 40	& 196 $\pm$ 5	&	2.6 $\pm$ 0.1		& 7.78 & 3.66\\
\\	
	& 1773 & 1781 &	2.49 & 60    &	115	$\pm$10	& 5.1$\pm$0.4		& 3.60	 	&	1.73 						\\
	\texttt{AIMD}  & 1773 & 1748 & 2.41 & 58 $\pm$ 15 &  /					& / &	5.0 $\pm$ 0.5				&  3.5 $\pm$ 0.5  \\	
\\
	& 2073 & 2068 & 2.51 & 70	 & 138 $\pm$ 5	& 4.1 $\pm$ 0.3 	& 5.25 &  	2.58			\\
\\
	& 2073 & 2080 &	2.69 & 100  & 110 $\pm$ 10	& 5.3 $\pm$ 0.3		& 	4.00 	&	2.00						\\
\\  
\hline
\\
\end{tabular}
\end{center}
\caption[Summary of the calculated properties for \ce{K2CO3}]{Thermodynamic averages and transport coefficients issued from all  MD simulations of $\mathbf{K_2CO_3}$. Diffusion coefficients estimated from the AIMD simulations are also mentioned (\texttt{AIMD}).}
\label{tantrans3}
\end{table}
\endgroup

\begingroup
\squeezetable
\begin{table}
\begin{center}
\begin{tabular}{l c c c c c c c c c c} 
		&$T$ 	& $\langle T \rangle$ & 	$n$ 		& $\langle P\rangle$ 	&	$\sigma$	&	$\eta$		&  ${D}_{\text{X}_1}$ 					   &  ${D}_{\text{X}_2}$ 	&${D}_{\text{CO}_3}$	\\ 
		&(K) 	&			(K)	      &(g/cm$^{3}$)	& (kbar)	    &  (S/m) 			& (mPa$\cdot$s) & ($10^{-9}$m$^2$/s) & ($10^{-9}$m$^2$/s) & ($10^{-9}$m$^2$/s)\\ \hline
\\
\ce{Li2CO3}--\ce{Na2CO3}\\
0.70: 0.30 	& 1100	& 1102	& 1.87& 0 &  330 $\pm$ 15 & 3.9 $\pm$ 0.3 & 4.55 & 3.68 & 1.39	 \\
	& 1173	& 1169 & 1.84 & 0 & 350 $\pm$ 10 & 3.4 $\pm$ 0.5 & 5.55 & 4.48 & 1.78\\		
\\
0.53: 0.47 (LNe)
	& 800	& 793	& 2.06 	& 0 & 84 $\pm$ 5 & 18 $\pm$ 2&  0.81 & 0.62 & 0.22	\\
	& 1000	& 979	& 1.95	& 0	& 208 $\pm$ 10	& 7.0 $\pm$ 1 &	 2.59 & 2.15 & 0.78 \\			
	& 1100  & 1089  & 1.90	& 0 & 265 $\pm$ 10	& 4.4 $\pm$ 0.2			  & 3.98  &	 3.41	  &	1.28 \\
\\
0.30: 0.70 
	& 1100	& 1088	& 1.95	& 0	& 9	& 4.6 $\pm$ 0.2	& 3.54	& 3.14 & 1.18 \\							
\\ \hline \\
\ce{Li2CO3}--\ce{K2CO3}\\	
0.70: 0.30 & 1100	& 1102 & 1.84 & 0 &	 326 $\pm$ 10 & 4.0 $\pm$ 0.3 & 4.54 &	3.69	 & 1.40	\\
			& 1173	& 1170 & 1.83 & 0 & 275 $\pm$ 15 & 3.1 $\pm$ 0.3 & 4.95 & 3.95 & 1.78				 \\
\\
0.62: 0.38 (LKe)  
			& 900	& 894  & 1.98  & 0 & 105 $\pm$ 5   & 10.5 $\pm$ 0.5	& 1.44 & 1.11 & 0.45 \\	  
			& 1100	& 1096 & 1.88  & 0 &	200 $\pm$ 10  & 3.9 $\pm$ 0.3 &	3.65 & 3.05 & 1.31     \\
\\					
0.50: 0.50
			& 1100	& 1082	& 1.89 & 0	& 186 $\pm$ 5	& 4.2 $\pm$ 0.2 & 3.23 & 2.93 & 1.25 \\
\\
0.43: 0.57 (LKe2)
			& 900	& 902	& 1.99	&	0	&	88 $\pm$ 5			&	10.5 $\pm$ 1			&	1.22 	& 1.12		& 0.42		\\				
			& 1100	& 1092	& 1.91	&	0	&	170 $\pm$ 10		&	3.9 $\pm$ 0.3 			&	2.99 	&	2.90 	&	1.20	\\							\\		
0.30: 0.70
			& 1100	& 1093	& 1.92 & 0	& 157 $\pm$ 5				& 	4.0   $\pm$ 0.2 & 	2.85	& 3.05	 & 1.24 \\							
\\ \hline  \\
\ce{Na2CO3}--\ce{K2CO3}\\
0.58: 0.42 	
	& 1000	& 995	& 2.02	& 0 & 124 $\pm$ 10 & 7.0 $\pm$ 0.5 & 1.89 & 1.81 & 0.70  \\
	& 1073 & 1070 & 1.99 & 0 & 148 $\pm$ 5  & 5.0 $\pm$ 0.3  & 2.62 & 2.59 & 1.00 \\	
\\							
0.50: 0.50 (NK)
	& 1040 & 1038 & 2.00 & 0 & 136 $\pm$ 5  & 6.4 $\pm$ 0.5  & 2.27 & 2.27 & 0.86 \\		
	& 1200 & 1185 & 1.91 & 0 & 195 $\pm$ 15 &  3.1 $\pm$ 0.4 & 4.00 & 4.08 & 1.60				 \\							
\\ \hline 
\end{tabular}
\end{center}
\caption[Summary of the calculated for binary mixtures]{Thermodynamic averages and transport coefficients issued from all  MD simulations of binary mixtures $\mathbf{Li_2CO_3}$--$\mathbf{Na_2CO_3}$, $\mathbf{Li_2CO_3}$-$\mathbf{K_2CO_3}$ and $\mathbf{Na_2CO_3}$-$\mathbf{K_2CO_3}$. The notation
X$_1$ (= Li or Na) and X$_2$ (= Na or K) are attributed in the same order as appearing in the composition name (first column).}
\label{tantrans4}
\end{table}
\endgroup

\begingroup
\squeezetable
\begin{table}
\begin{center}
\begin{tabular}{l c c c c c c c c c c c} 
		&$T$ 	& $\langle T \rangle$ & 	$n$ 		& $\langle P\rangle$ 	&	$\sigma$	&	$\eta$		&  ${D}_{\text{Li}}$ 					   &  ${D}_{\text{Na}}$ &  ${D}_{\text{K}}$	&${D}_{\text{CO}_3}$	\\ 
		&(K) 	&			(K)	      &(g/cm$^{3}$)	& (kbar)	    &  (S/m) 			& (mPa$\cdot$s) & ($10^{-9}$m$^2$/s)  & ($10^{-9}$m$^2$/s)  & ($10^{-9}$m$^2$/s) & ($10^{-9}$m$^2$/s)\\ \hline
\\
\ce{Li2CO3}--\ce{Na2CO3}--\ce{K2CO3}\\	
0.435: 0.315: 25 (LNKe)
	& 750 & 743 & 2.10 & 0 & 40 $\pm$ 5 & 35 $\pm$ 5	& 0.37 & 0.30 & 0.26 & 0.10 \\
	& 900  & 896 & 2.02 & 0 & 107 $\pm$ 10 & 10.7 $\pm$ 0.3 & 1.38 & 1.20 & 1.05 & 0.44 \\
	& 1050 & 1048 & 1.94 & 0 & 188 $\pm$ 10 & 4.9 $\pm$ 0.3 & 2.88 & 2.67 & 2.38 & 1.02 \\
\\
0.33: 0.33: 0.33 (LNK)	
	& 900	& 893	& 2.03 & 0 & 90 $\pm$ 10 & 13.7 $\pm$ 1	& 1.21 & 1.13 & 1.02 & 0.40 \\
	& 1073	& 1072 & 1.93  	& 0	& 176 $\pm$ 10 & 4.5 $\pm$ 0.3 & 3.01 & 2.90 & 2.68 & 1.13 \\
\\ \hline 
\end{tabular}
\end{center}
\caption[Summary of the calculated for ternary mixtures]{Thermodynamic averages and transport coefficients issued from all  MD simulations of ternary mixtures $\mathbf{Li_2CO_3}$--$\mathbf{Na_2CO_3}$--$\mathbf{K_2CO_3}$.}
\label{tantrans5}
\end{table}
\endgroup

\begingroup
\squeezetable
\begin{table}
\begin{tabular}{l c c c c c c} 
		&$T$ 	& $\langle T \rangle$ & 	$n$ 		& $\langle P\rangle$ 	&	$\gamma$ &	$\gamma_{\rm{Exp}}$	\\ 
		&(K) 	&			(K)	      &(g/cm$^{3}$)		& (kbar)	    & (mN/m) & (mN/m) \\ \hline
\\
\ce{Li2CO3}						        
	& 1100 & 1089 & 1.80 & -0.4 &	263 $\pm$ 5 & 240 \citep{Kojima2009} \\
	& 1140 & 1138 & 1.78 & -0.2	& 259 $\pm$ 3  & 239 \citep{Kojima2009} \\
	& 1200 & 1195 & 			   & -0.4 &	 256 $\pm$ 5 & 236 \citep{Kojima2009} \\	
\\
\hline
\\
\ce{Na2CO3}						        
	& 1140 & 1137 & 1.97 & -0.3 & 224 $\pm$ 5	&  211 \citep{Kojima2009} \\
	 & 1200 & 1203  & 1.94 & -0.2  & 219 		$\pm$ 4	& 208 \citep{Kojima2009} \\
\\ \hline \\
\ce{K2CO3}		& 1200 & 1168 &	1.89 & -0.3 &	178			$\pm$ 4	   &  167 \citep{Kojima2009} \\
	& 1240 & 1225 &	1.87 & -0.2	& 171  		$\pm$ 4	 & 165 \citep{Kojima2009} \\
\\ 
\hline	
\end{tabular}
\caption[Summary of the calculated surface tension for \ce{Li2CO3}, \ce{Na2CO3} and \ce{K2CO3}]{Surface tension from MD. For comparison values from the experimental literature are also reported ($\gamma_{\rm{Exp}}$). }
\label{tannstenatm}
\end{table}
\endgroup

\begingroup
\squeezetable
\begin{table}
\begin{tabular}{l c c c c c c} 
		&$T$ 	& $\langle T \rangle$ & 	$n$ 		& $\langle P\rangle$ 	&	$\gamma$ &	$\gamma_{\rm{Exp}}$	\\ 
		&(K) 	&			(K)	      &(g/cm$^{3}$)		& (kbar)	    & (mN/m) & (mN/m) \\ \hline
\\
\ce{Li2CO3}--\ce{Na2CO3}	\\							
0.70: 0.30 & 1173	& 1166 & 1.84 & -0.4 & 246 $\pm$ 3   & 228 \citep{Kojima2009}	\\				
\\
0.53:0.47 (LNe)	& 900	& 911	&  2.00		& -0.3	& 269 $\pm$ 5 & 249 \citep{Kojima2009}	\\
			& 	1100 &	1094 & 	1.90 	& -0.2	& 247 $\pm$ 3 & 231 \citep{Kojima2009}\\
\\ \hline \\
\ce{Li2CO3}--\ce{K2CO3}	\\							
0.70: 0.30 & 1173	& 1178 & 1.83 & -0.2 &  211	$\pm$ 3	&  195 \citep{Kojima2009} \\	
\\			
0.62: 0.38 (LKe) & 900 & 882	& 1.98	& -0.2	& 250 $\pm$ 3	& 207 \citep{Kojima2009}			 \\
		& 1100	& 1093	& 1.88	& -0.2	& 223 $\pm$ 3 & 193 \citep{Kojima2009}\\
\\ \hline \\	
\ce{Na2CO3}--\ce{K2CO3}\\							
0.50: 0.50 
	& 1040 & 1036 & 2.00	 & -0.2 & 212 $\pm$ 6 & 189 \citep{Kojima2009}\\
	& 1100 & 1093 & 1.97 & -0.3 & 204	 $\pm$ 3 & 185 \citep{Kojima2009} \\				 
	& 1200 & 1197 & 1.91 	 & -0.2 & 191 $\pm$ 5 	&  179 \citep{Janz1988} \\								
\\ \hline \\
\ce{Li2CO3}--\ce{Na2CO3}--\ce{K2CO3} \\
0.435: 0.315: 25 (LNKe)
	& 750	& 749 & 2.10	& -0.2	& 269 $\pm$ 5 & 223 \citep{Kojima2009}\\
	& 900 & 897 & 2.02 & 	-0.2 & 246 $\pm$ 2 & 213 \citep{Kojima2009}\\
	& 1200	& 1200	& 1.88 & -0.2	& 209 $\pm$ 3 & 194 \citep{Kojima2009}\\
\\	
0.33: 0.33: 0.33 (LNK)
	& 900   & 905   & 2.03	   & -0.2 & 239 $\pm$ 3    & 207 \citep{Kojima2009} \\ 		
	& 1100	& 1094	& 1.93 & -0.2 & 220  $\pm$ 4   & 194 \citep{Kojima2009} \\ 
\hline
\end{tabular}
\caption[Summary of the calculated surface tension for binary and ternary mixtures]{Surface tension from MD for binary and ternary mixtures. For comparison values from the experimental literature are also reported ($\gamma_{\rm{Exp}}$).}
\label{tannstenatm2}
\end{table}
\endgroup

\clearpage
\section*{CP2K input file example}
\noindent \texttt{\# CP2K input file example for Supplementary Material for \\
\# "Atomistic simulations of molten carbonates: \\
\# thermodynamic and transport properties  of the Li2CO3 - Na2CO3 - K2CO3 system" \\ 
@SET RESTART 1 \\ 
@SET LIBDIR /scratch/desmaele/cp2k/files\hspace*{0.5cm}\\ 
@SET SYSTEM K2CO3 \\ 
\& FORCE\_EVAL \\ 
\hspace*{0.5cm}METHOD QuickStep \\ 
\hspace*{0.5cm}STRESS\_TENSOR DIAGONAL\_ANALYTICAL \\ 
\hspace*{0.5cm}\& DFT \\ 
\hspace*{0.5cm}\hspace*{0.5cm}BASIS\_SET\_FILE\_NAME \$\{LIBDIR\}/BASIS\_MOLOPT \\ 
\hspace*{0.5cm}\hspace*{0.5cm}POTENTIAL\_FILE\_NAME \$\{LIBDIR\}/GTH\_POTENTIALS \\ 
\hspace*{0.5cm}\hspace*{0.5cm}\& MGRID \\ 
\hspace*{0.5cm}\hspace*{0.5cm}\hspace*{0.5cm}CUTOFF 700 \\ 
\hspace*{0.5cm}\hspace*{0.5cm}\hspace*{0.5cm}NGRIDS 6 \\ 
\hspace*{0.5cm}\hspace*{0.5cm}\hspace*{0.5cm}REL\_CUTOFF 40 \\ 
\hspace*{0.5cm}\hspace*{0.5cm}\& END MGRID \\ 
\hspace*{0.5cm}\hspace*{0.5cm}\& QS \\ 
\hspace*{0.5cm}\hspace*{0.5cm}\hspace*{0.5cm}EPS\_DEFAULT 1.0E-12 \\ 
\hspace*{0.5cm}\hspace*{0.5cm}\hspace*{0.5cm}MAP\_CONSISTENT TRUE \\ 
\hspace*{0.5cm}\hspace*{0.5cm}\hspace*{0.5cm}EXTRAPOLATION ASPC \\ 
\hspace*{0.5cm}\hspace*{0.5cm}\hspace*{0.5cm}EXTRAPOLATION\_ORDER 3 \\ 
\hspace*{0.5cm}\hspace*{0.5cm}\& END QS \\ 
\hspace*{0.5cm}\hspace*{0.5cm}\& SCF \\ 
\hspace*{0.5cm}\hspace*{0.5cm}\hspace*{0.5cm}MAX\_SCF 20 \\ 
\hspace*{0.5cm}\hspace*{0.5cm}\hspace*{0.5cm}SCF\_GUESS RESTART \\ 
\hspace*{0.5cm}\hspace*{0.5cm}\hspace*{0.5cm}EPS\_SCF 3.0E-7 \\ 
\hspace*{0.5cm}\hspace*{0.5cm}\hspace*{0.5cm}\& OUTER\_SCF \\ 
\hspace*{0.5cm}\hspace*{0.5cm}\hspace*{0.5cm}\hspace*{0.5cm}EPS\_SCF 3.0E-7 \\ 
\hspace*{0.5cm}\hspace*{0.5cm}\hspace*{0.5cm}\hspace*{0.5cm}MAX\_SCF 1000 \\ 
\hspace*{0.5cm}\hspace*{0.5cm}\hspace*{0.5cm}\& END OUTER\_SCF \\ 
\hspace*{0.5cm}\hspace*{0.5cm}\hspace*{0.5cm}\& OT ON \\ 
\hspace*{0.5cm}\hspace*{0.5cm}\hspace*{0.5cm}\hspace*{0.5cm}PRECONDITIONER FULL\_SINGLE\_INVERSE \\ 
\hspace*{0.5cm}\hspace*{0.5cm}\hspace*{0.5cm}\hspace*{0.5cm}MINIMIZER DIIS \\ 
\hspace*{0.5cm}\hspace*{0.5cm}\hspace*{0.5cm}\hspace*{0.5cm}N\_DIIS 6 \\ 
\hspace*{0.5cm}\hspace*{0.5cm}\hspace*{0.5cm}\& END OT \\ 
\hspace*{0.5cm}\hspace*{0.5cm}\hspace*{0.5cm}\& PRINT \\ 
\hspace*{0.5cm}\hspace*{0.5cm}\hspace*{0.5cm}\hspace*{0.5cm}\& RESTART\_HISTORY \\ 
\hspace*{0.5cm}\hspace*{0.5cm}\hspace*{0.5cm}\hspace*{0.5cm}\hspace*{0.5cm}BACKUP\_COPIES 0 \\ 
\hspace*{0.5cm}\hspace*{0.5cm}\hspace*{0.5cm}\hspace*{0.5cm}\& END RESTART\_HISTORY \\ 
\hspace*{0.5cm}\hspace*{0.5cm}\hspace*{0.5cm}\& END PRINT \\ 
\hspace*{0.5cm}\hspace*{0.5cm}\& END SCF \\ 
\hspace*{0.5cm}\hspace*{0.5cm}\& XC \\ 
\hspace*{0.5cm}\hspace*{0.5cm}\hspace*{0.5cm}\& XC\_FUNCTIONAL BLYP \\ 
\hspace*{0.5cm}\hspace*{0.5cm}\hspace*{0.5cm}\& END XC\_FUNCTIONAL \\ 
\hspace*{0.5cm}\hspace*{0.5cm}\hspace*{0.5cm}\& vdW\_POTENTIAL \\ 
\hspace*{0.5cm}\hspace*{0.5cm}\hspace*{0.5cm}\hspace*{0.5cm} DISPERSION\_FUNCTIONAL PAIR\_POTENTIAL \\ 
\hspace*{0.5cm}\hspace*{0.5cm}\hspace*{0.5cm}\hspace*{0.5cm} \& PAIR\_POTENTIAL \\ 
\hspace*{0.5cm}\hspace*{0.5cm}\hspace*{0.5cm}\hspace*{0.5cm}\hspace*{0.5cm}\hspace*{0.5cm}PARAMETER\_FILE\_NAME \$\{LIBDIR\}/dftd3.dat \\ 
\hspace*{0.5cm}\hspace*{0.5cm}\hspace*{0.5cm}\hspace*{0.5cm}\hspace*{0.5cm}\hspace*{0.5cm}TYPE DFTD3 \\ 
\hspace*{0.5cm}\hspace*{0.5cm}\hspace*{0.5cm}\hspace*{0.5cm}\hspace*{0.5cm}\hspace*{0.5cm}REFERENCE\_FUNCTIONAL BLYP \\ 
\hspace*{0.5cm}\hspace*{0.5cm}\hspace*{0.5cm}\hspace*{0.5cm}\hspace*{0.5cm}\hspace*{0.5cm}R\_CUTOFF 30.0 \\ 
\hspace*{0.5cm}\hspace*{0.5cm}\hspace*{0.5cm}\hspace*{0.5cm} \& END PAIR\_POTENTIAL \\ 
\hspace*{0.5cm}\hspace*{0.5cm}\hspace*{0.5cm}\& END vdW\_POTENTIAL \\ 
\hspace*{0.5cm}\hspace*{0.5cm}\hspace*{0.5cm}\& XC\_GRID \\ 
\hspace*{0.5cm}\hspace*{0.5cm}\hspace*{0.5cm}\hspace*{0.5cm}XC\_DERIV SPLINE3 \\ 
\hspace*{0.5cm}\hspace*{0.5cm}\hspace*{0.5cm}\& END XC\_GRID \\ 
\hspace*{0.5cm}\hspace*{0.5cm}\& END XC \\ 
\hspace*{0.5cm}\& END DFT \\ 
\hspace*{0.5cm}\& SUBSYS \\ 
\hspace*{0.5cm}\hspace*{0.5cm}\& CELL \\ 
\hspace*{0.5cm}\hspace*{0.5cm}\hspace*{0.5cm}ABC [angstrom] 16.0244	16.0244	16.0244 \\ 
\hspace*{0.5cm}\hspace*{0.5cm}\& END CELL \\ 
\hspace*{0.5cm}\hspace*{0.5cm}\& TOPOLOGY \\ 
\hspace*{0.5cm}\hspace*{0.5cm}\hspace*{0.5cm}COORD\_FILE\_FORMAT xyz \\ 
\hspace*{0.5cm}\hspace*{0.5cm}\hspace*{0.5cm}COORD\_FILE\_NAME ./k2co3.xyz \\ 
\hspace*{0.5cm}\hspace*{0.5cm}\hspace*{0.5cm}CONNECTIVITY OFF \\ 
\hspace*{0.5cm}\hspace*{0.5cm}\& END TOPOLOGY \\ 
\hspace*{0.5cm}\hspace*{0.5cm}\& KIND C \\ 
\hspace*{0.5cm}\hspace*{0.5cm}\hspace*{0.5cm}BASIS\_SET DZVP-MOLOPT-SR-GTH-q4\hspace*{0.5cm}\\ 
\hspace*{0.5cm}\hspace*{0.5cm}\hspace*{0.5cm}POTENTIAL GTH-BLYP-q4\hspace*{0.5cm}\\ 
\hspace*{0.5cm}\hspace*{0.5cm}\& END KIND \\ 
\hspace*{0.5cm}\hspace*{0.5cm}\& KIND O \\ 
\hspace*{0.5cm}\hspace*{0.5cm}\hspace*{0.5cm}BASIS\_SET DZVP-MOLOPT-SR-GTH-q6 \\ 
\hspace*{0.5cm}\hspace*{0.5cm}\hspace*{0.5cm}POTENTIAL GTH-BLYP-q6 \\ 
\hspace*{0.5cm}\hspace*{0.5cm}\& END KIND \\ 
\hspace*{0.5cm}\hspace*{0.5cm}\& KIND K \\ 
\hspace*{0.5cm}\hspace*{0.5cm}\hspace*{0.5cm}BASIS\_SET DZVP-MOLOPT-SR-GTH-q9 \\ 
\hspace*{0.5cm}\hspace*{0.5cm}\hspace*{0.5cm}POTENTIAL GTH-BLYP-q9 \\ 
\hspace*{0.5cm}\hspace*{0.5cm}\& END KIND \\ 
\hspace*{0.5cm}\& END SUBSYS \\ 
\& END FORCE\_EVAL \\ 
\& GLOBAL \\ 
\hspace*{0.5cm}PROJECT \$\{SYSTEM\} \\ 
\hspace*{0.5cm}RUN\_TYPE MD \\ 
\hspace*{0.5cm}PRINT\_LEVEL LOW \\ 
\hspace*{0.5cm}WALLTIME 86370 \\ 
\& END GLOBAL \\ 
\& MOTION \\ 
\hspace*{0.5cm}\& MD \\ 
\hspace*{0.5cm}\hspace*{0.5cm}ENSEMBLE NVT \\ 
\hspace*{0.5cm}\hspace*{0.5cm}STEPS 100000 \\ 
\hspace*{0.5cm}\hspace*{0.5cm}TIMESTEP 0.5 \\ 
\hspace*{0.5cm}\hspace*{0.5cm}TEMPERATURE 1190 \\ 
\hspace*{0.5cm}\hspace*{0.5cm}\& THERMOSTAT \\ 
\hspace*{0.5cm}\hspace*{0.5cm}\hspace*{0.5cm}TYPE CSVR \\ 
\hspace*{0.5cm}\hspace*{0.5cm}\hspace*{0.5cm}\& CSVR \\ 
\hspace*{0.5cm}\hspace*{0.5cm}\hspace*{0.5cm}\hspace*{0.5cm}TIMECON [fs] 1000 \\ 
\hspace*{0.5cm}\hspace*{0.5cm}\hspace*{0.5cm}\& END CSVR \\ 
\hspace*{0.5cm}\hspace*{0.5cm}\hspace*{0.5cm}\& NOSE \\ 
\hspace*{0.5cm}\hspace*{0.5cm}\hspace*{0.5cm}\hspace*{0.5cm}LENGTH 4 \\ 
\hspace*{0.5cm}\hspace*{0.5cm}\hspace*{0.5cm}\hspace*{0.5cm}YOSHIDA 9 \\ 
\hspace*{0.5cm}\hspace*{0.5cm}\hspace*{0.5cm}\hspace*{0.5cm}TIMECON [fs] 500.0 \\ 
\hspace*{0.5cm}\hspace*{0.5cm}\hspace*{0.5cm}\hspace*{0.5cm}MULTIPLE\_TIME\_STEPS 2 \\ 
\hspace*{0.5cm}\hspace*{0.5cm}\hspace*{0.5cm}\& END NOSE \\ 
\hspace*{0.5cm}\hspace*{0.5cm}\& END THERMOSTAT \\ 
\hspace*{0.5cm}\& END MD \\ 
\hspace*{0.5cm}\& PRINT \\ 
\hspace*{0.5cm}\hspace*{0.5cm}\& STRESS\hspace*{0.5cm}ON \\ 
\hspace*{0.5cm}\hspace*{0.5cm}\hspace*{0.5cm}FILENAME =\$\{SYSTEM\}.stress \\ 
\hspace*{0.5cm}\hspace*{0.5cm}\hspace*{0.5cm}\& EACH \\ 
\hspace*{0.5cm}\hspace*{0.5cm}\hspace*{0.5cm} MD 1 \\ 
\hspace*{0.5cm}\hspace*{0.5cm}\hspace*{0.5cm}\& END EACH \\ 
\hspace*{0.5cm}\hspace*{0.5cm}\& END STRESS \\ 
\hspace*{0.5cm}\hspace*{0.5cm}\& TRAJECTORY\hspace*{0.5cm}ON \\ 
\hspace*{0.5cm}\hspace*{0.5cm}\hspace*{0.5cm}FILENAME =\$\{SYSTEM\}.xyz \\ 
\hspace*{0.5cm}\hspace*{0.5cm}\hspace*{0.5cm}\& EACH \\ 
\hspace*{0.5cm}\hspace*{0.5cm}\hspace*{0.5cm}\hspace*{0.5cm}MD 1 \\ 
\hspace*{0.5cm}\hspace*{0.5cm}\hspace*{0.5cm}\& END EACH \\ 
\hspace*{0.5cm}\hspace*{0.5cm}\& END TRAJECTORY \\ 
\hspace*{0.5cm}\hspace*{0.5cm}\& VELOCITIES\hspace*{0.5cm}ON \\ 
\hspace*{0.5cm}\hspace*{0.5cm}\hspace*{0.5cm}FILENAME =\$\{SYSTEM\}.vel \\ 
\hspace*{0.5cm}\hspace*{0.5cm}\hspace*{0.5cm}\& EACH \\ 
\hspace*{0.5cm}\hspace*{0.5cm}\hspace*{0.5cm} MD 1 \\ 
\hspace*{0.5cm}\hspace*{0.5cm}\hspace*{0.5cm}\& END EACH \\ 
\hspace*{0.5cm}\hspace*{0.5cm}\& END VELOCITIES \\ 
\hspace*{0.5cm}\hspace*{0.5cm}\& FORCES\hspace*{0.5cm}ON \\ 
\hspace*{0.5cm}\hspace*{0.5cm}\hspace*{0.5cm}FILENAME =\$\{SYSTEM\}.force \\ 
\hspace*{0.5cm}\hspace*{0.5cm}\hspace*{0.5cm}\& EACH \\ 
\hspace*{0.5cm}\hspace*{0.5cm}\hspace*{0.5cm} MD 1 \\ 
\hspace*{0.5cm}\hspace*{0.5cm}\hspace*{0.5cm}\& END EACH \\ 
\hspace*{0.5cm}\hspace*{0.5cm}\& END FORCES \\ 
\hspace*{0.5cm}\hspace*{0.5cm}\& RESTART \\ 
\hspace*{0.5cm}\hspace*{0.5cm}\hspace*{0.5cm}FILENAME =\$\{SYSTEM\}.restart \\ 
\hspace*{0.5cm}\hspace*{0.5cm}\hspace*{0.5cm}\& EACH \\ 
\hspace*{0.5cm}\hspace*{0.5cm}\hspace*{0.5cm}\hspace*{0.5cm}MD 1 \\ 
\hspace*{0.5cm}\hspace*{0.5cm}\hspace*{0.5cm}\& END EACH \\ 
\hspace*{0.5cm}\hspace*{0.5cm}\& END RESTART \\ 
\hspace*{0.5cm}\& END PRINT \\ 
\& END MOTION \\ 
@if \$\{RESTART\} == 1 \\ 
\& EXT\_RESTART \\ 
\hspace*{0.5cm}RESTART\_FILE\_NAME \$\{SYSTEM\}.restart \\ 
\hspace*{0.5cm}RESTART\_COUNTERS\hspace*{0.5cm}\hspace*{0.5cm}T \\ 
\hspace*{0.5cm}RESTART\_AVERAGES\hspace*{0.5cm}\hspace*{0.5cm}T \\ 
\hspace*{0.5cm}RESTART\_POS\hspace*{0.5cm}\hspace*{0.5cm}\hspace*{0.5cm} T \\ 
\hspace*{0.5cm}RESTART\_VEL\hspace*{0.5cm}\hspace*{0.5cm}\hspace*{0.5cm} T \\ 
\hspace*{0.5cm}RESTART\_THERMOSTAT\hspace*{0.5cm} T \\ 
\& END EXT\_RESTART \\ 
@endif 
}

\clearpage
\bibliography{main}